\def\ispreprint{1}  

\if \ispreprint1
\documentclass[preprint,authoryear,5p]{elsarticle}  
\else
\documentclass[preprint,authoryear,12pt]{elsarticle}  
\fi

\usepackage[utf8]{inputenc}
\usepackage{lineno} 
\usepackage{amssymb}
\usepackage{amsmath}
\usepackage{graphicx,subcaption} 
\usepackage{hyperref} 
\usepackage{setspace}

\begin{document}

\title{Subsurface pulse, crater and ejecta asymmetry from oblique impacts into granular media}
\author[a1]{Bingcheng Suo\corref{cor1}}
\ead{bsuo@u.rochester.edu}
\author[a1]{A. C. Quillen}
\ead{alice.quillen@rochester.edu}
\author[a1]{Max Neiderbach}
\ead{mneiderb@u.rochester.edu}
\author[a1]{Luke O'Brient}
\ead{lobrient@u.rochester.edu}
\author[a1]{Abobakar Sediq Miakhel} 
\ead{amiakhel@u.rochester.edu}
\author[a1]{Nathan Skerrett}
\ead{nskerret@u.rochester.edu}
\author[a5]{J\'er\'emy Couturier}
\ead{jcouturi@ur.rochester.edu}
\author[a1]{Victor Lherm}
\ead{vlherm@ur.rochester.edu}
\author[a1]{Jiaxin Wang}
\ead{jwang198@u.rochester.edu}
\author[a4]{Hesam Askari}
\ead{askari@rochester.edu}
\author[a2]{Esteban Wright}
\ead{ewrigh10@umd.edu}
\author[a3]{Paul S\'anchez}
\ead{diego.sanchez-lana@colorado.edu}

\address[a1]{Department of Physics and Astronomy, University of Rochester, Rochester, NY 14627, USA}

\address[a5]{Department of Earth and Environmental Sciences, University of Rochester, 227 Hutchison Hall, Rochester, NY 14627, USA}

\address[a4]{Department of Mechanical Engineering, University of Rochester,  Rochester, NY 14627, USA}

\address[a2]{Institute for Physical Science and Technology, University of Maryland, College Park, USA}

\address[a3]{Colorado Center for Astrodynamics Research, The University of Colorado Boulder, 3775 Discovery Drive, 429 UCB - CCAR, Boulder, CO 80303, USA}

\cortext[cor1]{Corresponding author}

\begin{abstract}
We carry out experiments of 104 m/s velocity oblique impacts into a granular medium (sand). Impact craters have nearly round rims even at a grazing angle of about $10^\circ$, however, the strength of seismic pulses excited by the impact is dependent upon impact angle, and the ratio between uprange and downrange velocity peaks can be as large as 5, particularly at shallow depths.   
Crater slope, an offset between crater center and impact site, crater volume, azimuthal variation in ejection angle, seismic pulse shapes and subsurface flow direction are also sensitive to impact angle, but to a much lower degree than subsurface pulse strength.  Uprange and downrange pulse peak amplitudes can be estimated from the horizontal and vertical   components of the momentum imparted to the medium from the projectile.  

\end{abstract}

\maketitle

\if\ispreprint1
\else 
\linenumbers 
\fi






\section{Introduction}

Impacts on astronomical bodies often have projectile velocity vector that is not normal to the surface \citep{Gilbert_1893,shoemaker_1961}; with 
more than 50\% occurring at impact angles between 30 and 60$^\circ$ \citep{Melosh_2000}. Nevertheless, due to their cylindrical symmetry, laboratory impact experiments often focus on normal impacts \citep{tsimring05,goldman08,Matsue_2020,Quillen_2022,Cline_2022,Neiderbach_2023}.   
However, crater properties depend upon the impact angle.  For example,  the ratio of crater volume times substrate density to projectile mass, sometimes called `crater efficiency', is sensitive to the impact angle \citep{Gault_1978,Chapman_1986,Elbeshausen_2013,Michikami_2017,Takizawa_2020}.  
Except near grazing angles, oblique impact experiments and simulations show nearly round crater rims with subtle asymmetric deviations in crater shape  
\citep{Gault_1978,Anderson_2003,Anderson_2004,Wallis_2005,Raducan_2022}.
Ejecta mass, angle, and velocity distributions are sensitive to the azimuthal angle for oblique impacts \citep{Anderson_2003,Anderson_2004,Anderson_2006,Raducan_2022,Luo_2022}. 
Based on experiments into granular media, crater scaling laws for volume, diameter, and axis ratio in the gravity regime have recently been generalized or extended to take into account substrate slope and impact angle   \citep{Takizawa_2020}. 

Experimental studies of oblique impacts into a 2-dimensional photoelastic granular medium find that force-chain propagation in the horizontal (lateral) direction is relatively weak compared to that in the vertical direction \citep{Bester_2019}. 
Simulations in two dimensions of low-velocity oblique impacts find that there is a characteristic depth for disturbance within a granular substrate, which can be described with a skin depth \citep{Miklavcic_2022}.
Numerical simulations of low-velocity oblique impacts into a granular system showed that an empirical drag force model on the projectile is sensitive to impact angle 
\citep{Wang_2012}.  
\citet{Wright_2022,Wright_2020b} find that spherical projectiles, hitting sand at low impact velocity, ricochet depending upon the dimensionless Froude number that is related to the $\pi_2$ dimensionless parameter that is used to characterize impact crater dimensions \citep{housen11,Celik_2022}.  
Numerical simulations of low-g and low-velocity impacts into granular systems support Froude number scaling for oblique impacts \citep{Miklavcic_2023}. 

In a hypervelocity regime ($\gtrsim$ km/s) for impacts into solid rock \citep{Elbeshausen_2013,Michikami_2017,Takizawa_2020}, and for lower velocity impacts into sand  \citep{Takizawa_2020}, 
the ellipticity of the crater depends upon both impact angle and impact velocity. 
Crater size, volume, and shape are also sensitive to impact angle \citep{Gault_1978,Elbeshausen_2009,Elbeshausen_2013,Takizawa_2020}. 
Experiments and simulations measuring the angle of ejecta imply that the subsurface disturbance varies with azimuthal angle and is sensitive to impact angle \citep{Anderson_2003,Anderson_2004,Anderson_2006,Raducan_2022}.  Studies of hypervelocity impacts into solids have shown that asymmetries present in the shock wave produced during an oblique impact persist to late times, as measured in the far field \citep{Dahl_2001}. 
These studies imply that the momentum direction of the projectile affects the subsurface flow and the nature of a seismic disturbance excited by the impact. 

High-velocity impact craters into a solid can be divided into three sequential stages \citep{Melosh_1985, Melosh_1999}. In the first stage, the impact generates a shock that propagates radially outward from the site of impact and compresses material. 
A rarefaction wave propagates from the free surface, releasing the pressure in the compressed material.  This wave reduces the substrate velocity and changes the direction of the velocity vector \citep{Melosh_1985,Kurosawa_2019}. The deflected particle trajectories are curved so that they point upward toward the surface, generating the excavation flow of the second stage, and forming a transient crater.   An empirical analytical model that is frequently used to characterize flow during the excavation phase (e.g., \citealt{Croft_1981,Anderson_2006}) is Maxwell's Z-model \citep{Maxwell_1977}. 
In contrast with the longer time scale involved in crater excavation, energy, and momentum are rapidly transferred via shockwaves to the target material \citep{Melosh_1999}.
Lastly in the third stage, the crater is modified through even longer relaxation processes such as slumping and erosion.  

With embedded accelerometers, laboratory experiments of normal impacts into granular media have characterized the duration, strength, and decay rate of impact-excited seismic pulses 
\citep{yasui15,Matsue_2020,Quillen_2022,Neiderbach_2023}. 
At the time of peak acceleration, the seismic pulse vector orientation appears longitudinal (oriented along the radial vector from the impact site) \citep{yasui15,Quillen_2022}.
However, subsequently when the pulse velocity peaks, the subsurface velocity flow field resembles Maxwell's Z-model, leading to crater excavation  \citep{Neiderbach_2023}.  
Seismic pulse duration is significantly shorter than the crater excavation time-scale \citep{Quillen_2022,Neiderbach_2023}. 
Thus, laboratory experiments of normal impacts into granular media suggest that impacts into granular media exhibit three phases analogous to those of hypervelocity impacts into solids.  

Subsurface motions excited by laboratory impacts into solids differ from those excited by impacts into granular media in some ways. The duration 
of excited pulses is only a few microseconds in a solid (e.g, \citealt{Dahl_2001,Guldermeister_2017}) whereas the duration is longer, of order a ms, in a granular medium (e.g., \citealt{mcgarr69,yasui15,Matsue_2020,Quillen_2022}).
Whereas impact-excited seismic pulses traveling in a solid can exhibit a long coda and contain power at high frequencies \citep{mcgarr69,Dahl_2001}, those in granular systems tend to be comprised of a single, rapidly decaying pulse (e.g.,  \citealt{mcgarr69,yasui15,Matsue_2020,Quillen_2022,Neiderbach_2023}).
The difference in pulse duration in the two settings suggests that pulse duration is sensitive to the speed of sound or P-waves propagating through the medium.  An additional dependence of pulse duration on crater size (suggested by \citealt{Quillen_2022} with a seismic time-scale) would account for the similarity of the pulse durations in granular media in the lab for different experiments that have similar-sized craters.   

The detailed study of ejecta in oblique impacts into granular media \citep{Anderson_2003,Anderson_2004,Anderson_2006} inferred that there are asymmetries in the subsurface flow. 
Our goal in this manuscript is to measure and characterize the subsurface motions and their sensitivity to impact angle.  
Via tracking of ejecta, \citet{Anderson_2003,Anderson_2004,Anderson_2006} proposed variations of the Maxwell Z-model \citep{Maxwell_1977} for the subsurface flow field excited by oblique impacts.   
Through direct measurements of subsurface accelerations with accelerometers, we can test this type of flow model.

Even though most impacts occur at moderate impact angles, the paucity of elliptical craters on bodies such as the Moon and Mars implies that quite low impact angles,   $\theta_I \lesssim 10^\circ$ from horizontal, are required to form elliptical craters \citep{Bottke_2000,Collins_2011}.
As the angle required for an impact crater to be significantly asymmetric is sensitive to impact energy \citep{Collins_2011,Elbeshausen_2013,Takizawa_2020}, we are choosing to do experiments at a higher velocity, about 100 m/s, rather than the few m/s of our previous experiments \citep{Quillen_2022}.
Secondary craters formed by ejecta are typically at a lower velocity than the originating impact \citep{housen11}. 
Due to the different dynamical source populations, many of the crater-forming impacts on Transneptunian objects, such as (486958) Arrokoth, could have occurred at a few hundred m/s \citep{Mao_2021}.   

The goal of our experimental study is to better characterize the dynamics of oblique impacts into granular systems by measuring associated subsurface motions.  Our study will aid in predicting behavior of 
impacts on granular surfaces by landers \citep{Maurel_2018,Celik_2019,Ballouz_2021,Thuillet_2021}, and improve predictions for how impacts disturb the surfaces of astronomical bodies covered in and composed of granular materials. 
Understanding how seismicity is excited by oblique impacts is important in development of asteroid deflection strategies (e.g., \citealt{Rivkin_2021}).   By studying the subsurface motions, we hope to better understand the processes that cause asymmetry in crater shape and ejecta distributions. 

Following \citet{Elbeshausen_2013,Raducan_2022}, we denote the direction toward the projectile launcher as {\it uprange} and the opposite direction as {\it downrange}. 

\section{Experimental Methods}

Figure \ref{fig:expsetup} illustrates our experiments.
We carry out two types of experiments.
In a) we show impact experiments where we record acceleration in 3 axes from 4 embedded accelerometers. 
In b) we show how we measure crater profiles after impact by scanning a line laser across the crater. 
The surface is flat prior to impact in both types of experiments, and accelerometers are not embedded in the substrate in the experiments used to measure crater shape. 
The impact angle $\theta_I$ is shown in  Figure \ref{fig:expsetup} a) and is measured from the horizontal direction, so a low $\theta_I$ corresponds to a grazing impact.  

\subsection{Airsoft projectiles}

\begin{table}[ht]
    \centering
    \caption{Properties of the Projectiles}
    \begin{tabular}{lll}
    \hline
    Quantity &Symbol & Value \\
    \hline
     Mass     & $M_p$ & $0.20 \pm  0.002$ g \\
     Radius  &  $R_p$ & $2.98 \pm 0.005$ mm \\
     Density  & $\rho_p$ & 1.80 g cm$^{-3}$ \\
     Speed   & $v_{imp}$ & 104 $\pm 1 $ m/s \\
     Composition & \multicolumn{2}{l}{ Polylactic acid (PLA plastic) } \\
     \hline
    \end{tabular}
    \label{tab:airsoft_bbs}
\end{table}

The impact experiments use $M_p = 0.20$~g spherical projectiles (referred to as pellets or BBs) that are launched with an airsoft gun.
The projectile propellant is compressed carbon dioxide and comes in disposable cartridges.  The BB diameter is 5.95 $\pm$ 0.01 mm. 
The BBs are biodegradable and comprised of white polylactic acid (PLA). 
We have measured the velocity of the projectiles with high-speed video and find that the impact velocity lies within a narrow range; $v_{imp} = 103$ to 105 m/s. 
The velocity of the projectile is fixed and set by the propellant.  Consequently, in this study, we do not vary the impact velocity. 
The properties of the projectiles are summarized in Table \ref{tab:airsoft_bbs}.

We built a box to house the airsoft gun for two reasons. 
In an educational setting, it is important that something that looks like a firearm is not visible.  Secondly, to map out the subsurface pulse properties with a few accelerometers, we require the experiments to be repeatable.  By clamping the airsoft gun within the box, we can ensure that the BBs repeatably hit the same target position at the same angle. 

\subsection{Videos}
  
During each impact, we filmed with a Krontech Chronos 2.1 high-speed camera. An Arduino controls an actuator that is used to push the airsoft gun trigger so that it fires the BB.  The same Arduino is programmed to simultaneously trigger the recording of accelerometer signals by two oscilloscopes and the high-speed video camera. 

When doing experiments with the accelerometer array, videos from the high-speed video camera were
taken at 6265 frames per second (fps), with image frames 1920 $\times$ 120 pixels.
Since both high-speed video and accelerometers were
triggered simultaneously, 
the time of impact identified in the 6265 fps high-speed camera videos identifies the impact time in the accelerometer data. We estimate that the time of impact is accurate to about 0.2 ms.

Experiments used to measure crater profiles were taken along with high-speed videos at 1000 fps (with image frames 1920 $\times$ 1080 pixels) giving a wider field of view of the ejecta curtain compare to those taken in experiments with accelerometer arrays.  These videos are concatenated and available as supplemental video \texttt{Obliqueimpacts.mp4}. 
The impact was viewed with a near horizontal camera angle (about $7^\circ$ from horizontal) and was chosen to show the ejecta curtain.
These videos were used to verify the impact angle, determine whether the projectile
ricocheted and if so, measure the speed and angle of the projectile after rebound.  
These videos were also used to identify the site of impact with respect to the green laser target that illuminates the sand from above and which was also viewable during the laser scans.

Experiments used to measure crater profiles were filmed (after impact) with a regular video camera (Blackmagic Pocket Cinema Camera 4K) at 60 fps while scanning a line laser across the crater.  The camera viewing angle was $45^\circ$ from horizontal. Video frames from this camera are 3840 $\times$ 2160 pixels.  The same camera was also used to photograph the craters from above. 

High-speed videos for both sets of experiments were filmed with a bright (100,000 lumen) LED light at about $45^\circ$ from vertical.  Crater photographs viewed from above the crater were lit with the same light. 
To best show the crater rim we lit the crater with nearly horizontal light coming from the downrange side. 
Laser scan videos were done in ambient room light so that the laser line is clearly visible in each frame. 

The pixel scales in the 60 fps, 1000 fps videos and in the photographs are measured from a machined aluminum block placed on the surface that is 25.4mm long and wide.

Experiments measuring crater profiles were carried out early June 2023, and those using accelerometers were carried out during June and July 2023.

\begin{figure*}[ht]
    \centering
    \includegraphics[width = 3 truein ]{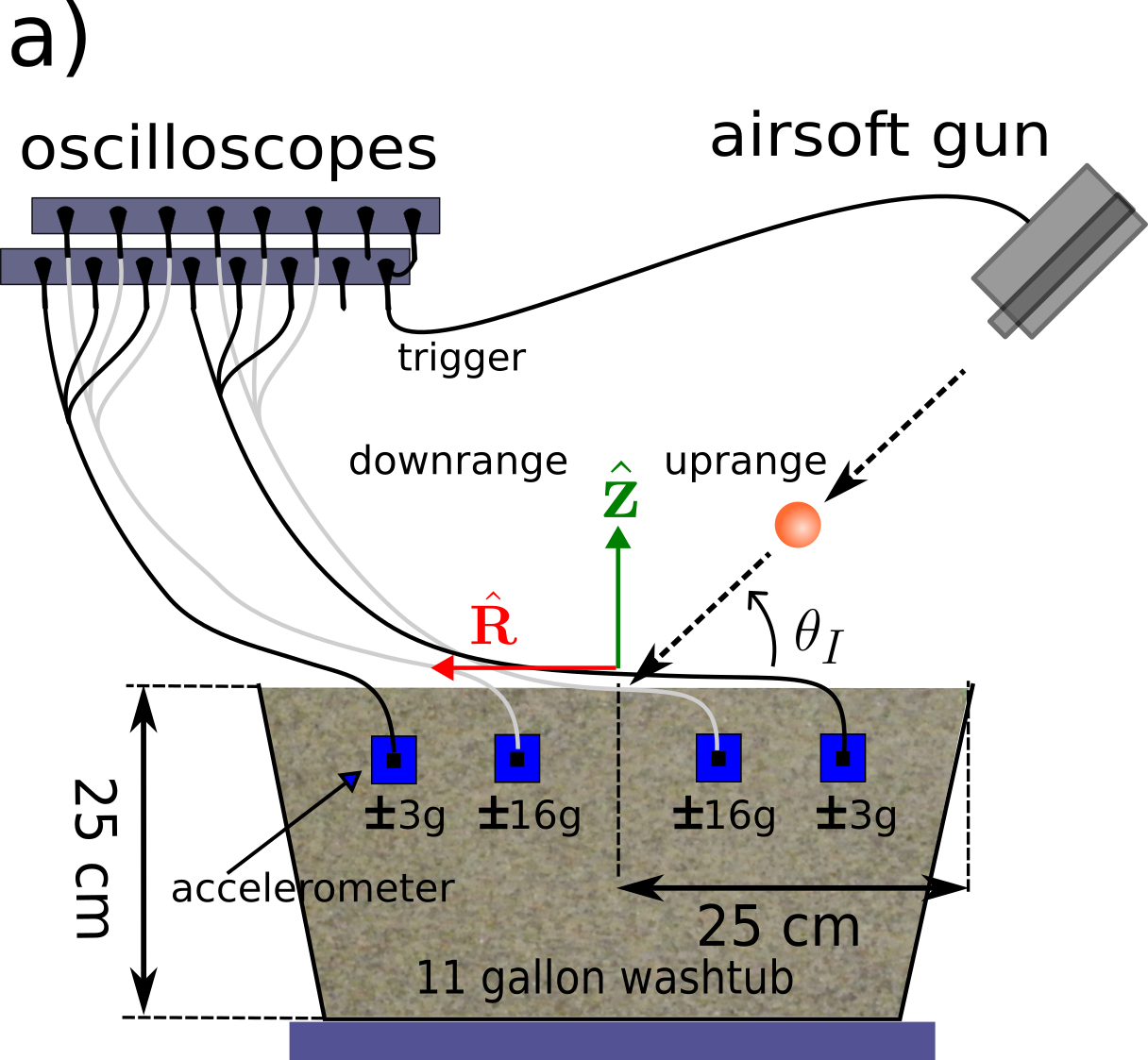}
    \includegraphics[width = 2.0 truein]{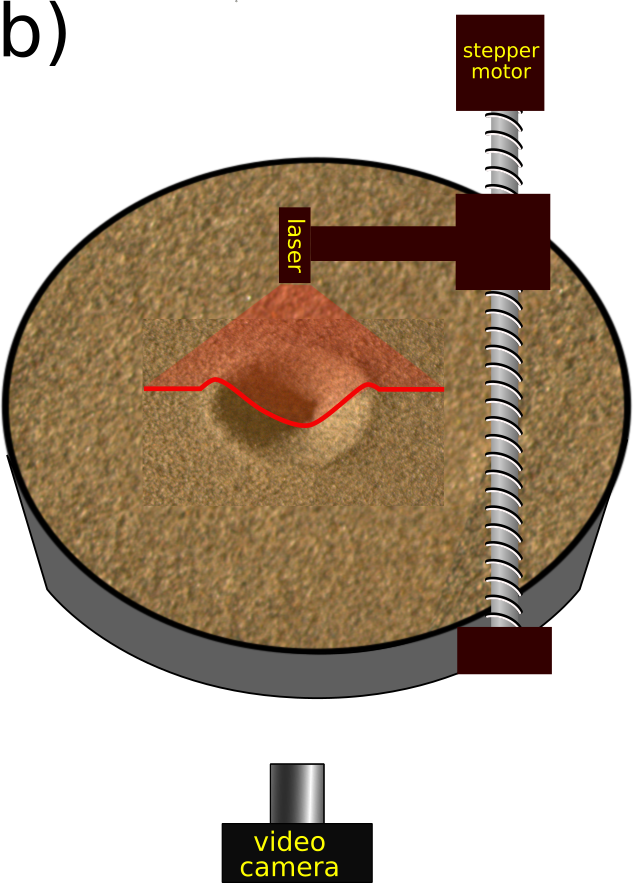}
    \caption{Illustrations of the experiments. a) We show the accelerometer array and the airsoft gun.  There are three oscilloscope channels per accelerometer because we record all three acceleration axes.  Accelerometer boards are shown larger than their actual scale.  b) We show the laser line that is used to measure crater profiles.  A stepper motor is used to slowly scan the laser line across the crater while the video camera records images. During the impact experiments, we also record high-speed video. }
    \label{fig:expsetup}
\end{figure*}

\subsection{Granular substrate target}

We use a galvanized 41.6 liter (11 gallons) washtub with a rim diameter of 50.2 cm and depth of 25 cm to hold our granular material.
The tub is filled with sand with a bulk density of $\rho_s = 1.5$\ g cm$^{-3}$.  
We use sand so that the ratio between projectile to grain radius is large ($\sim 10$), but the grains are not as small as in a powder which could be affected by aerodynamics and electrostatic phenomena. 
The grain semi-major axis mean value is $a_s \approx 0.3$~mm as measured in previous experiments  \citep{Wright_2020b}.

The sand in the tub is raked and leveled prior to every impact experiment to reduce local compaction caused by previous impacts.
Rake tongs are 10 cm long and 4 cm apart. 

Two green lasers are used to set the impact target point.  These two laser lines are mounted above the tub and mark the center of the tub.  Two additional red lasers are mounted on the airsoft gun to help us aim it.  The impact angle is measured using the red lasers and a large protractor prior to each impact.  

\begin{table}[ht]
    \centering
    \caption{Physical values and dimensionless numbers for experiments}
    \begin{tabular}{lcl}
    \hline
         Substrate & sand & \\
         Sand grain size & $a_s$ & 0.3 mm \\
         Substrate bulk density & $\rho_s$ & 1.5 g cm$^{-3}$  \\ 
         Washtub rim radius & $R_{\rm tub}$ & 25.1 cm\\
         Washtub depth & $H_{\rm tub}$ & 25 cm\\
         $\pi_2$ & $g R_p/v_{\rm imp}^2$ & $2.7\times 10^{-6}$ \\
         $Fr$ & $v_{\rm imp}/\sqrt{g R_p}$ & 608 \\
        $\pi_3$ (for $Y=500$ Pa) & $Y/(\rho_s v_{\rm imp}^2)$ & $3 \times 10^{-5}$ \\
         $\pi_4$ & $\rho_s/\rho_p$ & 0.83  \\
         $\tau_{ex}$ & $\sqrt{R_{cr}/g}$ & 60 ms \\
    \hline
    \end{tabular}
    \label{tab:exp_constants}
\end{table}

\subsection{Dimensionless parameters}

Experiment parameters are summarized in Table \ref{tab:exp_constants}
where we include the dimensionless parameters $\pi_2 \equiv g R_p/v_{imp}^2$ and $\pi_4 \equiv \rho_s/\rho_p$.   
We also compute dimensionless parameter $\pi_3 = \frac{Y}{\rho_s v_{imp}^2}$ using a bulk cohesion material strength of $Y = 500$ Pa based on measurements of regolith by 
\citet{Brisset_2022}. 
The Froude number is $Fr = \pi_2^{-\frac{1}{2}}$.
These dimensionless parameters are commonly used in crater scaling relationships \citep{housen11,Celik_2022}.
Here $R_p, \rho_p, M_p$ refer to projectile radius, density, and mass, $v_{imp}$ is projectile impact velocity, and $\rho_s$ is mean substrate density.  The gravitational acceleration is $g$. 
For a normal impact, we measure the distance 
from rim peak to rim peak giving diameter $D_{cr} = 7$ cm and a radius
$R_{cr} =3.5 $ cm.  This is used to estimate the time for transient crater formation $\tau_{ex} = \sqrt{R_{cr}/g}$ \citep{Housen_1983,Melosh_1985}.  A crater is in the strength regime if  $\pi_3^{1+ \mu/2} \pi_4^\nu/\pi_2 > 1$ with exponents $\mu \sim 0.4$ typical of granular systems and $\nu \sim 0.4$  \citep{housen11}.
For our experiments $\pi_3^{1.2} \pi_4^{0.4}/\pi_2 \sim 1 $  is similar to unity, so our experiments lie near the division between strength and gravity regimes for crater formation \citep{holsapple93,Scheeres_2010}. 

\subsection{Accelerometers and accelerometer placement}
\label{sec:acc_exp} 

The accelerometers are 5V-ready analog break-out boards 
and house a $\pm16$g (ADXL326)  or $\pm 3$g (ADXL335) triple-axis accelerometer Analog devices integrated circuits,  as described previously  \citep{Neiderbach_2023,Quillen_2022}. 
The dimensions of the accelerometer printed circuit boards (PCBs) are 19 mm $\times$ 19 mm $\times$ 3 mm.
The accelerations we measured are integrated over this area, not those experienced by individual sand grains. 
The accelerometer's 1600 Hz bandwidth upper limit for its x and y axes corresponds to a half period of 0.3 ms which is shorter than the width of the acceleration pulses seen in our experiments.  The bandwidth upper limit on the z-axis is lower at 550 Hz. 
The bandpass upper limits are frequencies at which the signal amplitude is reduced by 3 dB (the amplitude drops by a factor of 0.5) and approximately equal to the cutoff frequency of a low pass filter. We estimate the rms noise in the signal to be ranging from 0.07 to 0.14 $m/s^2$.
The output signals of all three axis outputs of the accelerometers were recorded with two 8-channel digital oscilloscopes (Picoscope model 4824A) with a 200kHz sampling rate.  

A long straight metal bar is used to level the surface after raking the sand. The accelerometers are then embedded into the substrate.
We orient the accelerometers so that their $+x$ axes points away from the impact site and their $+y$ axes point vertically up.  This leaves the lower bandwidth $z$ axis orientated in a direction that should give a weaker signal. 
Accelerometer calibration values were determined by measuring the voltage along each axis at six different cardinal orientations giving accelerations of $\pm$1g due to gravity. 

To ensure that the accelerometers are correctly spaced, are at the desired depth, and are correctly oriented, we individually placed each accelerometer in the sand.
Tweezers were used to embed the accelerometers.
The tweezer prongs are marked at centimeter intervals along their length so we can set the accelerometer depth.
The DC voltage levels of each accelerometer were monitored in all three axes during placement to monitor their orientation.
We compared the DC voltage levels of the accelerometer signals prior to impact to the calibration values and find that the accelerometers, once embedded, are typically within $10^\circ$ of the desired orientation.
We compute velocity as a function of time by numerically integrating the acceleration signal in all three components. The drift rate of velocity pulses, due to the rms noise in acceleration signals and integration errors, is around $10^{-4}$ $m/s$ per millisecond.
Because of the uncertainties in accelerometer orientation, it is more robust to use acceleration or velocity magnitudes than individual radial or vertical components. 
Because of the rapid decay of impact excited pulses as a function of on distance from the impact site \citep{Quillen_2022,Neiderbach_2023}, uncertainties in acceleration or velocity are dominated by a few mm errors in accelerometer placement with respect to the actual site of impact, rather than errors in orientation.   
Sometimes the projectile hits a wire during a ricochet causing a spurious spike in the accelerometer signals. 

Coordinate directions for the experiments differ from those used to describe the accelerometer.   We show cylindrical radius, $R$ from
the site of impact and $z$, giving height above the surface in Figure \ref{fig:expsetup}a.  Embedded accelerometers have $z<0$ and 
the distance in spherical coordinates from the site of impact $r = \sqrt{R^2 + z^2}$.  In both spherical coordinates and cylindrical coordinates, we use $\phi$ to represent the azimuthal angle. 
Accelerometer location with respect to the site of impact is measured from the position of the accelerometer integrated circuit which is located at the center of the PCB and is oriented vertically during the experiments.

For each individual impact experiment with accelerometer arrays, 4 accelerometers are embedded in the sand, as shown in Figure \ref{fig:expsetup}a.  Taking into account 3 axes per accelerometer and 2 trigger channels, we use 14 channels of the total 16 possible with our two 8-channel oscilloscopes. 

Each individual impact experiment is done with accelerometers at the same depth, with $|z| = 3,5$ or 7 cm. During each impact experiment we record 4 accelerometers, with 2 uprange and 2 downrange of the impact, 
as shown in Figure \ref{fig:expsetup}a.  Each uprange or downrange pair is at two radii, either at $R=6$ and 10 cm, $R=7$ and 11 cm, or $R=8$ and 12 cm.  The two positions
nearest the impact site are recorded with $\pm$16g accelerometers to prevent signals from over-saturation and the further two positions are recorded with $\pm$3g  accelerometers. 
Accelerometer positions are summarized in Table \ref{tab:acctemplate} in the form of templates. 
Each row in this table corresponds to a separate impact experiment as 4 accelerometers were used in each experiment. 
Impact experiments with accelerometers were 
carried out at 9 different impact angles, with $\theta_I = 10^\circ$ to and including 
90$^\circ$ in intervals of $10^\circ$. 
The list of experiments at each impact angle is summarized in Table \ref{tab:template}.
At $\theta_I = 20, 40$ and $60^\circ$ we did experiments
at three depths and 6 radii (as summarized with Template 1 in Table \ref{tab:acctemplate}). At other impact angles, we carried out 
fewer experiments (following Templates 2 and 3). 
Template 1 shows a relatively full sampling of subsurface positions at 3 depths and 6 radial distances at both uprange and downrange positions.  Sets of experiments using Template 1 are used to examine pulse velocity direction and peak values as a function of position. Accelerometer data from the different impact experiments at the same impact angle are combined to study the sensitivity of pulse amplitude with distance from the impact site, azimuthal angle, and depth.  
Templates 2 and 3, in addition to Template 1, are used to study the sensitivity of pulse strength and shape at specific positions as a function of impact angle.  

\begin{table}[htbp]
\caption{Accelerometer placement templates ($R,z,\phi$)}
\label{tab:acctemplate}
\begin{center}
\begin{tabular}{lccccc}
\hline
Template\!\! & $\pm3$g & $\pm 16$g & $\pm 16$g & $\pm 3$g \\
\hline
Template 1 & (10,-3,0) & (6,-3,0) & (-6,-3,180) & (-10,-3,180) \\
           & (11,-3,0) & (7,-3,0) & (-7,-3,180) & (-11,-3,180) \\
           & (12,-3,0) & (8,-3,0) & (-8,-3,180) & (-12,-3,180) \\
           & (10,-5,0) & (6,-5,0) & (-6,-5,180) & (-10,-5,180) \\
           & (11,-5,0) & (7,-5,0) & (-7,-5,180) & (-11,-5,180) \\
           & (12,-5,0) & (8,-5,0) & (-8,-5,180) & (-12,-5,180) \\
           & (10,-7,0) & (6,-7,0) & (-6,-7,180) & (-10,-7,180) \\
           & (11,-7,0) & (7,-7,0) & (-7,-7,180) & (-11,-7,180) \\
           & (12,-7,0) & (8,-7,0) & (-8,-7,180) & (-12,-7,180) \\
Template 2 & (10,-3,0) & (6,-3,0) & (-6,-3,180) & (-10,-3,180) \\
Template 3 & (10,-5,0) & (6,-5,0) & (-6,-5,180) & (-10,-5,180)\\
\hline
\end{tabular}
\end{center}
\par {\footnotesize
\begin{singlespace}
    Notes: In the top row we show which type of accelerometer is used at each position. Each row shows the position for placing the accelerometer $(R,z,\phi)$ in cylindrical coordinates and in units of cm and degrees. The azimuthal angle $\phi=0$ is a  position directly downrange of the impact site and $\phi=180$ is uprange of the impact site. The site of impact is at $(R,z) = (0,0)$. 
\end{singlespace}
}
\end{table}

\begin{table}
\caption{Impact experiments with accelerometers}
\label{tab:template}
\begin{center}
\begin{tabular}{l c c c c c c}
\hline
Impact angles $\theta_I$ & Templates & Figures \\
\hline
10,30,50,70$^\circ$ & 2,3  & \ref{fig:pulse_asym}, \ref{fig:pshape} \\
20,40,60$^\circ$ &  1 & \ref{fig:pulse_asym} -- \ref{fig:pulseshapes}, \ref{fig:rescale} \\
80$^\circ$ & 2,3  & \ref{fig:pulse_asym}, \ref{fig:pshape}, \ref{fig:rescale} \\
90$^\circ$ &  3 & \ref{fig:pulse_asym}, \ref{fig:pshape} \\
\hline
\end{tabular}\end{center}
\par {\footnotesize
\begin{singlespace}
At each impact angle, the number of experiments is based on the template, with templates listed in Table \ref{tab:acctemplate}. An impact experiment was done for each row in Table \ref{tab:acctemplate} for the given template. 
\end{singlespace}}

\end{table}

\subsection{Measuring crater profiles}

Experiments used for measuring crater profiles were done without embedding the accelerometers to reduce disturbance of the substrate prior to impact.  
After the impact we scanned the crater with a moving red line laser, as shown in Figure \ref{fig:expsetup}b.
The red line laser is mounted to a stepper motor-controlled worm drive and is oriented vertically so that the crater is illuminated with a vertical sheet of red light.  We drove the worm drive to move the laser line slowly and smoothly across the crater at 2~mm per second while filming at 60 fps.  

To measure crater profiles we use
a Cartesian coordinate system with $x,y$ coordinates in the 
horizontal plane and $+z$ upward.  The laser line lies in the $x,z$ plane. As the laser scans, it moves in the $y$ direction. The $+x$ direction is uprange, toward the airsoft gun. 
In each video frame, the horizontal direction is along the x-axis.  
The vertical direction in the camera images is converted to the 
$z$ coordinate by dividing by the cosine of the camera viewing angle ($\cos (45^\circ)$). The $z$ coordinate was then shifted 
so that the undisturbed surface outside the crater has $z=0$.  The $x$ and $y$ coordinates are shifted so that they are zero at the site of impact. We began each 60 fps video with a view of the green laser target which is also visible in the high-speed video.  We measure the site of impact from the 1000 fps high-speed video and use the distance between the impact site and the green target to estimate the site of impact in the 60 fps video.  This comparison allowed us to estimate the site of impact, which is not necessarily the same as the deepest point or center of the crater. 

From the 60 fps video of the scanned red laser, we extracted video frames every 1/30 of a second.
In each vertical column of each video frame, the location of the maximum red pixel gave us the position of the laser line in that pixel column. We fit a line to the laser line on the left and right sides, outside the crater, and subtracted it to make sure that crater depth was measured from the level of the undisturbed surface.  In each frame, we measured crater depth along a particular vertical $y$ coordinate. 
The values of the $y$ coordinate in each frame were determined by measuring the extent of laser line translation from one side of the crater to the other and assuming that the scan rate (set by the worm gear motor) is constant. 
By combining profiles from each video frame showing the laser line at a different position, we measure the crater depth as a function of the $x$,$y$ position. 

The local surface slope in degrees was computed from the $x$ and $y$ gradients of the $d(x,y)$ function giving crater depth.
With $\nabla d(x,y) = (\frac{\partial d}{\partial x},  \frac{\partial d}{\partial y})$ the slope is  $s = \arctan$  $  |\nabla d(x,y)|$.  

\begin{figure*}[htbp]\centering
\if \ispreprint1
\includegraphics[width = 6 truein, trim = 1 0 1 0, clip]{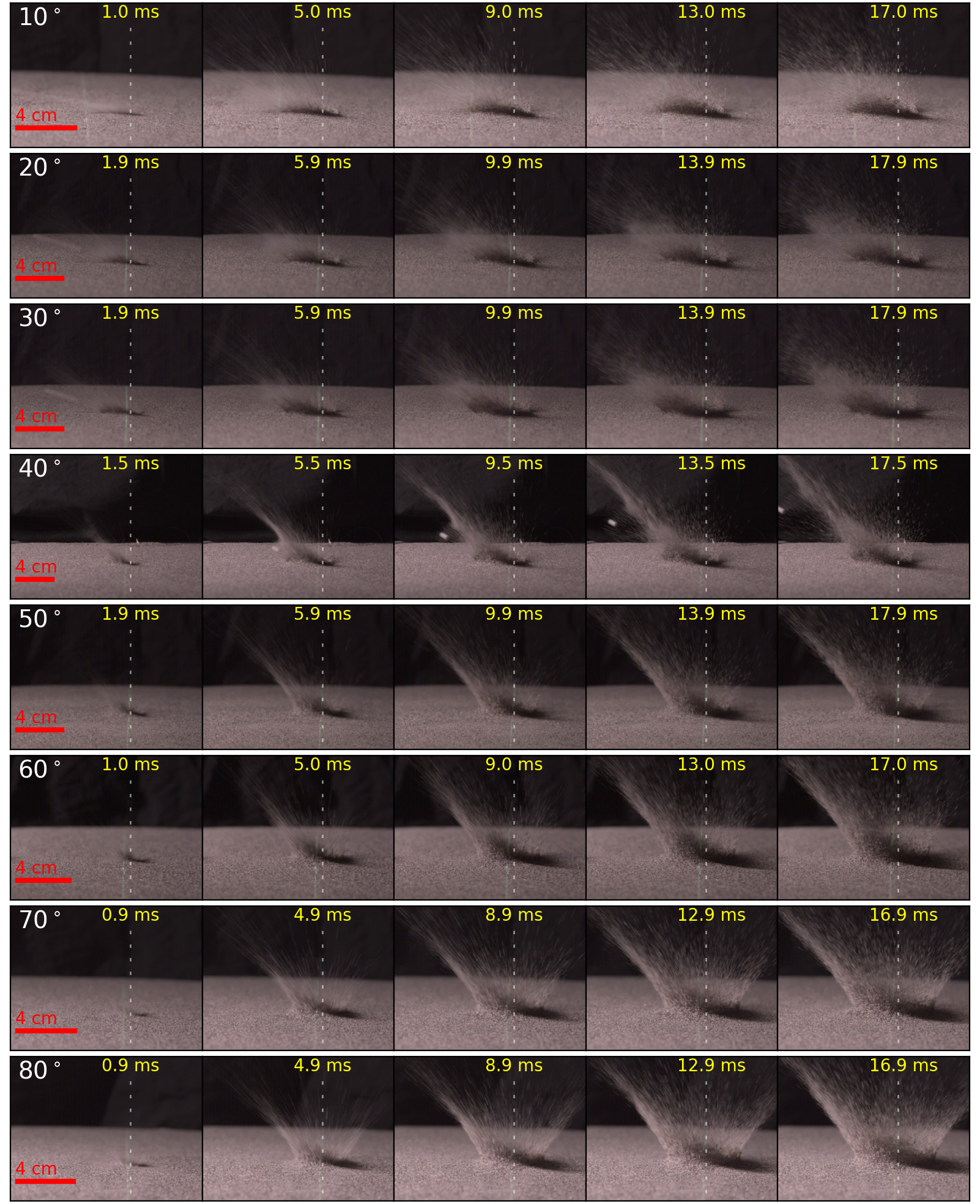}
\else
\includegraphics[width = 5 truein, trim = 1 0 1 0, clip]{A_ejecta.png}
\fi
\caption{Frames from 1000 fps videos showing the ejecta curtains for different impact angles.  In some cases, the projectile is seen ricocheting off the surface.  At the top of each frame, we label time from impact.  A 4~cm scale bar is shown in the left panels.  The projectile came from the right. The impact angle is labeled in the left panels. The angle of the ejecta curtain increases as a function of the impact angle. The ejecta curtains are asymmetric.  In each experiment, the site of impact lies on the dotted line. \label{fig:ejecta}}
\end{figure*}

\subsection{Ejecta angles}
\label{sec:hog}

While filming impacts, the target was brightly illuminated, 
nevertheless ejecta travels across more than a few pixels during each exposure in the 1000 fps high-speed videos. Because particles move more than a few pixels during a ms, ejecta particles appear as streaks that are aligned in the direction of motion. 
To measure the orientation of ejecta particle motions we compute local histograms of oriented gradients (HOG).
This type of histogram is commonly used in object recognition software \citep{Dalal_2005}.
In each 22$\times$22 pixel square cell (about 3.5$\times3.5$ mm) in a single video frame, we compute histograms of oriented gradients
with the \texttt{hog} routine that is part of the image processing python package \texttt{scikit-image}.
We use unsigned gradients so orientation angles lie between $[-90^\circ, 90^\circ]$.

\section{Experiments measuring crater shape and showing ejecta}

We carried out a series of impact experiments, at 8 different impact angles separated by $10^\circ$, to measure crater profiles.  
These same experiments were used to look at the ejecta curtains.  This set of experiments is shown 
in Figures \ref{fig:ejecta} -- \ref{fig:eff}.  
At each impact angle, a row of panels in Figure \ref{fig:ejecta}, \ref{fig:quiver}, \ref{fig:contour},  and \ref{fig:slopehist} shows the same impact experiment. 

\subsection{Ejecta curtains}
\label{sec:ejecta}

\begin{figure}[htbp]\centering

\if \ispreprint1
\includegraphics[width = 3.3 truein,trim = 0 10 0 10,clip]{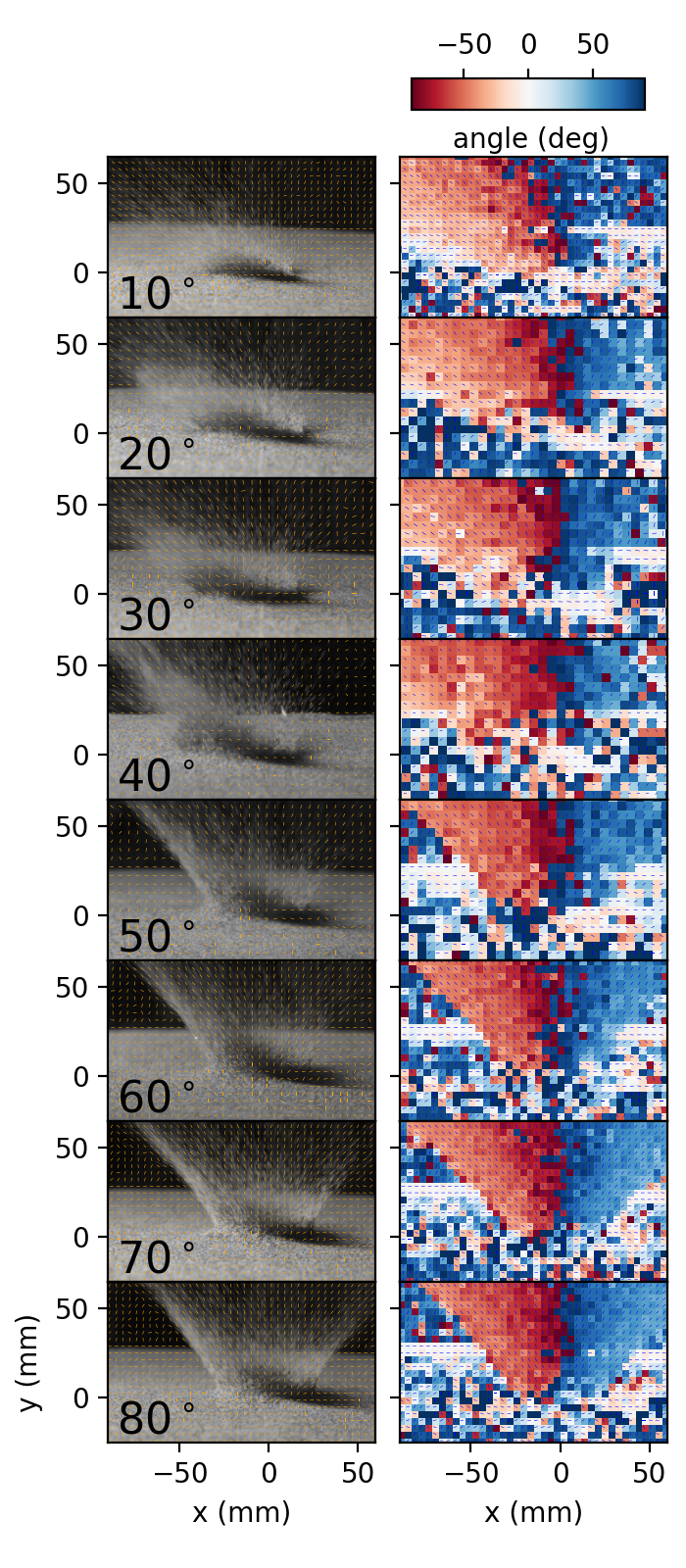}
\else
\includegraphics[width = 3.0 truein,trim = 0 10 0 10,clip]{A_quiver.png}
\fi
\caption{Histograms of oriented gradients are shown in right panels on ejecta curtain images 15 ms after impact. The impact angle is shown on the left panels.  Ejecta angle is shallower at grazing impact angles than at near-normal impact angles.  \label{fig:quiver}}
\end{figure}

Frames from 1000 fps videos are shown in Figure \ref{fig:ejecta}. 
Each row in this Figure shows an experiment at a different impact angle. The estimated time of each frame from the moment of impact (with an accuracy of about 1 ms) is shown on the top of each frame. The dashed white lines show the estimated horizontal coordinate of the estimated site of impact.  The projectile came from the right, so the uprange is to the right. 

As described in section \ref{sec:hog}, in image frames 15 ms after impact, the average direction for ejecta motion was computed using histograms of oriented gradients.  
These are plotted as tan segments on top of the original video frame in the left column of Figure \ref{fig:quiver}.  In the right column of 
Figure \ref{fig:quiver}, the orientation angle of each segment is shown in color with color-bar on the top.  White corresponds to a horizontal orientation.

Figure \ref{fig:ejecta} and \ref{fig:quiver} illustrate that the ejecta leaves the surface at a higher angle (from horizontal) for near-normal impacts than for grazing impacts.  The edge of the ejecta curtains appears more diffuse at grazing impact angles.  The ejecta angles are asymmetric with steeper ejecta angle uprange (to the right in Figure \ref{fig:ejecta} and toward the projectile launcher) than downrange (to the left and away from the projectile launcher).  The ejecta curtains are more massive on the downrange side.     
Using particle tracking methods on particles at all azimuthal angles, \citet{Anderson_2004} found that ejecta angle was $\sim 15^\circ$ lower downrange for a $\theta_I = 30^\circ$ impact into the sand than for a normal impact.   As our Figures \ref{fig:ejecta} and \ref{fig:quiver} primarily show ejected material downrange of the impact site, the trend we see of ejecta angle increasing with increasing impact angle is consistent with their study.

\subsection{Crater shapes}
\label{sec:shapes}

\begin{figure*}[htbp]\centering
\if \ispreprint1
\includegraphics[height = 8.5 truein,trim = 20 10 60 5,clip]{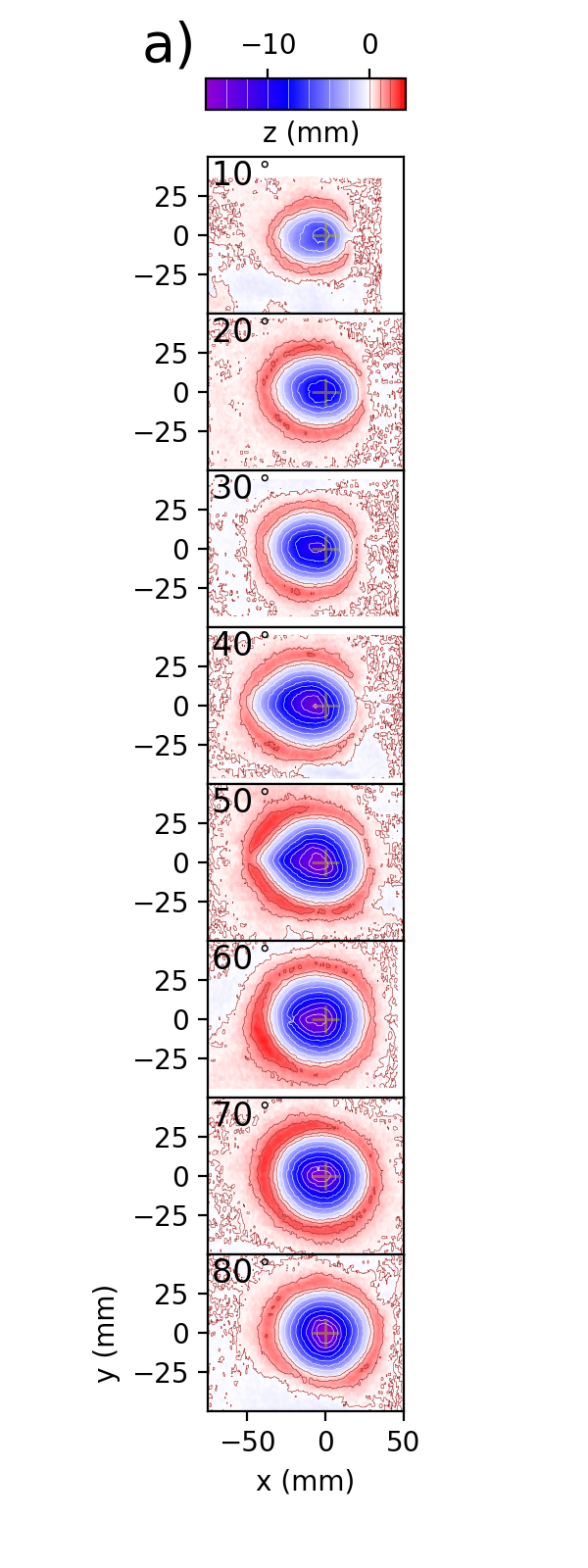}
\includegraphics[height = 8.5 truein,trim = 50 10 60 5,clip]{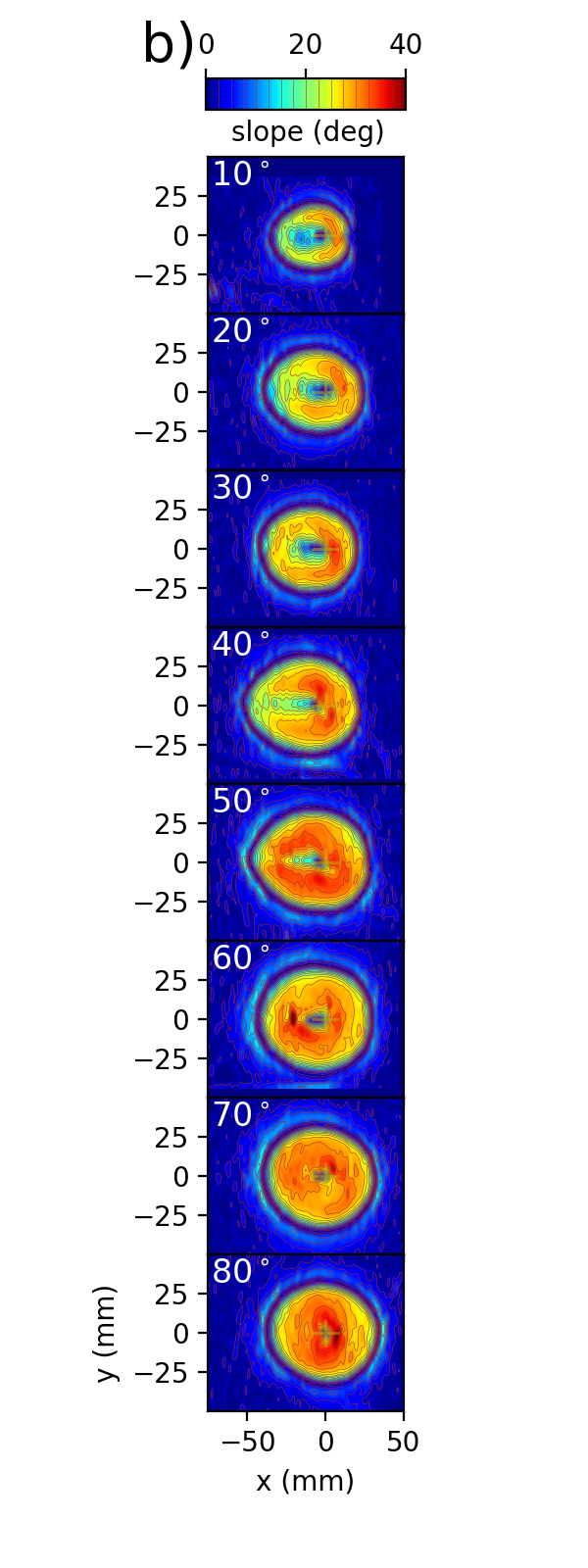}
\includegraphics[height = 8.5 truein,trim = 50 10 60 5,clip]{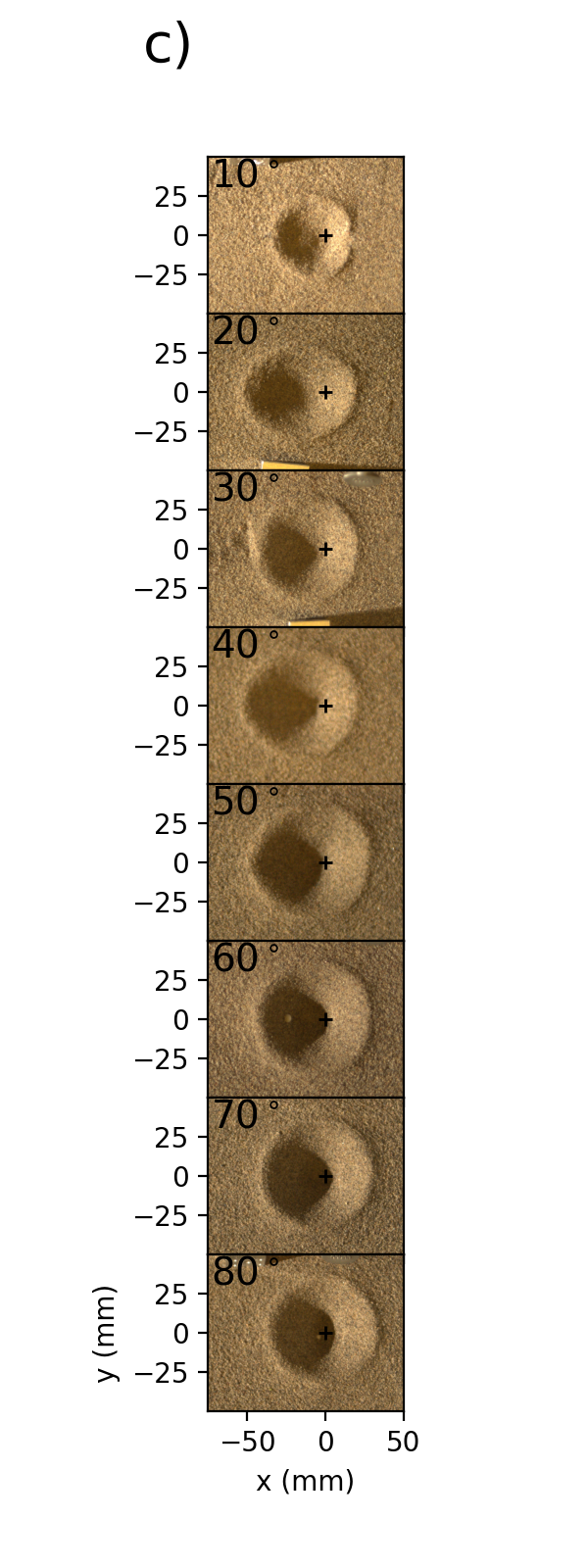}
\else
\includegraphics[height = 6 truein,trim = 30 10 60 5,clip]{A_contour.png}
\includegraphics[height = 6 truein,trim = 30 10 60 5,clip]{A_slope.png}
\includegraphics[height = 6 truein,trim = 30 10 60 5,clip]{A_still.png}
\fi
\caption{Crater shapes as a function of impact angle. In a) (the left panels) we show depth with a color bar to the top. In b) (the middle panels) we show the slope with a color bar at the top.  In c) (the right panels) we show photographs of the crater, taken from above.  Each row shows to a single impact at an impact angle which is shown on the upper left of each panel.  The projectile came from the right, so uprange is the +x direction. The site of impact is at the origin and marked with a plus sign in a), and c).  
 \label{fig:contour}}
\end{figure*}

For the same 8 experiments described in section \ref{sec:ejecta}, we measured crater profiles. 
The crater depth profiles are shown in the left column of Figure \ref{fig:contour}.
Local slopes are shown in the middle column of Figure \ref{fig:contour}.
After scanning the crater with the laser line, we took a single photograph of the crater from above. 
These photographs are shown in the right column in Figure \ref{fig:contour}. 

Features in the crater profiles and slopes 
are in some cases due to the projectile when it ricochets.  For example, a narrow low slope region along the major axis evident in the slope in Figure \ref{fig:contour}b at $\theta_I =40^\circ$ is probably due to the projectile which passed through the ejecta curtain as it ricocheted (see Figure \ref{fig:ejecta}).
Features in crater morphology associated with ricochet were also seen in simulations of hypervelocity impacts \citep{Elbeshausen_2013}. 

\begin{figure}[htbp]\centering
\includegraphics[width=3.2truein,trim = 0 0 0 0,clip]{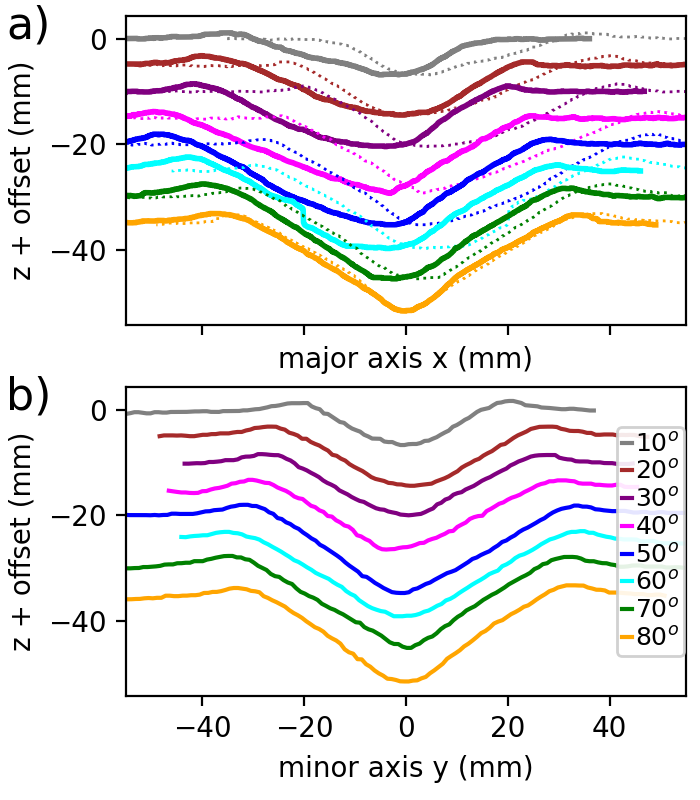}
\caption{Major and minor axis crater depth profiles.  a) Major axis crater profiles, with the projectile originating from the $+x$ direction. Solid lines show the profile, and dotted ones show a mirror image of the profile, reflected about $x=0$, the site of impact. Asymmetry in the profiles can  be seen by comparing dotted and solid lines in the top panel.   
Profiles have depth in mm but they are consecutively offset vertically by 5~mm and shown in order of impact angle with higher impact angles on the bottom.  
Crater symmetry depends upon impact angle with more asymmetric craters at lower impact angle. After impact in the experiment with $\theta_I=60^\circ$, the projectile itself lay inside the crater which is why the major axis profile has a bump at $x\approx -20$ mm.    
b) Similar to a) except showing minor axis profiles.  The key shows the impact angle, $\theta_I$, for profiles in both a) and b). 
\label{fig:profiles}}
\end{figure}

\begin{figure}[htbp]\centering
\includegraphics[width=3.5truein,trim = 0 20 0 0, clip]{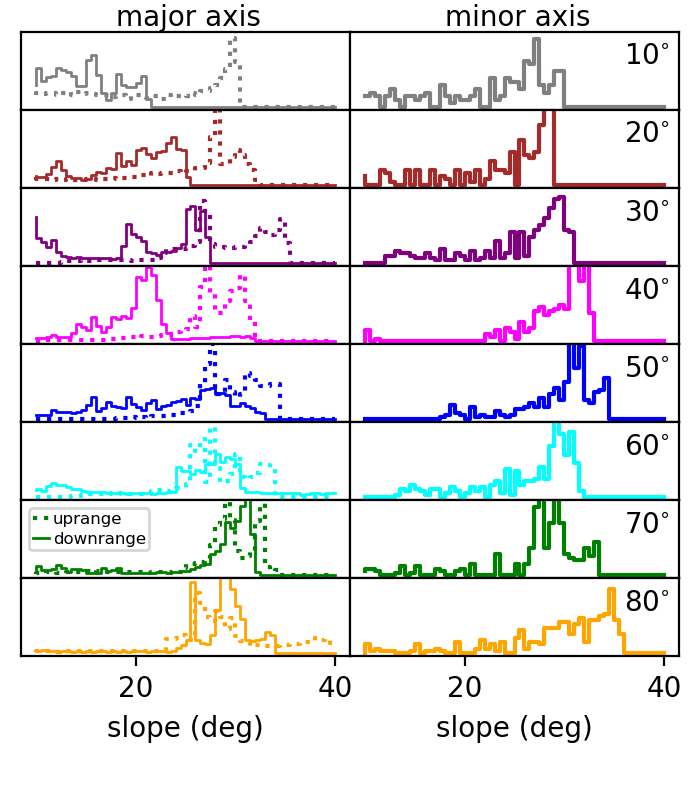}
\caption{Histograms of slopes are computed along the major and minor axes of craters made at different impact angles.  The left panel shows histograms along the major axis with solid lines showing histograms for the downrange side and dotted lines showing histograms for the uprange side.  The right panels show histograms of slopes along the minor axis. Impact angles are labeled on the upper right in each panel of the right column. The surface slopes within the crater are steeper at higher impact angles. At low-impact angles ($\theta_I$ below $40^\circ$), the slope is lower on the downrange side.  \label{fig:slopehist} }
\end{figure}

Figure \ref{fig:profiles} shows major and minor axis profiles and Figure \ref{fig:slopehist} shows histograms of slope values along the major and minor axes.  In Figure \ref{fig:profiles}  the minor axis profiles for different impact angles are shown in the bottom panel and the major axis profiles are shown in the top panel.  Uprange is to the right. Profiles are offset vertically so that they can be compared.  In the left panels and with dotted lines, we show a horizontally reflected version of the major axis profile to make it clear that the oblique impact craters are not symmetrical about the site of impact which is approximately at $x=0$. 
Shallow slopes on the downrange side of a crater caused by an oblique impact were previously seen in the high velocity impacts into pumice dust by \citet{Gault_1978}. 

To characterize the distribution of slopes within the craters,  in Figure \ref{fig:slopehist} we show histograms of slopes.  In this figure, each row shows a different impact angle.  Dotted and solid lines in the left column show histograms of the slope along the crater major axis uprange and downrange, respectively.  
The right column shows histograms of slope values taken along the crater minor axis.  The histograms show that the surface slopes tend to be higher at higher impact angles.  For low impact angles $\theta_I<50^\circ$, the uprange side of the crater is steeper than the downrange side. 


Figure \ref{fig:profiles} illustrates that the crater profiles have regions with nearly constant slopes.   A cone with a point directed down is a surface with a constant slope, whereas a bowl-shaped surface has a shallow slope at the bottom and steep slopes along its rim.  
When are impact craters more nearly conical rather than bowl-shaped? We find that for $\theta_I\gtrsim 40^\circ$, the craters are more sharp-bottomed while for more grazing impact angles, the bottoms are more round. 

Based on shock models for simple (non-complex) craters into a solid, we expect that craters should be bowl-shaped \citep{Melosh_1989}.  This is confirmed via experiments of impacts into solids \citep{Turtle_2005}.
However, a bowl-shaped crater has a high surface slope just inside its rim.
In a granular material, if the transient crater is initially bowl-shaped, the surface slope would be above the static angle of repose of the medium, near but within the crater rim. For our sand, the static angle of repose (the angle of the steepest sand pile that is stable) is approximately $35^\circ$.  After excavation of a transient crater in a granular system, material near the rim slides downward toward the crater center (as seen in videos from above, E. Wright private communication).  
The collapse near the rim would give a more conical-shaped crater with a slope near the static angle of repose \citep{Yamamoto_2006}.   However, we see an impact angle dependent crater slope distribution and uprange/downrange asymmetry in crater slope (as seen in the histograms of Figure \ref{fig:slopehist}), so while crater slopes are near the static angle of repose, we also see variations in slope. Perhaps a late phase of deformation (e.g., \citealt{Neiderbach_2023}) further disturbs the substrate material, causing the material to slide further and reducing the slope to below the static angle of repose (e.g., \citealt{Carrigy_1970}). 
The final slope angle would be more similar to a lower angle,  called the dynamic angle of repose,  which is measured on a granular system following a landslide \citep{Kleinhans_2011}. 
If this occurs preferentially on the downrange side and for grazing impacts, then we might also account for the shallower slopes seen in these two settings. Alternatively, crater excavation in a granular system may differ from that in a solid, giving a transient crater that is not bowl-shaped, particularly on the downrange side of grazing impacts where we see the shallowest slopes. 

In oblique impacts into sand,  \citet{Anderson_2004} found that the ejecta angle on the uprange side was nearly equal to that of a normal impact. 
In Figure \ref{fig:slopehist} we see no strong relation between the uprange crater slope and impact angle. Assuming that a higher ejection angle gives a higher crater slope, this is consistent with uprange ejection angle being insensitive to impact angle (and supporting the results by \citealt{Anderson_2004}). 

\begin{figure}[htbp]\centering
\includegraphics[width=3.2truein]{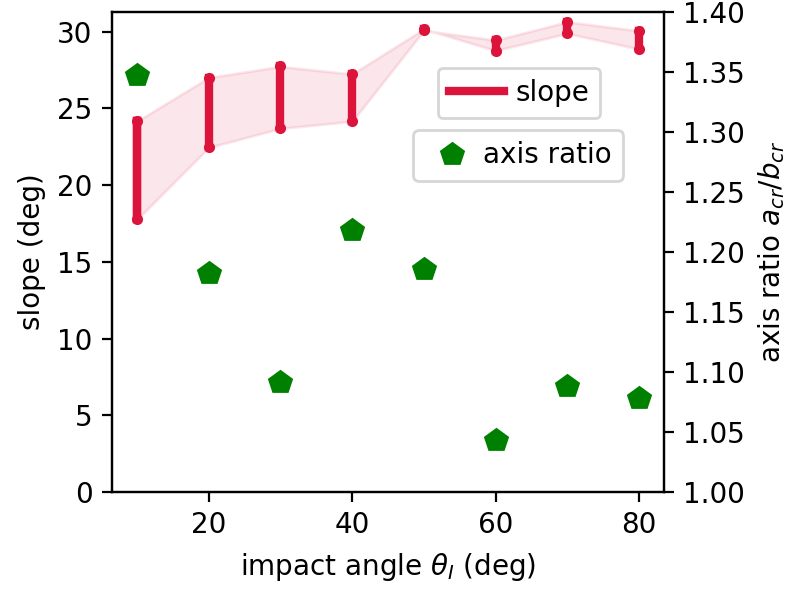}
\caption{Median slopes and crater axis ratio as a function of impact angle. 
The red bars show the uprange and downrange median values (with the axis scale on the left) for the crater slope computed along the crater major axis.  A slope of 0 is a flat surface. 
Crater slope increases with impact angle and an asymmetry in slope is seen primarily at lower impact angles. Green pentagons show crater axis ratios with axis scale on the right.  As expected, near-normal impacts have rounder craters. 
\label{fig:med_slope} }
\end{figure}

In Figure \ref{fig:med_slope} the red bars extend between the slopes of  
uprange and downrange median
slopes, measured along the crater's major axis. The widths of the bars illustrate that the downrange crater sides tend to be shallower than the downrange sides at a low impact angle, $\theta_I < 40^\circ$.  The figure also shows the weaker sensitivity of the uprange slope (the top of the red bars) to the impact angle.  In this plot we have also plotted the crater axis ratio, using the major and minor axes measured from rim peak to rim peak.  While normal craters are rounder, there was a significant scatter in the crater axis ratios. No strong variation in crater ellipticity is evident near $\theta_I = 50^\circ$ which divides impacts that ricocheted from those that did not ricochet. 

\begin{table*}[!htb]
\if \ispreprint1
\else\tiny
\fi
    \centering
     \caption{Crater and ricochet properties}
    \label{tab:crater}
    \begin{tabular}{llllllllll}
    \hline
Impact angle $\theta_I$ (deg)     
& 10 & 20  & 30 & 40 & 50 & 60  & 70 & 80 \\
\hline
Crater major axis $2a_{cr}$ (mm)      
& 56.6 & 64.2 & 62. & 74.1 & 77.9 & 72.8 & 72.7 & 71.4 \\
Crater axis ratio  $a_{cr}/b_{cr}$  &              
1.35 & 1.18 & 1.09 & 1.22 & 1.19 & 1.04 & 1.09 & 1.08\\
Distance $d_{ai}$ (mm)          
& 6.3 & 8.2 & 11.1 & 14.9 & 10.5 & 6.8 & 3.8 & 2.6\\
Maximum crater depth (mm)        
& 6.8 & 9.5 & 10.5 & 14.3 & 15.3&  14.7 & 15.5 & 16.7 \\
Crater volume $V_{cr}$ (cm$^{-3}$) 
& 3.3 & 7.0 &   8.5 & 13.2 & 15.3 & 14.6&  13.3&  14.7 \\
Crater efficiency   $\pi_V$
& 25  & 52 &   64 &   99&    115 &  109 &   100 &  110 \\
Median slope uprange (deg) & 
24 &  27 &  28 &  27 &  30 &  29 &  30 &  29 \\
Median slope downrange (deg) & 18 & 22 & 24 & 24 & 30 & 29 & 31 & 30\\
Ricochet speed $v_{Ric}$ (m/s)       
& 83 & 46 & 31 &  7.8 & 0.8 & - & - & -  \\
Ricochet angle $\theta_{Ric}$ (deg)   
 &10 &   16 &   17.5 &  28 &   12 &   - &   - &  -  \\
Horizontal momentum lost $\Delta p_x/(M_pv_{imp})$ & 0.12 & 0.51& 0.58& 0.70 & 0.64 &  0.50& 0.34& 0.17 \\ 
Vertical momentum change $\Delta p_z/(M_pv_{imp})$  & 0.31& 0.46& 0.59&  0.68 & 0.77 &  0.87 &  0.94 & 0.98 \\
Fraction of kinetic energy lost $\Delta E/(0.5 M_p v_{imp}^2)$ & 0.36& 0.80& 0.91& 0.99&   1 &  1 & 1 &   1   \\
         \hline   
    \end{tabular}  
\end{table*}

\subsection{Crater and ricochet measurements}

Crater properties measured from the crater profiles ( shown in Figure \ref{fig:contour}) are listed in Table \ref{tab:crater}.
The crater's major axis is the distance between uprange and downrange rim peaks. 
The crater axis ratio is the ratio of the crater length to width, measured from rim to rim.  
The distance $d_{ai}$ is the distance between
the midpoint of uprange and downrange rim peaks and the site of impact (which we estimated using high-speed videos).   We list the maximum depth, with zero corresponding to the undisturbed surface level well outside the crater. 
We measured crater volume $V_{cr}$ by integrating the depth profile within the zero-level contour just inside the crater rim. 
Crater efficiency, denoted $\pi_V$, for each impact is computed from the crater volume $V_{cr}$, substrate density $\rho_s$, and projectile mass $M_p$
(following \citealt{housen11,Elbeshausen_2009})
\begin{equation}
\pi_V  \equiv \frac{\rho_s V_{cr}}{M_p}. \label{eqn:pi_V}
\end{equation}

\begin{figure}[!ht]\centering
\includegraphics[width = 3.2 truein]{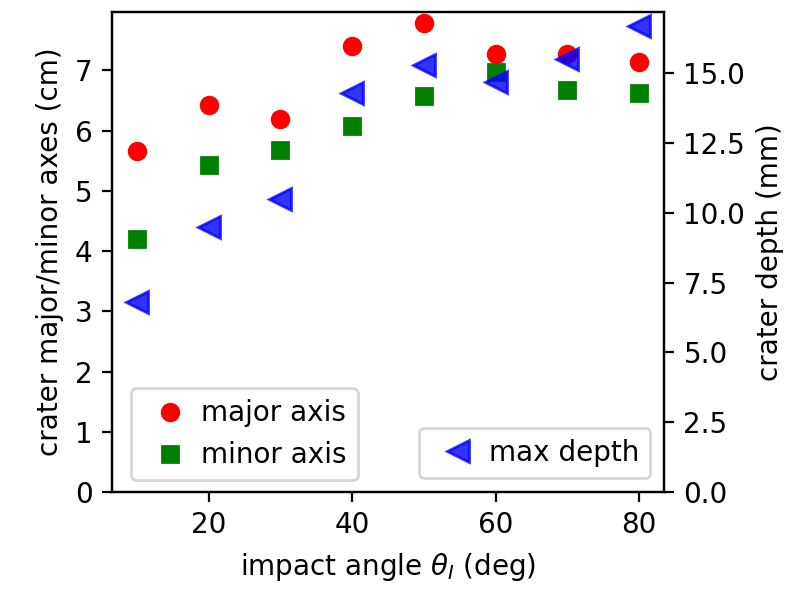}
\caption{Major and minor axis crater lengths as a function of impact angle. 
These are measured from rim to rim and 
are plotted as red circles and green squares with an axis scale on the left in cm.  Blue triangles show maximum crater depth, with respect to the undisturbed level surface prior to impact. 
Crater depth is in mm with axis scale on the right. 
These quantities all increase with impact angle. 
\label{fig:majorminor}}
\end{figure}

\begin{figure}[!ht]\centering
\includegraphics[width = 3.2 truein]{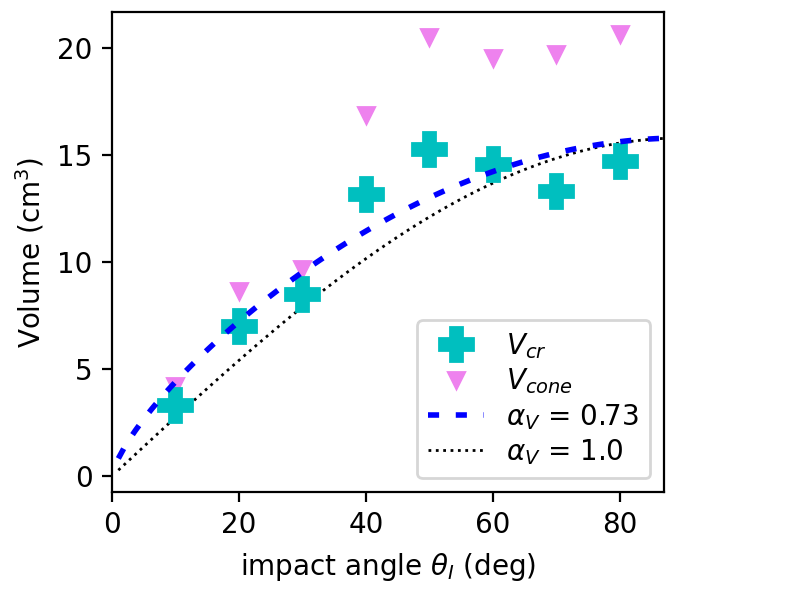}
\caption{Crater volume $V_{cr}$ in cm$^3$ as a function of impact angle is shown with cyan crosses. The crater volume is measured from the crater profiles shown in Figure \ref{fig:profiles}. 
With dotted and dashed lines, 
we show the function in equation \ref{eqn:alpha_V} for crater volume with exponent $\alpha_V$ values shown in the key.  Violet triangles show the volume of a conical shape  $V_{cone}$ computed from the crater's semi-major and minor axes and its maximum depth. 
\label{fig:vol}}
\end{figure}

\begin{figure}[htbp]\centering
\includegraphics[width = 3.2 truein]{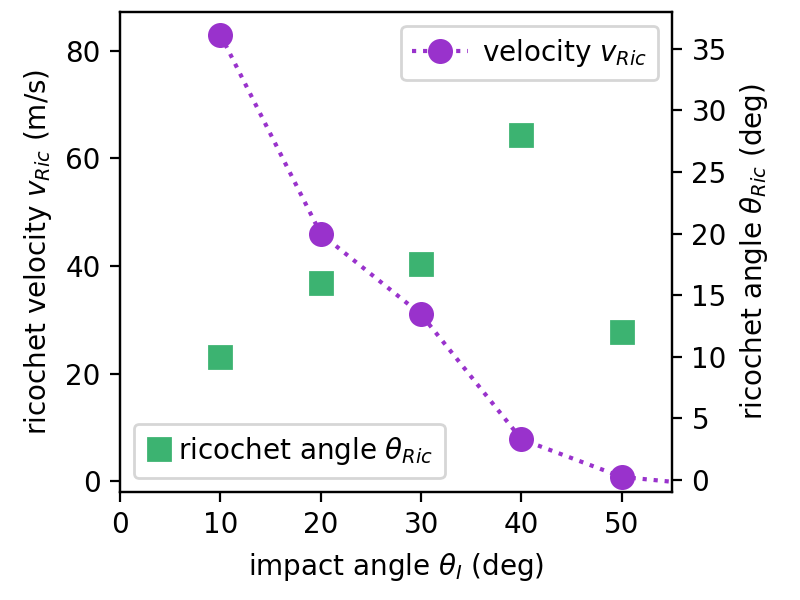}
\caption{Ricochet angle and velocity as a function of impact angle. 
The ricochet angle is shown with green squares and with an axis scale on the right.  
Velocity is shown with violet dots and with an axis scale on the left.  
Past an impact angle of $50^\circ$, the projectile did not ricochet. Instead, the projectile remained in the crater or was embedded after impact.  Ricochet velocity is high at grazing impact angles. \label{fig:ricochet} }
\end{figure}

\begin{figure}[htbp]\centering
\includegraphics[width = 3 truein]{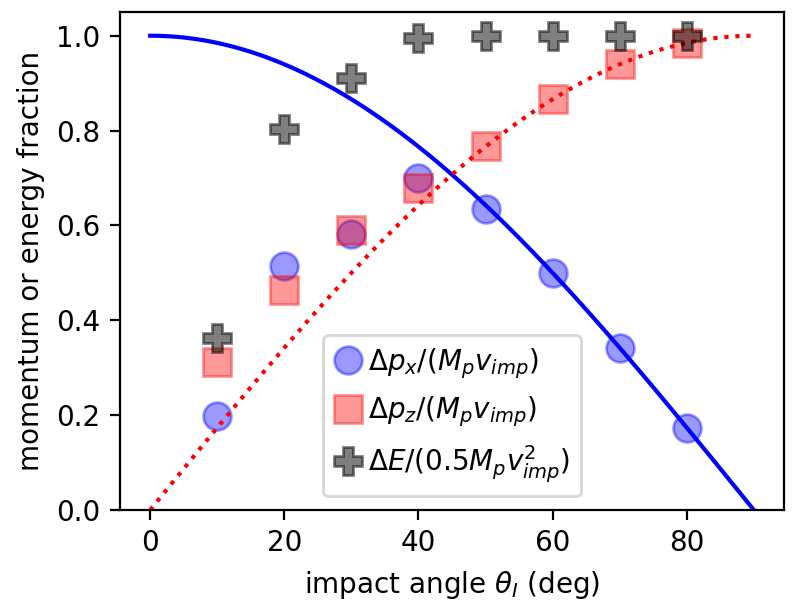}
\caption{Change in horizontal and vertical momentum components of the projectile before and after ricochet.  For impact angle $\theta_I>50^\circ$ the projectile remained within the impact crater.   The solid blue line and dotted red lines show the initial horizontal and vertical components, respectively, of projectile momentum divided by the total projectile momentum $M_p v_{imp}$. These are equal to $\cos \theta_I$ and $\sin \theta_I$.   The solid blue circle and solid red square shows horizontal and vertical components of the momentum change of the projectile, taking into account measurements of momentum of the projectile after ricochet for $\theta_I \le 50^\circ$.   As solid grey crosses, we plot the fraction of kinetic energy lost by the projectile. 
\label{fig:momchange}}
\end{figure}

\begin{figure}[htbp]\centering
\includegraphics[width = 3.2 truein]{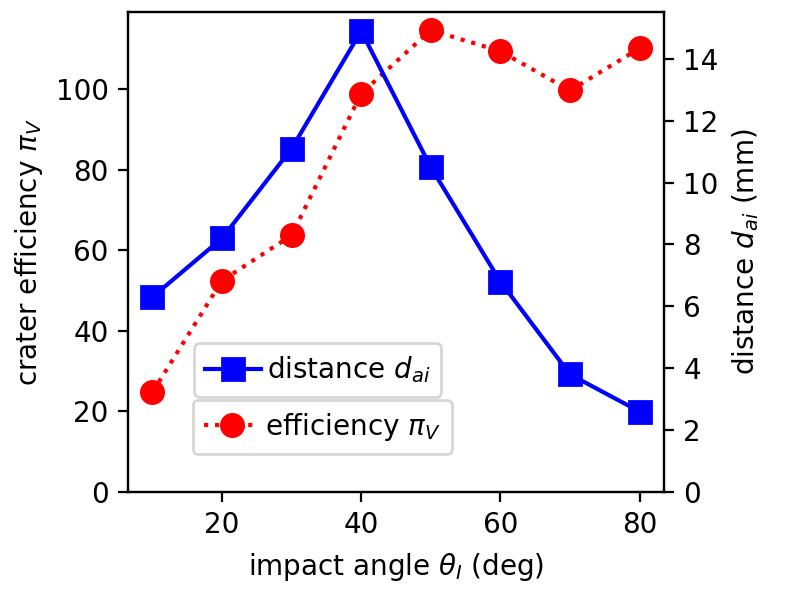}
\caption{Crater efficiency (red dots, axis scale on left) and distance between impact point and crater center (blue squares, axis scale on the right) as a function of impact angle.  Crater efficiency is the dimensionless quantity defined in equation \ref{eqn:pi_V}. Past an impact angle of $\theta_I = 40^\circ$ the distance between the impact site and the center of the crater $d_{ai}$ decreases to near zero for a normal impact.  The distance $d_{ai}$ is also low at grazing angles. 
\label{fig:eff}}
\end{figure}

High-speed videos from the same impact experiments used to measure crater profiles are used to measure the projectile velocity $V_{Ric}$ and angle $\theta_{Ric}$ (from horizontal) after ricochet.
Uncertainty in the crater and ricochet measurements are reflected in the specified decimal precision of the measurements listed in Table \ref{tab:crater}.
In Figures \ref{fig:profiles} --  \ref{fig:eff} we plot measurements from Table \ref{tab:crater}.



Figure \ref{fig:majorminor} shows crater major and minor axis lengths as a function of impact angle.  
The same figure shows the maximum crater depth.
Figure \ref{fig:vol} shows crater volume as a function of impact angle.  Crater major and minor axes, depth and volume increase with impact angle, confirming the simulation studies by \citet{Elbeshausen_2009} 
and consistent with experimental work by \citet{Gault_1978,Michikami_2017}.
\citet{Gault_1978} found that crater volume is a proportional normal component of the projectile momentum, $V_{cr} \propto \sin \theta_I$.   
\citet{Elbeshausen_2009} fit crater volume to a function 
\begin{equation}
    V_{cr}(\theta_I) = V_{cr}(90^\circ) \sin(\theta_I)^{\alpha_V} \label{eqn:alpha_V}
\end{equation}
and found exponent $\alpha_V \approx 0.89$ for their frictionless simulations. 
In Figure \ref{fig:vol}
we have plotted black dotted and dashed blue lines giving a sine dependence of crater volume on impact angle in the form used by \citet{Elbeshausen_2009} and in equation \ref{eqn:alpha_V}.  The best fit (via least squared minimization), shown with the dashed line, has
$V_{cr}(90^\circ) = 15.8 \pm 0.9$, $\alpha_V = 0.73 \pm 0.03$ with uncertainties based on the diagonals of the covariance matrix scaled by the variance of the residuals. 
We concur with \citet{Elbeshausen_2009} that an exponent $\alpha_V<1$ gives a better approximation for the dependence of crater volume on impact angle. 

We compute a volume for a conical shape $V_{cone} = \frac{\pi}{3} a_{cr} b_{cr} d_{max}$ using the crater semi-major and semi-minor axes $a_{cr}$, $b_{cr}$ and maximum depth $d_{max}$. The conical shape has sides with a constant slope, and a top surface with the area of an ellipse with major and minor axes the same as our craters. The volumes of the associated conical shape computed from our crater measurements  are plotted with violet triangles as a function of impact angle in Figure \ref{fig:vol}.  
The volume of half an ellipsoid $\frac{2\pi}{3} a_{cr} b_{cr} d_{max}$ with semi-axes $a_{cr},b_{cr}$ and $d_{max}$ is twice as large as $V_{cone}$.
Figure \ref{fig:vol} shows that an estimate for the crater volume based on a conical shape is similar to the measured volumes.  Such an ellipsoid gives a worse overestimate for the crater volume than that based on a conical shape.  This is not surprising as the craters have regions with nearly constant surface slope (see Figure \ref{fig:profiles}).

Ricochet velocity and angle are shown for impact angle $\theta_I\le 50^\circ$ in Figure \ref{fig:ricochet}.
The velocity is near the impact velocity at grazing impact angle and drops to near zero at $\theta_I = 50^\circ$.  The Froude number for our impacts is high, so we expect larger impact angles than the $\sim 30^\circ$ limit at $Fr \sim 15$ \citep{Wright_2020b}
for ricochet.  Only for $\theta_I\gtrsim 50^\circ$ does all the projectile energy go into crater formation as at these higher impact angles the projectile does not ricochet.
This is reflected in the major and minor axis lengths, crater depth, crater volume and efficiency, $\pi_V$,  which level off past $\theta_I \sim 40^\circ$, as seen in Figures \ref{fig:majorminor} and \ref{fig:vol}, above,  and Figure \ref{fig:eff}, below.   Crater volume and efficiency could be poorly fit by a curve proportional to $\sin \theta_I$ in part because the projectile does not carry away momentum or energy at $\theta_I > 50^\circ$.  

Using the ricochet velocity and angle we compute the change in the horizontal and vertical components of projectile momentum;
\begin{align}
\Delta p_x &= M_p v_{imp} \left[ \cos(\theta_I)- \frac{V_{Ric}}{v_{imp}} \cos (\theta_{Ric}) \right] \nonumber \\
\Delta p_z & = M_p v_{imp}
\left[\sin(\theta_I)+ \frac{V_{Ric}}{v_{imp}} \sin (\theta_{Ric} )  \right]. \label{eqn:Deltap}
\end{align}
Here $\Delta p_x,\Delta p_z$ 
are the horizontal and vertical components of the momentum 
from the projectile imparted to the granular substrate upon impact.  
We compute $\Delta p_x,\Delta p_z$ using the ricochet velocity and angle measured for experiments where the projectile ricocheted ($\theta_I \le 50^\circ$) and with values listed in Table \ref{tab:crater}).
These are plotted in Figure \ref{fig:momchange} in units of $M_p v_{imp}$, the initial projectile momentum,  and as a function of impact angle.   We also plot the fraction of kinetic energy lost by the projectile to the substrate 
\begin{equation}
    \frac{\Delta E}{0.5 M_p v_{imp}^2} = 1 - \frac{v_{Ric}^2}{v_{imp}^2}. 
\end{equation}
The momentum components and fraction of kinetic energy lost to the substrate are also listed in Table \ref{tab:crater}.

Figure \ref{fig:momchange} shows that 
for grazing impact angles, the vertical component of momentum imparted to the medium is similar to that estimated from the initial projectile's vertical velocity component (the red squares are near the red dotted line for $\theta_I \le 50^\circ$).
However, when the projectile ricochets,   the horizontal component of momentum imparted to the medium is well below that estimated using $\cos \theta_I$  (the blue dots are well below the blue solid line at a low impact angle). 
The projectile lost all its energy to ricochet after $\theta_I=50^\circ$ and in that respect, it resembles the crater volume as a function of impact angle plotted in Figure \ref{fig:vol}. However, the fraction of energy lost by the projectile does not drop as rapidly for $\theta_I< 50^\circ$ as the crater volume plotted in Figure \ref{fig:vol}, so energy scaling alone does not account for the dependence of crater volume upon impact angle.  Our experiments are not in a hypervelocity regime, so at least some of the decrease in crater efficiency ($\pi_V$; equation \ref{eqn:pi_V}) at grazing angle compared to normal impacts is due to energy and momentum carried away by the projectile as it ricochets.  
Nevertheless, we do see a deviation from a sine function (equation \ref{eqn:alpha_V} with $\alpha_V=1$) corresponding to a dependence on the z component of momentum, similar to that found by \citet{Elbeshausen_2009}.

In low velocity impacts the projectile's spin varies while it interacts with a granular medium \citep{Wright_2020b}. 
Some of the sensitivity of crater efficiency in our experiments to impact angle could be due to the angular momentum and rotational energy carried by a projectile that ricochets.


\subsection{The difference between the site of impact and the crater center}

Figure \ref{fig:eff} shows the distance between the impact site and crater center (denoted $d_{ai}$) with blue squares. 
This distance reaches a peak at $\theta_I \sim 40^\circ$ and is lower at both higher and lower impact angles. 
The change in behavior might be related to the variation in the fraction of energy that is imparted to the medium due to ricochet.  With red dots Figure \ref{fig:eff} also shows crater efficiency $\pi_V$ (defined in equation \ref{eqn:pi_V}) which plateaus past $\theta_I\sim 50^\circ$, supporting this connection. 

A difference between the impact point and center of the crater (as measured from the intersection of the crater's major and minor axes)
was noted previously by \citet{Anderson_2003} in high-velocity oblique experiments into the sand.  If a seismic source is generated at the impact point, then variations in both ejection angle and the distance of the resulting crater rim from the point of impact can be linked via a flow model that has azimuthal asymmetry in the strength of a propagating pulse   \citet{Anderson_2004,Anderson_2006}.
A model with a migrating flow center 
\citep{Anderson_2006} would be sensitive to the distance $d_{ai}$ between the crater center and the site of the impact that we measured and shown in Figure \ref{fig:eff}.

\section{Subsurface seismic pulses}
In this section, we discuss measurements from experiments using the accelerometers.

\subsection{Asymmetry in seismic pulse strengths}
\label{sec:upeak}

\begin{figure}[htbp]\centering
\includegraphics[width = 3.4truein, trim = 10 0 0 0 , clip]{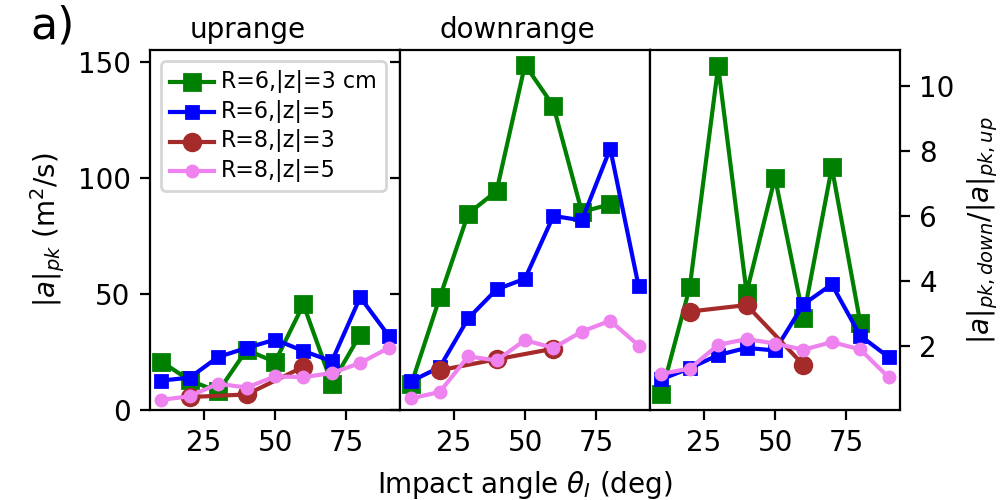}
\includegraphics[width = 3.4truein,trim = 10 0 0 0 , clip]{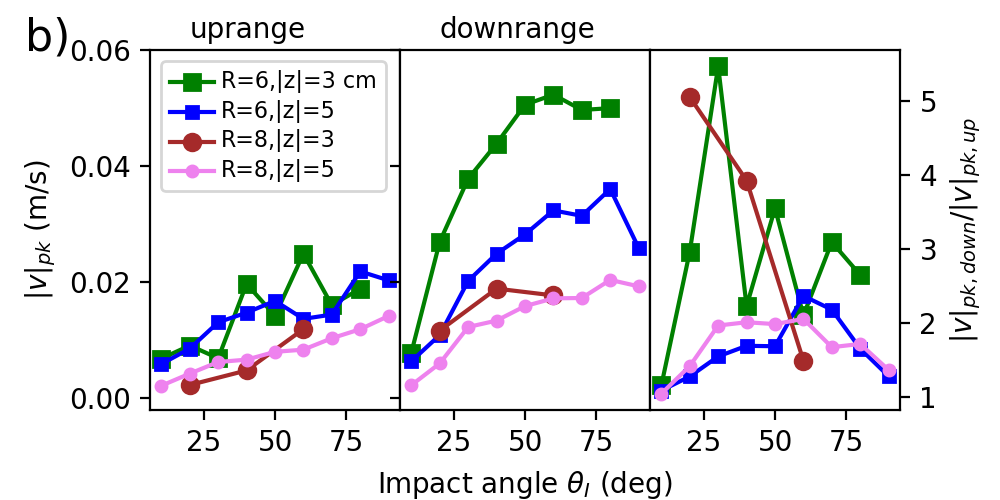}
\caption{a) We show peak acceleration magnitudes as a function of impact angle.  The left panel shows peak values for accelerometers located uprange of the impact site.
The middle panel shows peak values for accelerometers located downrange of the impact site.
The scale for the left and middle panels is on the left vertical axis. The right panel, with a vertical axis scale on the right, shows the ratio of downrange to uprange peak accelerations.  The peak acceleration values are shown at 4 different locations with radius $R$ and depth $|z|$ listed in the key.
b) Similar to a) except for peak velocity magnitudes. 
There is a strong asymmetry in pulse strengths, with pulse strengths much higher on the downrange side than on the uprange side of impact.   This is particularly noticeable in the velocities of the shallow accelerometers.  The scatter in the ratio is primarily due to scatter in the weaker accelerometer signals on the uprange side. }
\label{fig:pulse_asym}
\end{figure}

In Figure \ref{fig:pulse_asym} we compare the strength of seismic pulses, which is measured from peak accelerations, downrange of the impact site to those uprange of the impact site. We plot peak velocity and acceleration magnitudes as a function of impact angle for accelerometers at 4 different locations. 
Except for near-normal impacts, 
pulse peak heights, seen in acceleration (Figure \ref{fig:pulse_asym}a) and in velocity (Figure \ref{fig:pulse_asym}b) are higher on the downrange side than on the uprange side. 
The ratio of uprange to downrange peak velocity is particularly high $|v|_{pk,down}/|v|_{pk,up} \sim 5$ at low impact angle for the shallow accelerometers at $|z|=3$ cm.
For deeper accelerometers at $|z|=5$ cm, the ratio of uprange and downrange 
peak velocity and peak acceleration are a maximum at intermediate impact angles.   

As shown in Figure \ref{fig:momchange}, the horizontal component of momentum that is imparted to the medium is relatively low at grazing impact angles because of the projectile ricochets. 
Nevertheless, the ejecta curtain and crater profiles are most asymmetric at a grazing impact angle.  The ratio of up and downrange pulse strengths is highest at the shallower accelerometers,  
suggesting that the energy in the subsurface flow field is more strongly concentrated at shallower depths for grazing impacts than for near-normal impacts. The shock model by \citet{Kurosawa_2019} could predict such an effect, as at a shallow depth, post-rarefaction streamlines with a deflected velocity vector can converge onto locations that previously had radial streamlines, enhancing the pulse strength.  

For different impact angles we measured small (less than 20\% percent) variations in crater slope and axis ratio (as shown in Figure \ref{fig:med_slope}),  ejecta angle (visible in Figures \ref{fig:ejecta} and \ref{fig:quiver}, and crater shape, as seen in Figure \ref{fig:contour}.  However here we find much larger (factors of 2 to 5) differences in the strengths of the uprange and downrange seismic pulses.   The accelerometer positions are well outside the crater radius so the asymmetry in pulse strength persists as it travels. 

Large (factor of 2) asymmetry in seismic stress was previously measured with piezoelectric sensors by \citet{Dahl_2001} for oblique hypervelocity impacts into solid aluminum. Like \citet{Dahl_2001}, we see a significant asymmetry between uprange and downrange seismic pulse strengths in oblique impacts, though our experiments are into the sand and our impact velocity is lower than theirs. They measured pulses on either side of a solid aluminum block caused by a 6 km/s impact. 

The momentum flux carried by a pulse depends on its velocity amplitude. 
Perhaps the uprange/downrange subsurface pulse height asymmetry seen in our oblique impact experiments is caused by the momentum direction of the projectile. This would imply that the vertically downward propagating pulse should be stronger than the laterally propagating one for a normal impact.  Downward propagating pulses were measured to be about a factor of 2 stronger than laterally propagating pulses in lower velocity normal impact experiments into millet \citep{Quillen_2022}, confirming this expectation.   The strong uprange/downrange asymmetry in pulse strength seen in our oblique impact experiments suggests that the projectile momentum vector influences the subsurface flow field. Given that it may be difficult to infer or constrain the projectile direction from the crater shape, the strong subsurface pulse strength asymmetry is remarkable. 

\begin{figure*}[htbp]
    \centering
    \if \ispreprint1
    \includegraphics[scale=0.7, trim=15mm 0mm 0mm 0mm]{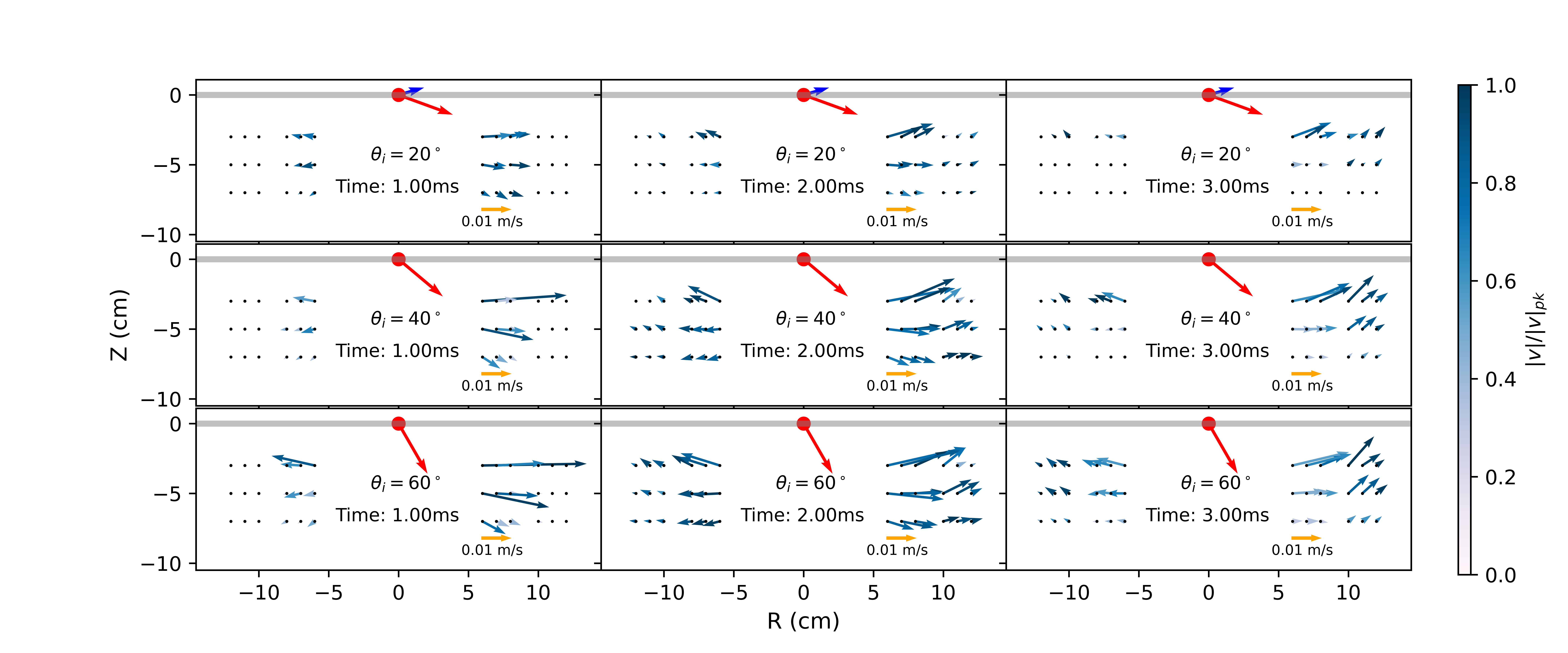}
    \else
    \includegraphics[scale=0.6, trim=30mm 0mm 0mm 0mm]{oblique_quiver_angle.png}
    \fi
    \caption{Subsurface velocity flow fields for different impact angles. Each column shows a different time after impact and each row shows a different impact angle.  The length of the blue arrows shows the velocity magnitude at the position of each accelerometer.  The scale of these arrows is shown with the small gold arrows on the bottom of each panel with a length corresponding to 0.01 m/s. The blue velocity vectors have a shade that depends on the velocity magnitude with a color bar on the right that gives the ratio of velocity with respect to the peak value (during the experiment) of each accelerometer signal.
    The impact point is shown with a red dot. The initial projectile momentum is indicated by the red arrows at the site of impact.   Accelerometer placement positions are shown as black dots. The thick grey line on top of each panel shows the level substrate surface prior to impact.  The velocities are higher on the downrange side (on the right) and at later times vectors increasingly point upward. 
    }
    \label{fig:ob_ray}
\end{figure*}

\subsection{Subsurface ray angles}

We examine the subsurface flow field at impact angles of $\theta_I = 20$,  40, and $60^\circ$.  For these impact angles, we have many impact experiments giving us information at a number of subsurface accelerometer positions. 
From the signals from each accelerometer, the direction of the velocity at uprange and downrange positions (at  $\phi=0$ or $\pi$) is computed using the ratio of the $R$ and $z$ velocity components in cylindrical coordinates.
Ray angles are shown with vectors in Figure \ref{fig:ob_ray}. Each column is at a different time after impact and each row is at a different impact angle. 
The vector lengths are related to the value of the pulse velocity in m/s. A gold arrow on the bottom of each panel is shown to present the 0.01 m/s scale.
Points and vectors are shown at the position of each accelerometer.
The horizontal axes show radius $R$ but with positive values corresponding to downrange positions and negative values corresponding to uprange positions.  
The vertical axes show depth. 
For each accelerometer signal, we computed the maximum velocity amplitude $|v|_{max}$ during the experiment.  The colors of the vectors depend on the velocity divided by this maximum value.  The color bar on the right relates color to pulse velocity.      
The projectile velocity direction is shown with a red arrow at the position of impact. The little blue arrows shown near the origin in the top row show the direction of the projectile after it ricocheted in the impact angle $\theta_I=20^\circ$ impacts.  
The length of this arrow is scaled with respect to the red arrow showing projectile velocity prior to impact.  
The red arrow, showing the impact direction, is not on the same scale as the subsurface velocity vectors. 
We include a supplemental video, denoted \texttt{ray\_angles}, which shows the ray angles evolving in time.

Note that at each impact angle, we used accelerometer data from 9 separate impact experiments (via Template 1 in Table \ref{tab:acctemplate}) to construct Figure \ref{fig:ob_ray} as only 4 accelerometers were used in each individual impact experiment.  We infer that experiments at the same impact angle are similar because neighboring accelerometers have similar pulse strengths and directions.  

Figure \ref{fig:ob_ray} shows a strong asymmetry in the strength of subsurface motions.  Pulse velocities are larger for more normal impacts, which is consistent with their larger crater volumes. 
At later times velocities are higher at shallower depths. 
As was true for normal impacts \citep{Neiderbach_2023}, velocities are initially nearly radial (as shown in the leftmost column) and pointed upward at later times (as shown in the rightmost column).  At later times (the rightmost column) the velocity is primarily high in the shallower accelerometers.  The ray angles suggest that the pulse initially propagates radially, and then tilts upward.  Streamlines change direction, tilting toward the surface.  This type of time-dependent flow field was predicted with a shock model by \citet{Kurosawa_2019}. Figure \ref{fig:ob_ray} suggests that impacts in granular systems exhibit similar behavior.  Velocity appears to accumulate near the surface at later times.  Within the crater radius, the flow would eventually launch the ejecta curtain.  The accelerometers in our experiments are outside the crater radius, so particles are not lofted off the surface past $R\sim 3.5$ cm, instead the flow seen in Figure \ref{fig:ob_ray} later causes the surface to move and deform, as described by \citet{Neiderbach_2023}. 

\subsection{The Maxwell Z-model}

As the Maxwell Z-model \cite{Maxwell_1977} was used by \citet{Anderson_2004} to describe the subsurface excavation flow field (for oblique impacts), based on measured ejecta angles, we introduce it here. 
We will discuss this model in context with the downrange/uprange asymmetry in pulse peak strengths (mentioned previously in section \ref{sec:upeak})
and the subsurface ray angles (shown in the previous section). 

A Maxwell Z-model has a flow velocity 
\begin{align}
  u_r(r,\vartheta,t) &= \frac{a(t) }{r^{Z}} \nonumber \\
  u_\vartheta(r,\vartheta,t) &= \frac{a(t)}{r^Z} (Z-2) \frac{\sin \vartheta }{ 1+ \cos\vartheta } , \label{eqn:Maxwell}
\end{align}
where we have given velocity components in spherical coordinates $(r, \varphi, \phi)$.  
Here $\vartheta = 0$ along the negative $z$ axis below the surface. The velocity field satisfies $\nabla \cdot {\bf u} = 0$ so is incompressible. The prefactor $a(t)$ specifies the time dependence of the velocity field. 

Streamlines are described with the radius $R_s$ where the streamline intersects the surface.  The radius of a streamline with $R_s$ as a function of $R_s$ and $\vartheta$ is  
\begin{equation}
    r(\vartheta,R_s) = R_s (1-\cos \vartheta)^\frac{1}{Z-2}.
\end{equation}
With exponent $Z=2$, the flow is radial; the velocity vector ${\bf u} \propto \hat {\bf r}$ where $\hat {\bf r}$ is the radial unit vector. 

At the surface where $\vartheta=\pi/2$, 
the horizontal and vertical velocity components are 
\begin{align}
u_r(r, \pi/2, t)  &= \frac{a(t)}{r^Z}\nonumber \\
u_z = u_\vartheta(r, \pi/2, t)  &= (Z-2)u_r.
\end{align}
The ratio between $u_r$ and $u_z$ at the surface gives the ejecta angle
\begin{align}
\tan \vartheta_{ej} = Z-2.
\label{eqn:theta_ej}
\end{align}

The Maxwell Z-model has the nice property that a single parameter $Z$ gives a direct connection between subsurface flow, ejecta angle, and the radial decay of the flow field.  This connection was leveraged by \citet{Anderson_2004} to relate measured ejecta angles to subsurface flow models. Unfortunately, the Z-model does not take into account the time dependence of the flow. If the ejection angle or the velocity angle at subsurface positions varies with time, the model must be modified (e.g., \citealt{Anderson_2006}). 



Equation \ref{eqn:theta_ej} implies that ejecta angle $\vartheta_{ej}$  for a Maxwell Z-model is not sensitive to the prefactor $a(t)$, only to the exponent $Z$.   
\citet{Anderson_2004} explored a Z-model with an azimuthally varying exponent; 
\begin{align}
Z(\phi)  = Z_0(1 + A_\phi \cos \phi).  \label{eqn:Aphi}
\end{align}
Using equation \ref{eqn:theta_ej},  variations in the ejecta angle of 20\% between $\phi=0$ and $\phi=\pi$ corresponding to up and downrange, would require a model amplitude $A_\phi \sim 0.1 $.  How large a velocity asymmetry would this give?
The ratio of downrange to uprange velocity radial velocity component 
\begin{align*}
\frac{u_r(r,\vartheta, \pi)}{u_r(r,\vartheta,0)}  & \sim \frac{r^{Z_0(1 + A_\phi)}}{ r^{Z_0(1-A_\phi)}} \\
& = r^{2Z_0 A_\phi} .
\end{align*}
For $Z_0 \sim 3$ and $A_\phi \sim 0.1$ this gives 
$\frac{u_r(r,\vartheta, \pi)}{u_r(r,\vartheta,0)}  \propto r^{0.6} $.  We expect that the ratio would increase with increasing radius.  Due to the power law dependence of the flow model on $r$, radial variations exceed angle variations in the velocity vector so we expect that the downrange to uprange ratio of velocity amplitude should approximately scale the same way as the downrange to uprange ratio of the radial velocity components. 
However, the ratio of peak velocity amplitudes,  as shown in Figure \ref{fig:pulse_asym}, decreases with increasing radius, suggesting that the subsurface flow is more symmetric as it travels away from the impact, rather than becoming more asymmetric with increasing radius.  
Plots similar to those shown in Figure \ref{fig:pulse_asym} but using peak radial velocity and accelerometer components, were similar to those shown in Figure \ref{fig:pulse_asym}. 

Note that in Figure \ref{fig:pulse_asym} we have plotted peak acceleration and velocity amplitudes. However, these peaks do not occur at the same time when comparing two accelerometers at different positions in the same experiment. Furthermore, the peaks do not necessarily occur at the same time when comparing two accelerometers at the same position but from experiments at different impact angles.  
Unfortunately, the Maxwell Z-model does not take into account the time dependence of pulse propagation, making it more challenging to adopt it when subsurface pulse strength and direction are time-dependent (though see \citealt{Kurosawa_2019} who related shock structure to the subsequent excavation flow for impacts into solids). 

In summary, the Maxwell Z-model predicts that flow velocity decays rapidly with distance from the impact point.  As the exponent $Z$ also determines the crater ejecta angle, a downrange/uprange asymmetry in the ejecta angle in oblique impacts can be interpreted in terms of azimuthal variation in the exponent $Z$ \citep{Anderson_2004}.   As the associated flow field is likely to be decaying rapidly with distance from the site of impact $r$ we would expect an associated difference in the strength of uprange and downrange pulses.  We find that downrange pulse peak velocities can be 2 to 5 times larger than uprange pulse peak velocities for oblique impacts (as shown in Figure \ref{fig:pulse_asym}) confirming this expectation. However, pulse strength asymmetry seems to decay with distance, contrary to what would be expected with a simple variant of the Maxwell Z-model (described with equation \ref{eqn:Aphi}).  

\begin{figure}[htbp]
\centering
    \includegraphics[scale=0.6, trim=18mm 0mm 0mm 0mm]{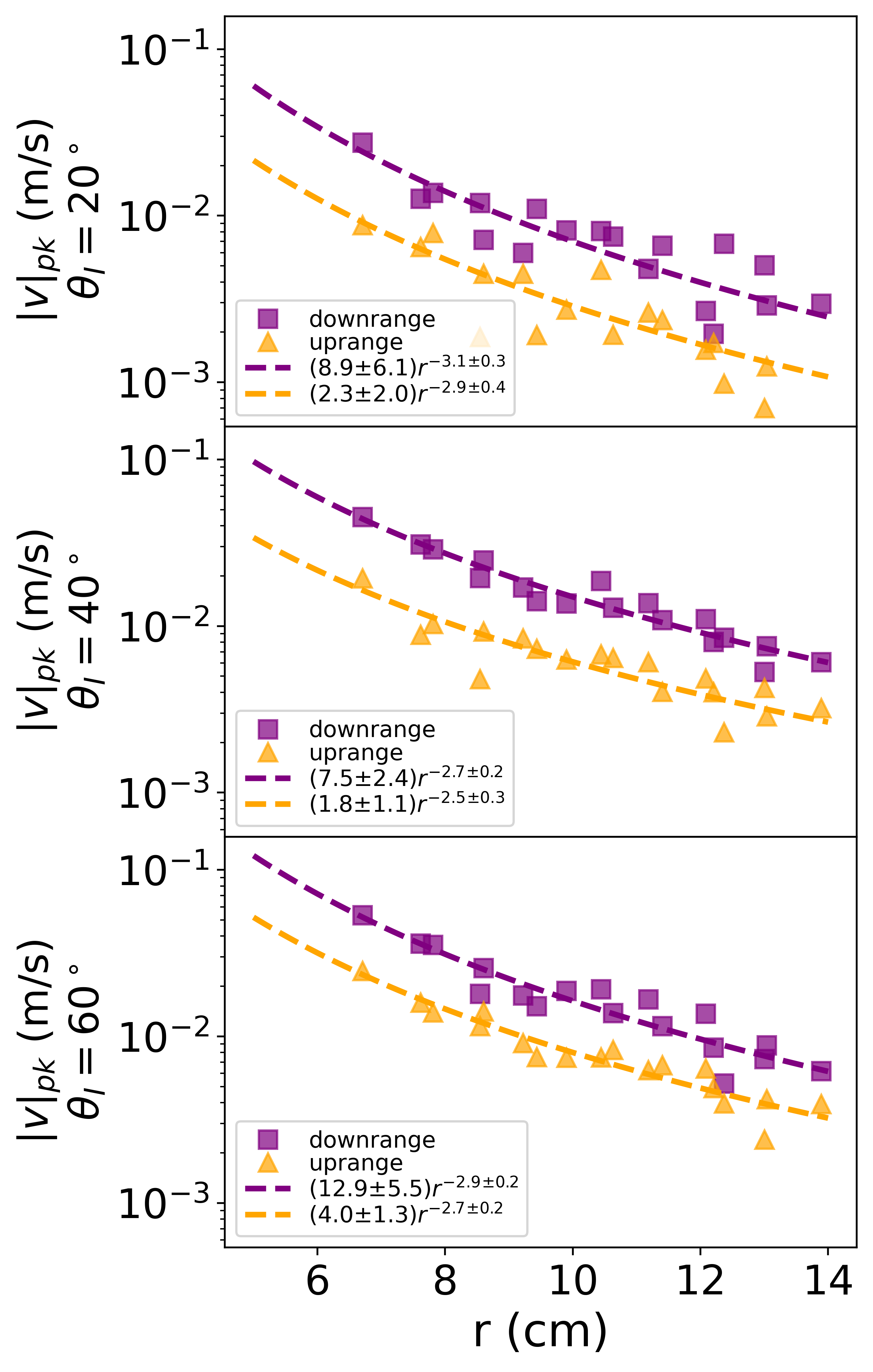}
    \caption{Peak velocity amplitude as a function of distance $r$ from the site of impact.  Each panel shows experiments at a different impact angle, $\theta_I = $ 20, 40, and 60$^\circ$ from top to bottom. Downrange data points are represented as purple squares while uprange data points are show with gold triangles.  Each point is a measurement from a single accelerometer. Power-law fits are shown with dashed lines in the color matching the points they fit.  The coefficients and exponents of the power-law model are shown in the keys. The vertical scales are the same in all three panels. We find that there is no significant difference between the radial decay rate (given by the exponents of the fit) between uprange and downrange pulse peaks.  There is no strong dependence of the radial decay rate of pulse strength on impact angle. }
    \label{fig:vfit}
\end{figure}

\subsection{Decay of pulse peak velocities}
\label{sec:decay}

In this section, we examine whether the rate that pulse strengths decay is different on the uprange and downrange sides.  
We use accelerometer signals for experiments with impact angles $\theta_I= $ 20, 40, and 60$^\circ$ as these were covered with the largest number of accelerometer positions.  
As done in previous studies (e.g., \citealt{Quillen_2022,Neiderbach_2023}) we compute the magnitude of the vector as a function of time in each accelerometer and then select the peak value.  This is done for each accelerometer.  The position of the accelerometer is taken from the accelerometer placements listed in Table \ref{tab:acctemplate}, giving the distance $r=\sqrt{R^2+z^2}$ of the accelerometer from the site of impact. The resulting peak velocities $|v|_{pk}$ are shown in Figure \ref{fig:vfit} as a function of distance from the impact site.  Each panel shows data for a different impact angle.  The violet squares and gold triangles show velocities downrange and uprange, respectively, of the impact site.  

To the data points shown in Figure \ref{fig:vfit} we fit power-law functions in the form $|v|_{pk}(r) = B r^\beta$, to find coefficients $B$ and $\beta$ at each impact angle. Via least squares minimization, we fit uprange and downrange points separately. 
The best-fitting power-law curves are shown with dashed lines, with gold and violet lines giving uprange and downrange fits, respectively.  The exponents $\beta$ and scaling factors $B$ for each fit are printed in the key on the lower left side of each panel. Uncertainties in the coefficients are estimated from the scatter in the points from the best-fitting line.  Figure \ref{fig:vfit} shows that there is no significant difference between uprange and downrange pulse strength decay rates.  The uprange and downrange exponents are similar at all three impact angles.   The asymmetry in the pulse strengths is seen in the differences in the $B$ coefficients of the fits.  We find that the pulse strength is more strongly dependent upon the azimuthal angle than the pulse decay rate.  
This implies that the prefactor of a  Maxwell Z-model for the flow should be dependent upon the azimuthal angle $\phi$.  
As the ejecta angle also depends on the impact angle, an associated Maxwell Z-model might also require a $\phi$ dependent exponent.  

The exponents measured from the fits are near -3.  This is near but somewhat steeper than the -2.5 exponent predicted for the spherically symmetric propagation model by \citet{Quillen_2022}  (giving $v_{pk} \propto r^{-2.5}, a_{pk} \propto r^{-3}$) and is shallower than the exponents (which are lower than -3.5) measured for the decay of surface particle velocity by \citet{Neiderbach_2023} (listed in their Table 4).   The -2.5 exponent is predicted for a momentum-conserving pulse that broadens with duration $\Delta t \propto  t^{-1/2}$ as it travels and assuming a constant propagation velocity $v_P$ \citep{Quillen_2022}.  With no pulse broadening a similar model would give $v_{pk} \propto r^{-3}$, as found here and would also give $a_{pk} \propto r^{-3}$ which is consistent with normal impact  experiments \citep{yasui15,Matsue_2020,Quillen_2022}. 

We also examined the peak radial component of velocity $v_r$ (in spherical coordinates) as a function of distance from impact $r$.   However, the scatter in these points was larger than that in $|v|_{pk}$ (shown in Figure \ref{fig:vfit}) so the fits to these points were less certain. 

\begin{figure}[htbp]
    \centering
    \includegraphics[width=3.6truein, trim = 0mm 2mm 0mm 0mm,clip]{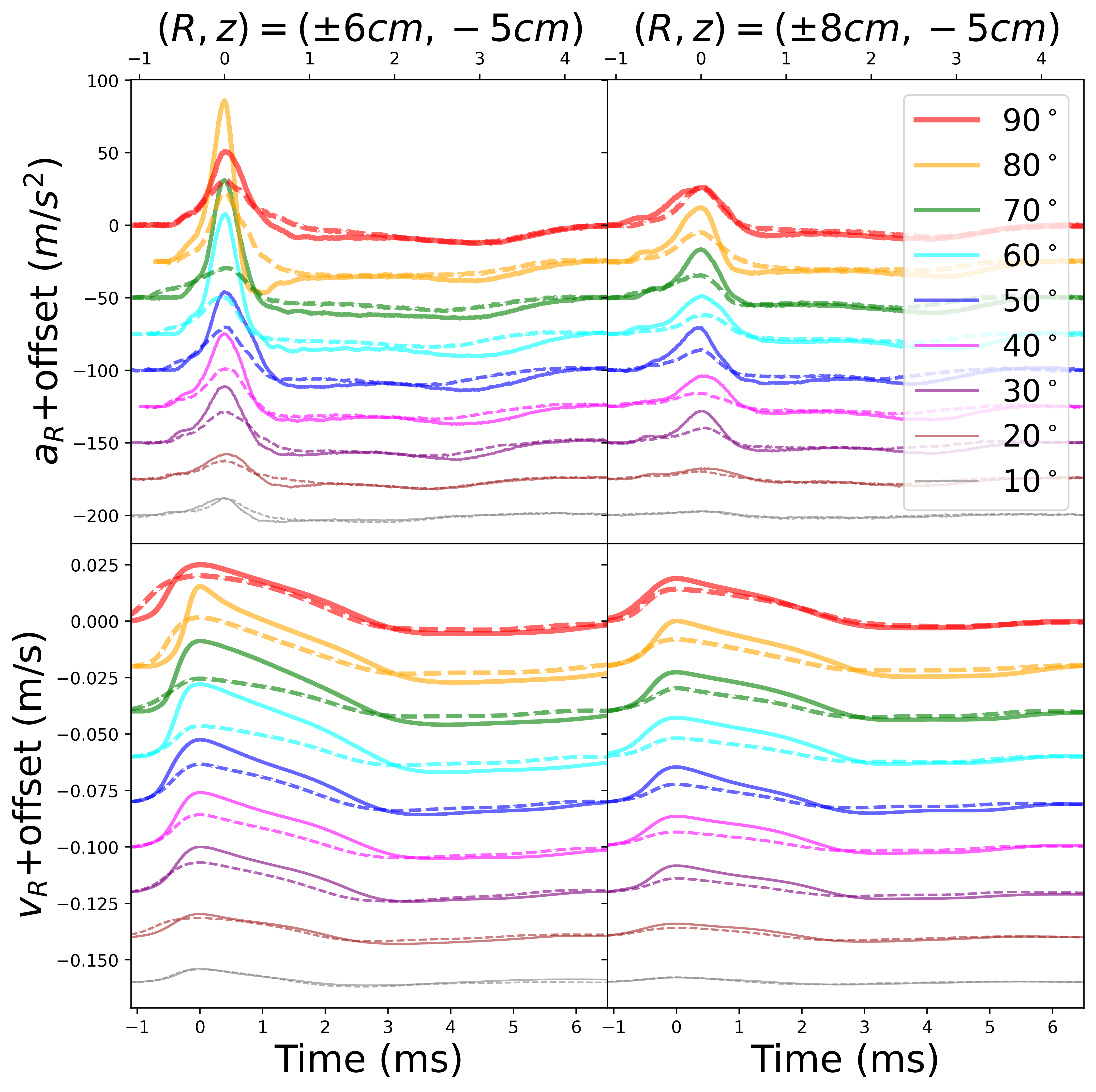}
    \caption{Cylindrical radial acceleration and velocity components as a function of time and for different impact angles. In each panel, we show an uprange and a downrange accelerometer at the same depth and radial distance from the impact point. The different color lines show different impact angles, with key in the upper right panel. The solid lines indicate downrange pulses while the dashed lines represent those uprange. To facilitate comparison between pulse shapes, peaks are aligned in each panel by offsetting the x axis so that peaks occur at $t = 0$ ms, and pulse profiles are offset vertically.  The top two panels show accelerations and the bottom two panels show velocities.  The left two panels show accelerometers at $R=6$ cm and the right tow panels column show accelerometers at $R=8$ cm.  
    }
    \label{fig:pshape}
\end{figure}

\begin{figure*}[htbp] 
\centering 
\if \ispreprint1
\includegraphics[width = 3.5truein ]{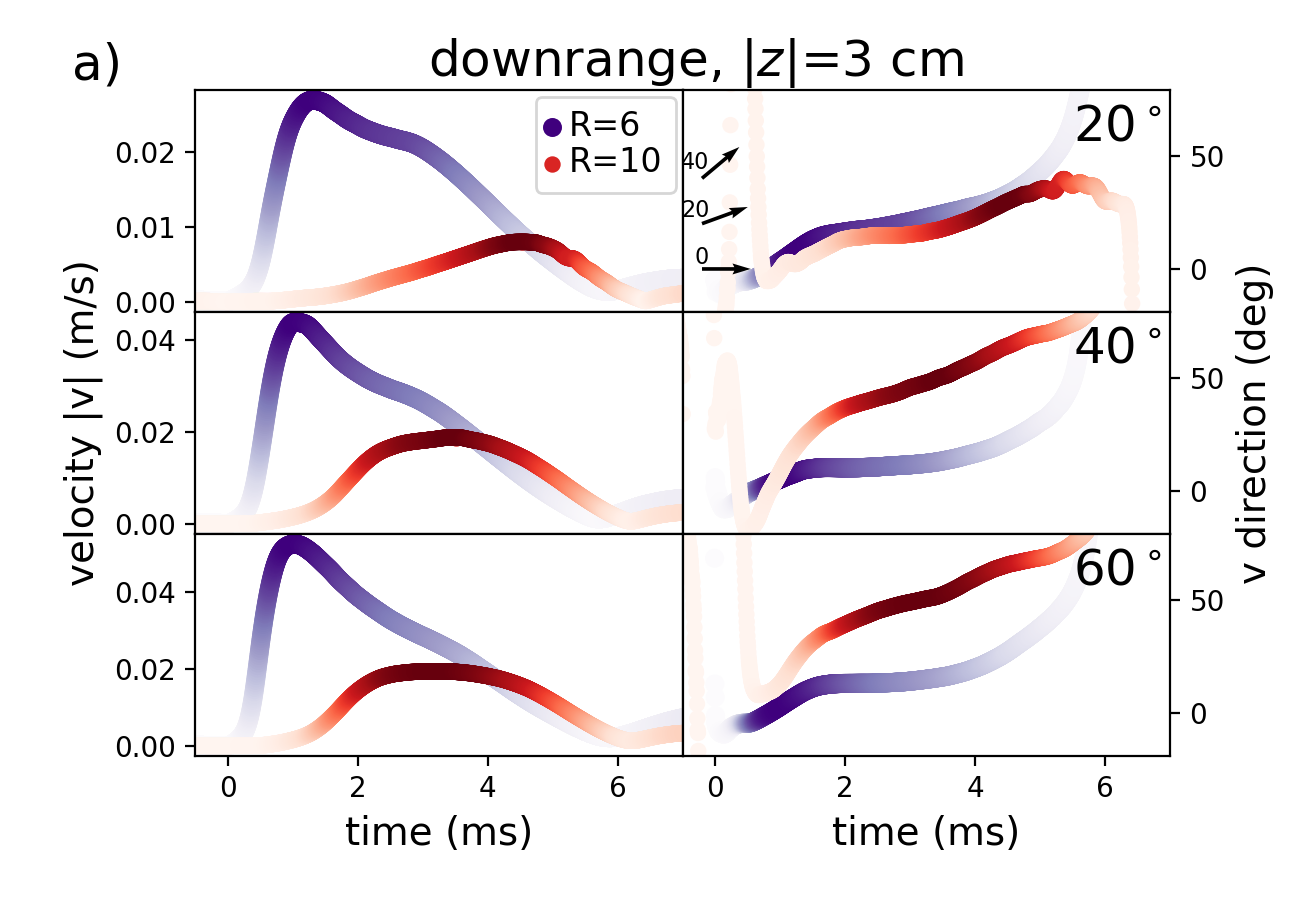}
\includegraphics[width = 3.5truein ]{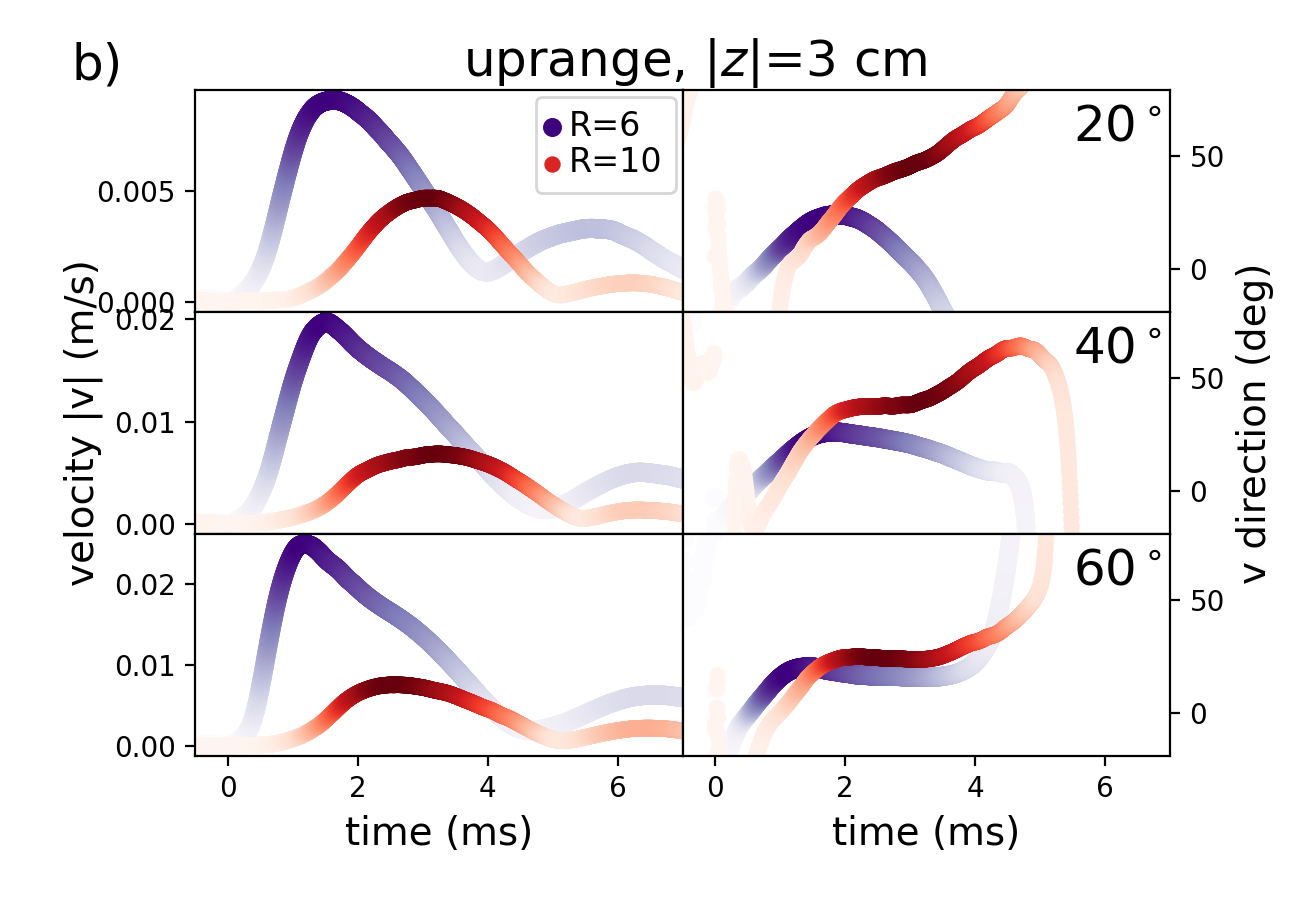}
\includegraphics[width = 3.5truein ]{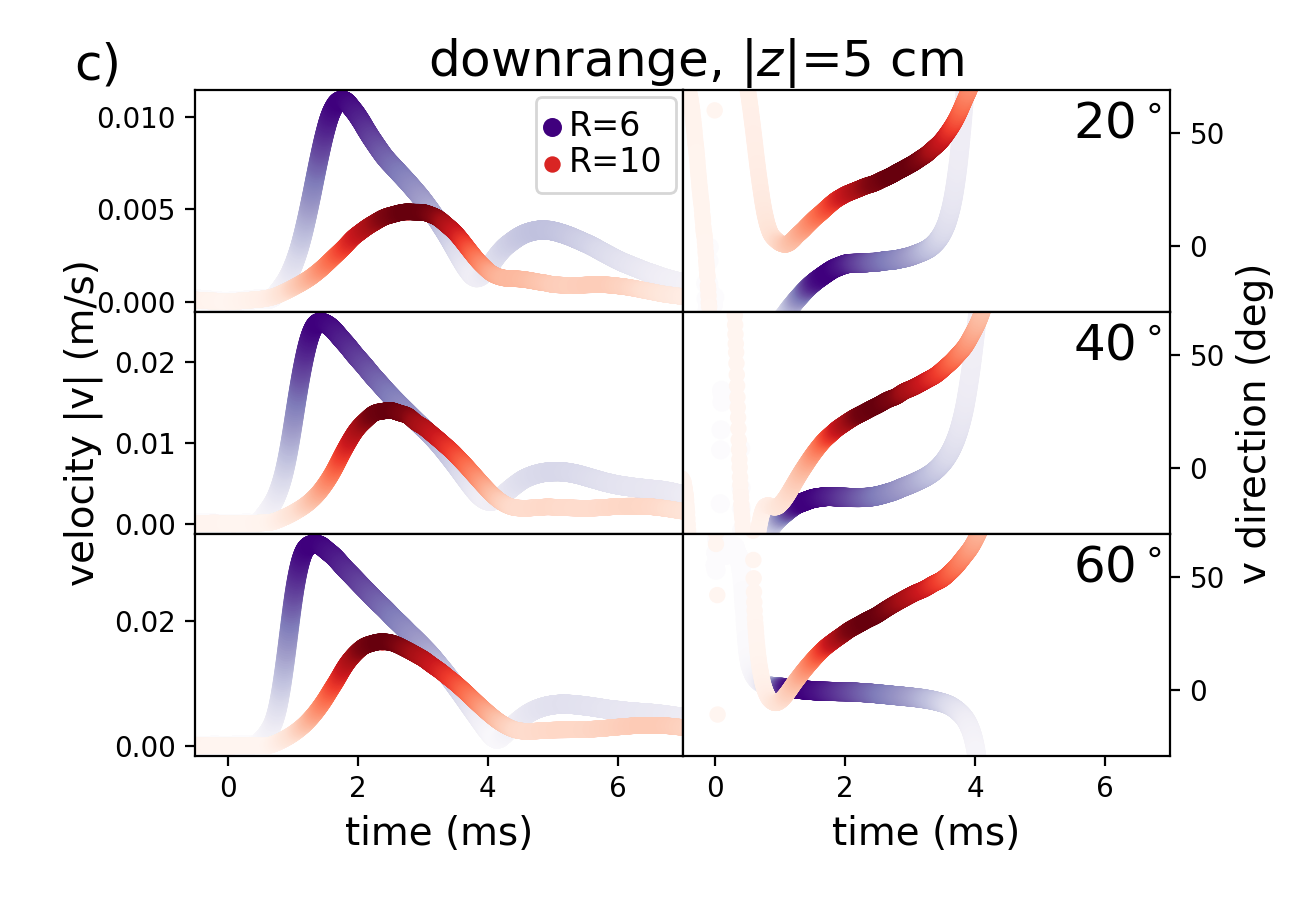}
\includegraphics[width = 3.5truein ]{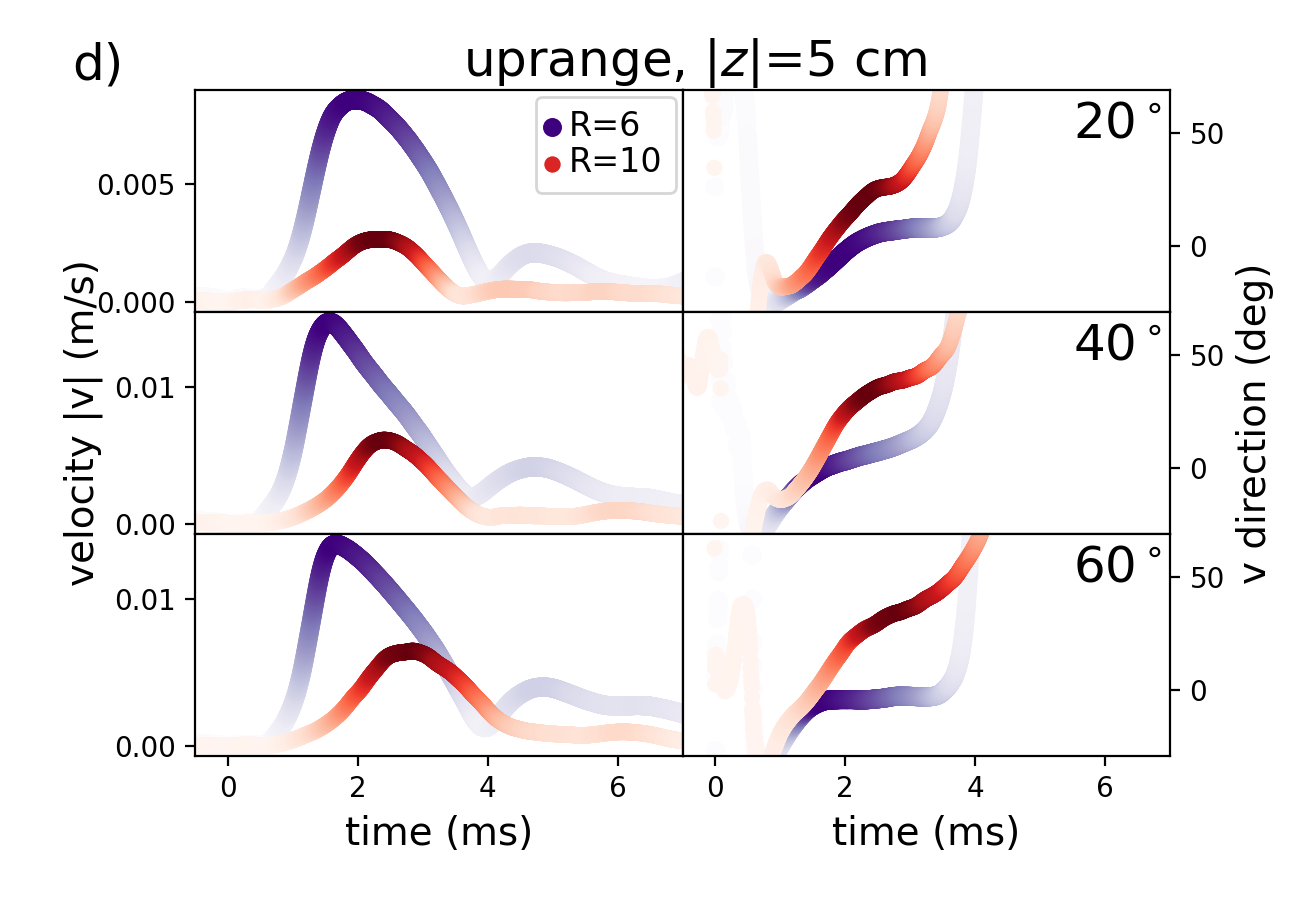}
\else
\includegraphics[width = 2.5truein ]{Rpulse_downrange.png}
\includegraphics[width = 2.5truein ]{Rpulse_uprange.png}
\includegraphics[width = 2.5truein ]{Rpulse_downrange2.png}
\includegraphics[width = 2.5truein ]{Rpulse_uprange2.png}
\fi
\caption{Seismic pulse velocity and velocity direction as a function of time. 
a) In the left column we show velocity amplitude as a function of time from accelerometers at $R=6$ cm (thick line in blue)
and at $R=10$ cm (thick line in red) with $R$ the radius in cylindrical coordinates from the site of impact. The right column shows the velocity angle computed from the same accelerometers. An angle of zero corresponds to a horizontal direction and $90^\circ$ corresponds to moving upward.   The blue lines are slightly thicker and above the red ones in the left column.  In the right column, the red lines are on top of or above the blue lines.  
Lines have shade related to the strength of the velocity amplitude.  
Here the accelerometers are downrange of the impact site and at a depth of 3 cm.
The top row shows accelerometers from a single impact experiment at an impact angle of $\theta_I = 20^\circ$.  In the middle and bottom rows $\theta_I = 40^\circ$ and $60^\circ$, respectively.   The small arrows on the top right panel show the direction of motion at three angles. 
b) Similar to a) except for accelerometers uprange of the impact site. 
c) Similar to a) except the accelerometers are at a depth of 5 cm.
d) Similar to a) except the accelerometers are at a depth of 5 cm and are uprange.
\label{fig:pulseshapes}}
\end{figure*}

\subsection{Pulse shape and duration, and ray angles}

In this subsection, we examine the time-dependent behavior of the subsurface motions.  

Figure \ref{fig:pshape} shows the cylindrical radial component of acceleration $a_R $ and velocity $v_R$ as a function of time from accelerometers at the same depth (5 cm) and for experiments at different impact angles.
The top row shows $a_R$ and the bottom row shows $v_R$.  
The left column shows signals from accelerometers at cylindrical radius $R=6$ cm whereas those in the right column are at $8$ cm.  Solid lines and dotted lines show quantities from accelerometers downrange and uprange of the impact site, respectively. 
Signals are offset vertically (by 20 m/s$^2$ for acceleration and by 0.02 m/s for velocity) so that they can be compared, with impact angles labeled in the key in the upper right panel.  To facilitate comparison between pulses, we also shifted the horizontal positions of each pulse so that the peaks occur at $t = 0$.  

The pulse durations, measured from the accelerations  in Figure \ref{fig:pshape}, 
range from 0.3 to 0.65 ms (FWHM). 
These durations are short, in comparison to the time for crater excavation, listed in Table \ref{tab:crater}.  The time for a pressure wave to cross the projectile is $\lesssim 6\mu$s, (based on the elastic modulus of a few GPa for PLA plastic; \citealt{Farah_2016}) and is more than an order of magnitude shorter than our pulse durations.  Pulse duration is approximately consistent with a seismic source time $t_s \sim R_{cr}/v_P \sim 0.7$ ms estimated from a crater radius of 3.5 cm and a propagation speed of $v_P = 50$ m/s (see discussions by \citealt{Gudkova_2011,Quillen_2022} on possible scaling relations for pulse duration).  
There is no strong dependence of pulse duration on impact angle.  Pulse strength is weaker at a lower impact angle, consistent with the dependence of crater volume and crater efficiency on impact angle (as shown in Figures \ref{fig:vol} and \ref{fig:eff}) which were measured from the crater profiles.  

Figure \ref{fig:pshape} shows that uprange pulses tend to be weaker than downrange pulses, as expected from the asymmetries in the peak values shown in Figure \ref{fig:pulse_asym}.   We attribute the difference between the uprange and downrange accelerometer in the normal impact to errors in the accelerometer placement with respect to the site of impact. 

  
Material is launched off the surface if the vertical component of acceleration exceeds 1g. 
In the top two panels of Figure \ref{fig:pshape}, we see that pulse peak acceleration pulse strengths are well above 1g,  even though the accelerometers are outside the crater radius where surface material does not join the ejecta curtain.  The crater excavation time can be estimated from the ejecta curtain snapshots in Figure \ref{fig:ejecta} and exceeds 20 ms. This length of time is at least an order of magnitude longer than the seismic pulse durations. We infer that seismic pulse acceleration must be reduced near the surface.  
Indeed, for normal impacts, accelerometers placed on the surface showed weaker motions that took place during a longer time interval, compared to those even a few cm deep. The longer duration surface motions outside the crater were closer to the crater excavation time \citep{Neiderbach_2023}. 

In Figure \ref{fig:pulseshapes} we compare pulse shapes and directions as a function of time from pairs of accelerometers that are recorded in the same impact experiment. In each subfigure, each row shows a different impact angle, from top to bottom $\theta_I = 20$, 40, and 60$^\circ$.  In Figures \ref{fig:pulseshapes} a) and b) accelerometers are at a depth of 3~cm and in Figures and \ref{fig:pulseshapes} c) and d) accelerometers are deeper, at a depth of 5~cm.  The left columns show velocity amplitude and the right columns show the direction of the velocity vector with an angle of $0^\circ$ corresponding to a horizontal vector. 
Vector directions are illustrated in the top right panel of  Figure \ref{fig:pulseshapes}a.
The vector directions are predominantly above horizontal and also above the direction of $\hat r$ from the site of impact which gives a direction of -27 and -40$^\circ$ at $R=6$ and $|z| = 3$ and 5 cm respectively, and -17 and -27$^\circ$ at  $R=10$ and $|z| = 3$ and 5 cm, respectively.  
In Figure \ref{fig:pulseshapes} the horizontal time axes are the same in all panels. 
Thicker blue lines show accelerometers at $R=6$ cm and thinner red lines show those at $R=10$ cm from the site of impact.  Lines are shaded according to the strength of the velocity magnitude. 

A comparison between Figures \ref{fig:pulseshapes} a) and c) and between b) and d) show that pulses are longer duration nearer the surface. 
A comparison between the top and bottom panels of a) shows that near the surface pulses are particularly long for grazing impacts.  This suggests that the induced flow field is shallower for the grazing impacts than the nearly normal ones. In all subfigures, by comparing $R=6$ cm pulses with those at $R=10$ cm that are more distant from the site of impact and at the same depth,  we see that pulses are smoother (less triangular) further from the site of impact. Smoothing of pulse shape was interpreted in terms of a diffusive model for momentum by \citet{Quillen_2022} but only in the context of pulses propagating in one dimension. 

Examination of any of the subfigures in Figures \ref{fig:pulseshapes} shows that the pulse peaks are later at $R=10$ cm than at $R=6$ cm. However, the delay between the pulse peaks is longer at a shallower depth (comparing a) to c) or comparing b) to d).   The speed of pulse travel may be slower at shallower depths.  For these pulses, the pressure amplitude is only higher than hydrostatic pressure within about 6 cm of the site of impact and a depth of about 5 cm (as estimated via $P_{pk} \sim \rho_s |v|_{pk} v_P)$ \citealt{Quillen_2022}). 
A pulse travel speed of about 50 m/s in the same medium was estimated by \citet{Quillen_2022}. 
At 50 m/s it takes only about 1 ms to travel about 4 cm (the distance between the accelerometers) and this is approximately consistent with the delays between the deeper accelerometers seen in c) and d).  
Comparison of Figures \ref{fig:pulseshapes}
a) and b) or b) and d) implies that there is no significant difference between uprange and downrange pulse arrival times.  

The velocity ray angles (shown in the right columns in each subfigure) tend to increase with time, but primarily for accelerometers more distant from the site of impact.  This tendency for upward flow at later times is also visible in Figure \ref{fig:ob_ray} showing ray angles. Differences in velocity directions among different radial distances seem larger on the downrange side than on the uprange side of impact.

A Maxwell Z-model gives a fixed angle for flow at each subsurface location. 
The variations in ray angle as a function of time shown in the right columns of Figure \ref{fig:pulseshapes} imply that a static Maxwell Z-model would not describe the directions. To match the ray angles with a variant of the Maxwell Z-model, a time-dependent exponent or flow center (or possibly both) would be required.   
While the Maxwell Z-model roughly characterizes flow directions in excavation flows, it would be difficult to modify it to match the time-dependent phenomena seen here that are also a function of impact angle. 

In Figure \ref{fig:pulseshapes} vertical scales for the left panels are not the same because the pulse heights differ.  
The pulse shapes in the different subfigures are similar, suggesting that the primary difference between uprange and downrange pulses and between flows generated by different impact angles is in the pulse amplitudes.   
Perhaps a model for the time-dependent flow field for a single impact could be extended to match a larger set of oblique impacts by adjusting the amplitude of the velocity field.  

\begin{figure}[htbp]
\centering
\includegraphics[width=3.3truein]{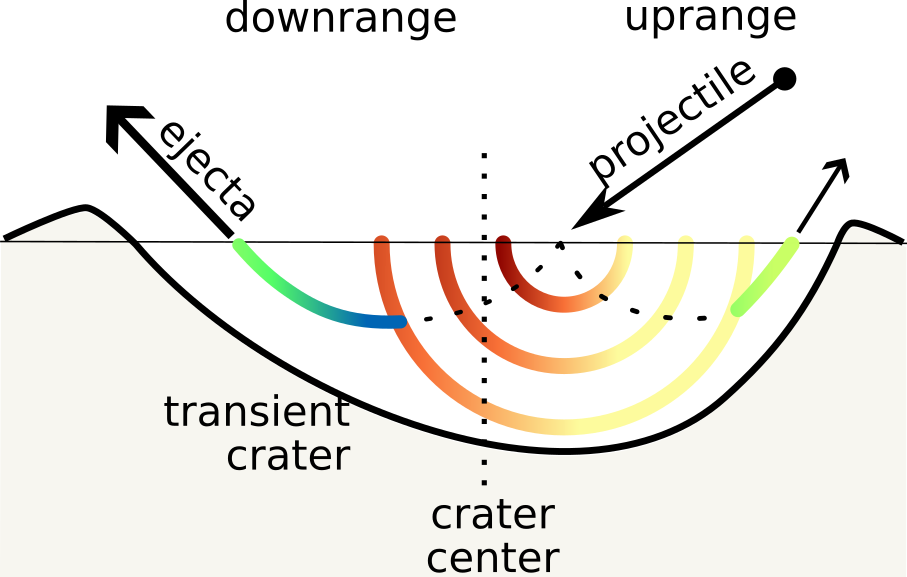}
\caption{An illustration of an excited pulse and an excavation flow caused by an oblique impact. A compressive seismic pulse is launched by the impact, but it is stronger on the downrange side than on the uprange side.
This phase is shown in red, orange and yellow and with darker hues representing a stronger pulse amplitude. As the pressure is released, the direction of flow tilts toward the surface, and the flow resembles a Maxwell Z-model.  This phase of the flow is shown in blue and green.  When the pulse reaches the surface it launches ejecta but more ejecta is launched on the downrange side. The resulting crater is lopsided with the crater center offset from the site of impact. }
\label{fig:crater_illust}
\end{figure}

In Figure \ref{fig:crater_illust}, we illustrate the phenomena seen in our experiments.  The projectile launches a compressive pulse that propagates away from the site of impact.  Because the medium exerts a drag force on the projectile, the pulse is stronger on the downrange side than on the uprange side,  as illustrated by the different hues.  The arcs are darker red on the downrange side indicating that the pulse is stronger on this side.   As the pressure from the pulse is released, the velocity in the pulse changes direction and points upward.  The streamlines are shown in green.  When the pulse reaches the surface, it launches ejecta, but with more ejecta launched on the downrange side than on the uprange side.  Because the pulse is stronger on the downrange side, excavation takes longer on that side. The resulting crater has a center that is offset from the site of impact. Outside the crater radius, the flow is similar but upon reaching the surface, there is plastic deformation rather than ejecta launched \citep{Neiderbach_2023}. 
Seismic energy is dissipated at the surface and not reflected back into the medium. 

The illustration of Figure \ref{fig:crater_illust} does not show a depth or time dependence for the flow field center,  which also could be present.  Nor does it capture complex near-surface behavior where short-duration pulses broaden to better match the crater excavation time \citep{Neiderbach_2023}. 
This illustration combines phenomena seen in our experiments and in measurements of ejecta curtains of oblique impacts \citep{Anderson_2003,Anderson_2004,Anderson_2006} with ideas from the shock and flow field model by \citet{Kurosawa_2019} for an impact into a solid. 

\subsection{Scaling subsurface pulse amplitude with projectile momentum}
\label{sec:rescale}

\begin{figure}[htbp]
    \centering
    \includegraphics[width=3.3truein,trim = 0 5 0 0, clip]{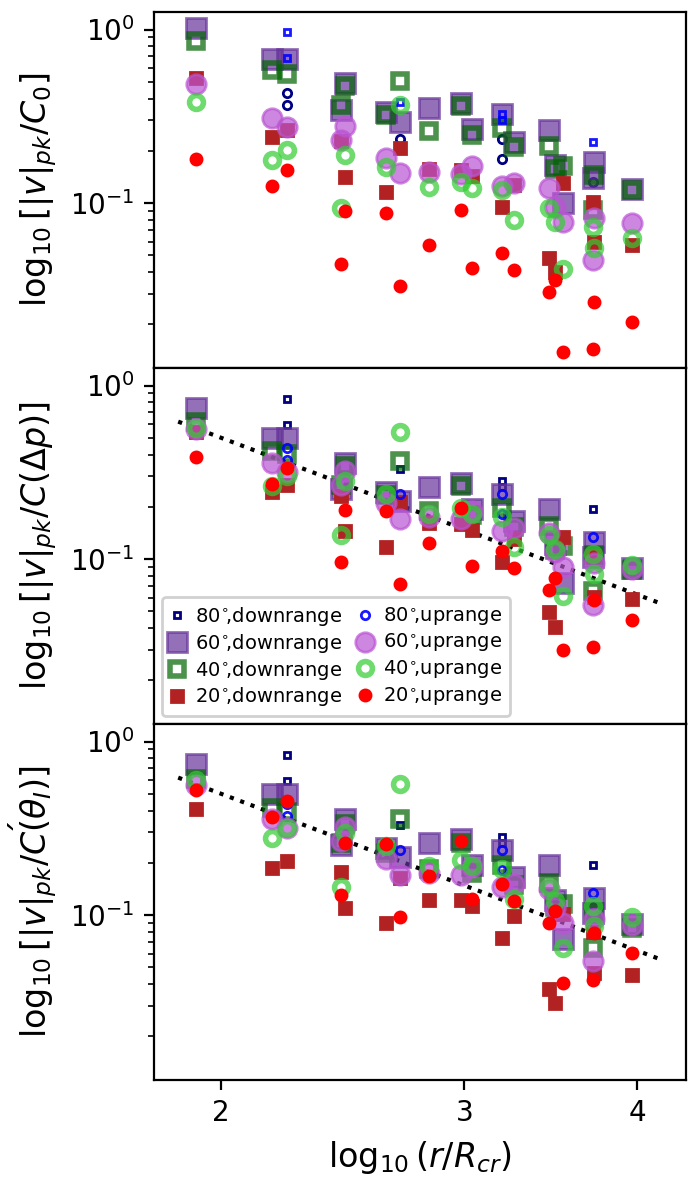}
    \caption{Peak velocity amplitude as a function of distance from the site of impact.  In all panels, we plot points for experiments at impact angle $\theta_I = 20,40,60$ and $80^\circ$.  Squares show downrange accelerometer locations and circles show uprange locations.   In the top panel, the peak velocity is normalized with the momentum of the projectile using equation \ref{eqn:vcr}. 
    In the middle panel, the peak velocity is normalized with a factor that depends upon the components of momentum imparted to the medium and takes into account the momentum carried away by ricochet (given in equation \ref{eqn:cases}). 
    The bottom panel, the peak velocity is normalized with a factor that only depends on the initial projectile momentum and impact angle. For normalization in the middle and bottom panels, 
    we assume that the uprange pulse velocity peak only depends on the z-component of momentum but the downrange peak velocity depends on both horizontal and vertical momentum components.  
    The scatter is reduced in the middle and bottom panels. 
    The dotted lines show a power law with an exponent of -3.
    }
    \label{fig:rescale}
\end{figure}

In the top panel of Figure \ref{fig:rescale}  we plot peak velocity amplitudes as a function of distance from impact site $r$ for the impact experiments with impact angle $\theta_I = 20, 40,60$ and $80^\circ$. 
Except for the impacts at $80^\circ$, these are the same values shown in Figure \ref{fig:vfit} but they are plotted together on the same plot. Uprange accelerometer locations are shown with dots or circles, whereas downrange locations are shown with squares. A key for all panels is shown in the middle panel. 
The x-axis shows the log of $r/R_{cr}$
where we used $R_{cr}=3.5$ cm based on the crater radius for a normal impact.

In the top panel, we plot 
$|v|_{pk}/C_0$ where the normalization factor 
\begin{equation}
    C_0 = \frac{M_p v_{imp}}{\rho_s 2 \pi R_{cr}^3} 
\label{eqn:vcr}
\end{equation}
is a dimensional estimate for the peak pulse velocity at the crater radius that is based on the projectile momentum (equation 50 by \citealt{Quillen_2022} with dimensionless coefficient $B_{\rm eff} =1$).   
In the middle panel of Figure \ref{fig:rescale}, we normalize the peak velocities with 
\begin{equation}
C(\Delta {\bf p}) = 
\frac{1}{\rho_s 2 \pi R_{cr}^3} \times
\begin{cases}
    \Delta p_z, & \text{if uprange}\\
    \Delta p_z + \Delta p_x, & \text{if downrange}
\end{cases}. \label{eqn:cases}
\end{equation}
This replaces projectile momentum magnitude with a function that depends on $\Delta p_x$ and $\Delta p_y$,  the components of the momentum imparted by the projectile into the substrate (see equation \ref{eqn:Deltap}).  These are listed in Table \ref{tab:crater} and take into account the momentum that is carried away if the projectile ricochets.
For a normal impact $M_p v_{imp} = \Delta p_z$  and the normalization coefficient reduces to that used in equation \ref{eqn:vcr}; $B(\Delta {\bf p}) = C_0$. 
The ansatz made in choosing the normalization factors in equation \ref{eqn:cases} is that the strength of the uprange pulses and flow field is set by the vertical component of the momentum that is imparted to the medium, but the downrange pulse strengths are influenced by both horizontal and vertical components.  
In the bottom panel of Figure \ref{fig:rescale},  we renormalize with a factor that depends on the projectile momentum and impact angle  (and does not take into account possible ricochet) 
\begin{equation}
C'(\theta_I) = 
C_0 \times 
\begin{cases}
        \sin \theta_I, & \text{if uprange}\\
        \sin\theta_I + \cos \theta_I, & \text{if downrange}
\end{cases}. \label{eqn:cases2}
\end{equation}
Our simple choices of normalization reduce the scatter in the second and third panels in Figure \ref{fig:rescale}. 

The dotted line in the second panel of Figure \ref{fig:rescale} shows the line 
\begin{equation}
|v|_{pk}(r) = 4 C(\Delta {\bf p}) \left(\frac{r}{R_{cr}} \right)^{-3} . \label{eqn:obscale}
\end{equation}
The dotted line in the third panel is similar, showing 
\begin{equation}
|v|_{pk}(r) = 
4 C'(\theta_I) \left(\frac{r}{R_{cr}} \right)^{-3}. \label{eqn:obscale2}
\end{equation}
The exponent is slightly steeper than the -2.5 exponent used by \citet{Quillen_2022}.  A line with a slope of -2.5 would be about as good a match to the points shown here if the scaling factor is lower, 2.4 instead of 4.
The standard deviations of residuals from the dotted lines in the second and third panels of Figure \ref{fig:rescale} are equivalent.  
The coefficient of 4 in equations  \ref{eqn:obscale} and \ref{eqn:obscale2} is similar to that used for matching normal impacts \citep{Quillen_2022}.
Equations \ref{eqn:obscale} and \ref{eqn:obscale2} give an estimate for subsurface peak pulse velocity up or downrange from the impact site. Here  $R_{cr}$ is the crater radius for a normal impact and can be estimated using crater scaling laws.  

If the projectile does not ricochet, it is straightforward to estimate the momentum transferred to the medium at the time of impact $\Delta {\bf p}$.  However, at grazing angles, equation \ref{eqn:obscale} is not useful unless the momentum of the projectile after ricochet can be predicted. 
A number of recent studies have focused on whether or not projectiles ricochet \citep{Wright_2020b,Wright_2022,Miklavcic_2022,Miklavcic_2023}, but only a few studies have measured the fraction of momentum carried away by the projectile when it does ricochet (e.g.,  \citealt{Wright_2022}).  Nevertheless, we suspect that projectile ricochet affects the strength and angular dependence of the excited subsurface seismic disturbance, particularly in the velocity regime studied here. 
In the absence of models for ricochet, equation \ref{eqn:obscale2} serves and is as good a match to the peak velocities as a function of $r$.  A model that predicts pulse peak velocities as a function of azimuthal angle and depth as well as $r$ might further reduce the scatter.   

\section{Formation of the Sky Crater on Arrokoth}

\begin{figure}[htbp]
    \centering
    \includegraphics[width=3truein]{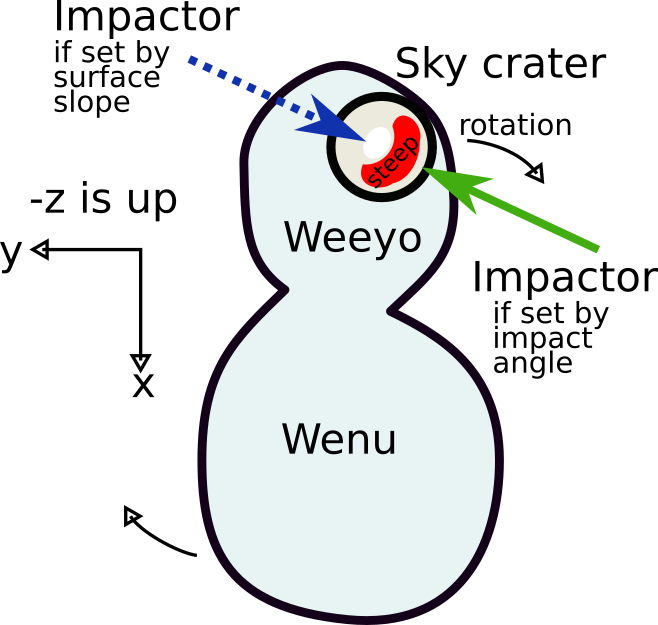}
    \caption{Illustration of the Sky crater on Arrokoth. One side of the crater, shown in red, has steeper surface slopes than the other, as shown in Figure 8 by \citet{Keane_2022}. If the steeper side of the crater is due to the impact angle, we suspect the projectile came from the right.  If the stepper side is due to the initial surface slope, the projectile could have come from the left. We follow \citet{Keane_2022} for the Cartesian coordinate system. }
    \label{fig:sky}
\end{figure}

(486958) Arrokoth (formerly 2014 MU69) is a bilobate cold classical Transneptunian object that was observed as part of the New Horizons extended mission \citep{Spencer_2020}.  
We discuss the response of the body to an impact that formed Arrokoth’s largest crater, the Sky Crater (formerly “Maryland”) on Weeyo, the smaller of Arrokoth's two lobes.
The impact that formed the Sky crater is estimated to be at a speed of about 0.5 km/s and with a projectile diameter of about 1~km \citep{McKinnon_2022}, (see their Figure 3). 
The body's mass density is not well constrained but a low value of 500 kg m$^{-3}$ is considered plausible (e.g., \citealt{Spencer_2020}). 
The low density suggests that the body is porous \citep{Spencer_2020}. Images of the crater suggest that it has a conical shape,  typical of craters in granular media.  As Arrokoth could be a granular system that experienced an impact at a velocity similar to our experiments, we discuss the formation of the Sky Crater in the context of what we have learned from studying oblique impacts. 


Stress associated with the impact that formed the Sky Crater could have caused the narrow neck between Weeyo and Wenu, the two lobes, to break \citep{Hirabayashi_2020}, however, \citet{McKinnon_2022} argued that this was unlikely. By taking into account the direction-dependent strength and decay rate of an oblique impact-generated seismic pulse, we can independently estimate the stress in Arrokoth's neck to reexamine this issue.   

Following \citet{Keane_2022}, because of Arrokoth's unusual bilobate shape, a Cartesian coordinate system is used to describe its features.  The z-axis is aligned with Arrokoth's spin axis with a positive direction defined by the right-hand rule. The x-axis is defined to be perpendicular to the z-axis and aligned with Arrokoth's long axis and with a positive x pointed toward the larger lobe. 
The surface slope at locations on Arrokoth's surface is defined as the angle between the surface normal vector and the acceleration vector due to self-gravity and rotation (centrifugal acceleration). In the Sky Crater, the surface slope is higher on the -y side of the crater, as shown in Figure 8 by \citet{Keane_2022}, and illustrated with the red region in Figure \ref{fig:sky}. 

We consider the possibility that variations within the Sky crater's slope are related to impact angle. 
As steeper crater slopes tend to be found on the uprange side of an oblique impact (as discussed in section \ref{sec:shapes}), the high slopes on the -y side of Sky crater would suggest that the bearing direction of the impactor was in approximately the +y direction, as shown with the green arrow in Figure \ref{fig:sky} (and it came from the -y direction), giving an impact that decreased the body's spin. 

Crater slope is also sensitive to the substrate's initial slope \citep{Takizawa_2020}.  Because Arrokoth has a flattened shape (is quite thin in the $\pm z$ direction), points with high $|z|$ values, in the center of the smaller lobe, are downhill of the points at the outer edge of this lobe.  If the projectile comes from the upslope direction, it bulldozes material in front of it, making the downrange crater side steeper than the uprange side \citep{Takizawa_2020}.
If the crater slope asymmetry is due to the initial substrate slope, because the -y side of the crater is steeper, the projectile could have originated from the shallower and higher side, coming from the +y side, agreeing with the dotted blue arrow for the projectile shown in Figure \ref{fig:sky}. 
Independent of whether the steeper side of the crater is due to the initial slope of the surface or because the impact was oblique, we suspect that the projectile  came along a tangential direction. 

The distance of crater center from the neck is about $L=11$ km. The crater diameter is about 7 km in diameter with a depth of about 0.5 km \citep{Spencer_2020}.
\citet{McKinnon_2022} estimate a momentum of the projectile is $\Delta p = 5 \times 10^{13}$ kg m s$^{-1}$. 
They estimate the travel time to the neck as $\Delta t = L/v_P = 100$ s, using a travel speed of 100 m/s adopted by \citet{cooper74,McKinnon_2022}.  They estimate the stress on the neck as $\sigma \sim \frac{\Delta p }{\Delta t A_N} \sim 15$ kPa where $A_N = 30.5$ km$^2$ is the cross sectional area at the neck.  
This potentially gives a stress higher than a few kPa which could exceed the tensile or compressive strength at the neck.   Based on the total kinetic energy that could be imparted to Weeyo, \citet{McKinnon_2022} argued that the impact would only crush material at the neck. 

We consider a seismic pulse propagating from the crater toward Arrokoth's neck.  In our experiments, we see little energy reflected from the surface, rather pulse energy seems to go into launching ejecta, or plastic deformation outside the crater radius \citep{Neiderbach_2023}. Thus the pressure in the pulse would primarily depend upon the distance from the site of impact. We do not expect focusing of seimic energy at Arrokoth's neck.   
We approximate the vertical and horizontal components of projectile momentum transferred to Weeyo $\Delta p_y = \Delta p_z \sim \sqrt{2}\Delta p$ assuming a $45^\circ$ impact angle. Using Eq. \ref{eqn:cases} and Eq. \ref{eqn:obscale}, we estimate that the pulse peak velocity magnitude at the neck region is $|v|_{pk} \sim 6.7$ cm/s. The peak pressure at the neck we estimate from the peak velocity,  $P_{pk} \sim \rho_s v_P v_{pk}$ \citep{Quillen_2022}. Applying a P-wave pulse travel speed of $v_P = 100$ m/s adopted by \citet{McKinnon_2022}, we obtain pulse peak pressure $P_{pk}$ at the neck to be $\sim 3$ kPa. This is lower than that estimated by \citet{McKinnon_2022} as we have taken into account attenuation, based on the decay rates of pulses seen in our experiments. However, this stress value exceeds both tensile and compressive strength estimates for granular systems \citep{Brisset_2022}. The impact could have caused deformation throughout Weeyo if it was a granular system. 




Because of major axis slope asymmetry, we can estimate the bearing of the impactor that made the Sky Crater.  However, it is more difficult to estimate the impact angle because the slopes are likely to be sensitive to the static and dynamic angles of repose of the granular medium and we don't know how the angles of repose of Arrokoth's material compares to our laboratory sand.  
Furthermore, crater shape would be affected by inhomogeneity in the material properties, and Arrokoth's crater could be shallower than those in our lab because it has had more time to slump and relax and because much of the ejecta curtain escaped \citep{Mao_2021} rather than added to the crater rim.

We support the finding by \citet{McKinnon_2022} that the Sky Crater impact would at most partly crush or plastically deform material in Arrokoth's neck.
We have primarily discussed the formation of the Sky Crater to illustrate some of the issues involved in interpreting subsurface pulse propagation in astronomical bodies in context with what our experiments show in granular systems under laboratory conditions. 

\section{Summary and Discussion}

We have carried out a series of experiments of oblique impacts of airsoft BBs at a projectile velocity of about 104 m/s into the sand.    
Even at grazing impact angles as low as  $10^\circ$, the craters are nearly round. 
Evidence that the crater was formed by an oblique impact with an impact angle below $45^\circ $ can be inferred from variations in crater slope, with an uprange surface slope about $10^\circ$ higher than that downrange.  An oblique impact could also be inferred from variations in rim height and ejecta distribution, as the downrange side has a higher rim and a thicker ejecta blanket.   The sensitivity of ejecta angle with azimuthal direction is consistent with studies tracking ejecta \citep{Anderson_2004,Anderson_2006}. While differences in crater slope, volume, and ejecta are subtle, seismic pulses detected below the surface with accelerometers exhibit a remarkably large asymmetry between uprange and downrange positions.  We find that pulse peak velocity has a ratio of downrange to uprange amplitude as large as 5.  
This ratio is particularly large at shallow depths. 

We confirm prior studies \citep{Elbeshausen_2009,Takizawa_2020} finding that crater efficiency (as measured from the crater volume) is lower for grazing impacts than normal impacts, but in our experiments, this is in part due to the energy carried away by the projectile as it ricochets at impact angles below $\theta_I \sim 50^\circ$.

We considered a Maxwell Z-model for the velocity peak values at different subsurface locations. However, we find that time-dependent angle and velocity amplitude variations make it difficult for a simple modification of the Maxwell Z-model to match our subsurface pulse properties. 
Pulse shape (as a function of time) and direction are remarkably similar among experiments (shown in  Figure \ref{fig:pulseshapes}), suggesting that it may be possible to rescale subsurface velocity amplitudes and relate time-dependent models for flow at one impact angle to others. 
We succeeded in reducing scatter in plots of peak velocity amplitude versus distance from impact site via a normalization factor that depends on the projectile momentum and its direction. 

Our picture for the subsurface impact excited seismic pulse and generated excavation flow in granular systems, shown in Figure \ref{fig:crater_illust}, is remarkably similar to that developed for impact craters into solids, e.g., \citet{Melosh_1985,Kurosawa_2019}. However, analogies for shocks and rarefaction waves are challenging to describe in a granular system. 
The transition between elastic behavior typical of a solid, which we describe as subsurface pulse propagation and granular flow, which we describe as an excavation flow, perhaps could be described with a non-linear continuum model (e.g., \citealt{Agarwal_2021}), with discrete element simulations (e.g., \citealt{Miklavcic_2023,Sanchez_2022}) or modifications of semi-analytical models developed for impacts into fluids \citep{Lherm_2023}.

Impact-generated motions in granular systems are particularly challenging to model as they encompass elastic wave propagation through a granular medium and granular flow.  Numerical models for them are potentially powerful as they can cover low-g environments that are not accessible in our lab.  Ricochet in granular systems likely scales to low-g environments using the dimensionless Froude number \citep{Wright_2022, Miklavcic_2023}. 
In contrast, we suspect that pulse propagation in granular systems is sensitive to the pressure amplitude in the pulse \citep{Quillen_2022, Sanchez_2022}.  In low-strength materials, crater excavation should scale with the dimensionless $\pi_2$ parameter \citep{housen11} which is related to the Froude number.   The transition between subsurface seismic pulse propagation and crater excavation combines these two different physical scaling scenarios.   
Recent simulations of granular systems have been successful at predicting ricochet \citep{Miklavcic_2022, Miklavcic_2023} and propagation of seismic pulses through granular columns \citep{Sanchez_2022}.  
Similar numerical studies, developed to match subsurface response,  could improve upon our understanding of how pulses travel and how flow is driven by these pulses in granular systems so that we can extend and improve software used for modeling impacts on asteroids and other bodies in the Solar system (e.g., \citealt{Raducan_2022}). 

The largest uncertainties in our experiments are due to differences between the intended and actual accelerometer location with respect to the impact site.  Future experiments could improve upon techniques for accelerometer emplacement and increase the number of accelerometers so that the full 3-dimensional flow field in a single experiment can be characterized.  Of particular interest for future experiments are regions where the flow is least understood.  This might be where the transition from elastic phenomena to granular flow takes place,  near and just below the surface where the transition from narrow subsurface seismic pulses becomes a slower crater excavation flow. 
As astrophysical objects are not homogeneous,  future studies could also explore subsurface motions excited by impacts in polydisperse granular media. 

\section*{Acknowledgements}

We are grateful to Mokin Lee for discussing his work on simulations of oblique impacts.

\noindent
This work has been supported by 
NASA grant  80NSSC21K0143.  

\section*{Data availability and Supplemental Videos}

Datasets and analysis scripts related to this article can be found at \url{https://github.com/URGranularLab/Oblique_impact}, hosted at GitHub.

\begin{figure}[htbp]\centering
\includegraphics[width=3truein]{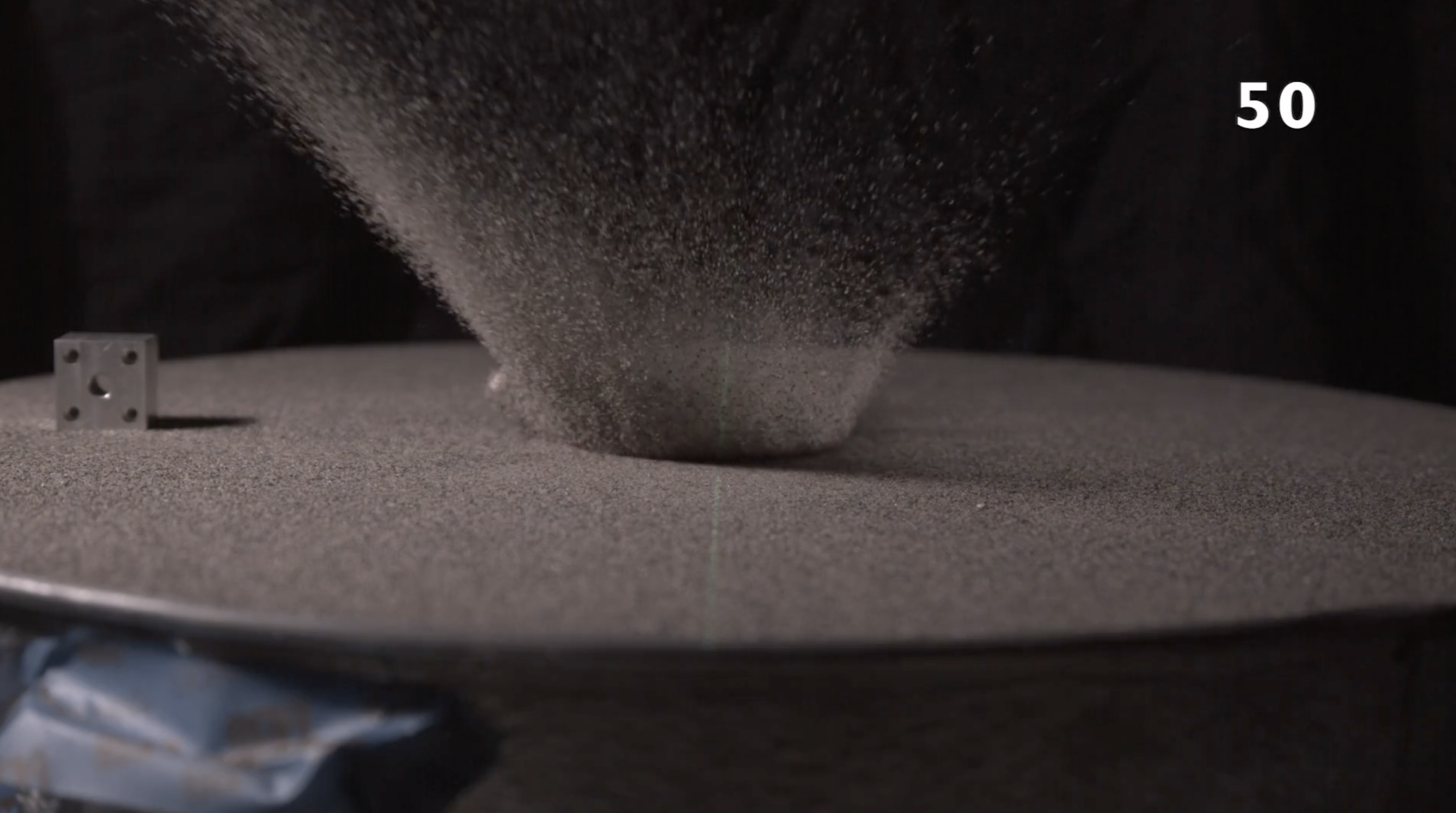}
\caption{
Supplemental video \texttt{Obliqueimpacts.mp4}.
This video is a concatenation of 
high-speed videos at 1000 fps at 8 different impact angles.  The impact angle $\theta_I$ (from horizontal) in degrees is shown on the upper right in degrees for each impact.  The aluminum block gives the scale and has long dimension 2.54 cm. 
Get this movie here: \url{https://drive.google.com/file/d/1mMJmNE8pUu3A_qhFcitfoJNXRcd17B-7/view?usp=drive_link}
}
\end{figure}

\begin{figure}[htbp]\centering
\includegraphics[width=3truein]{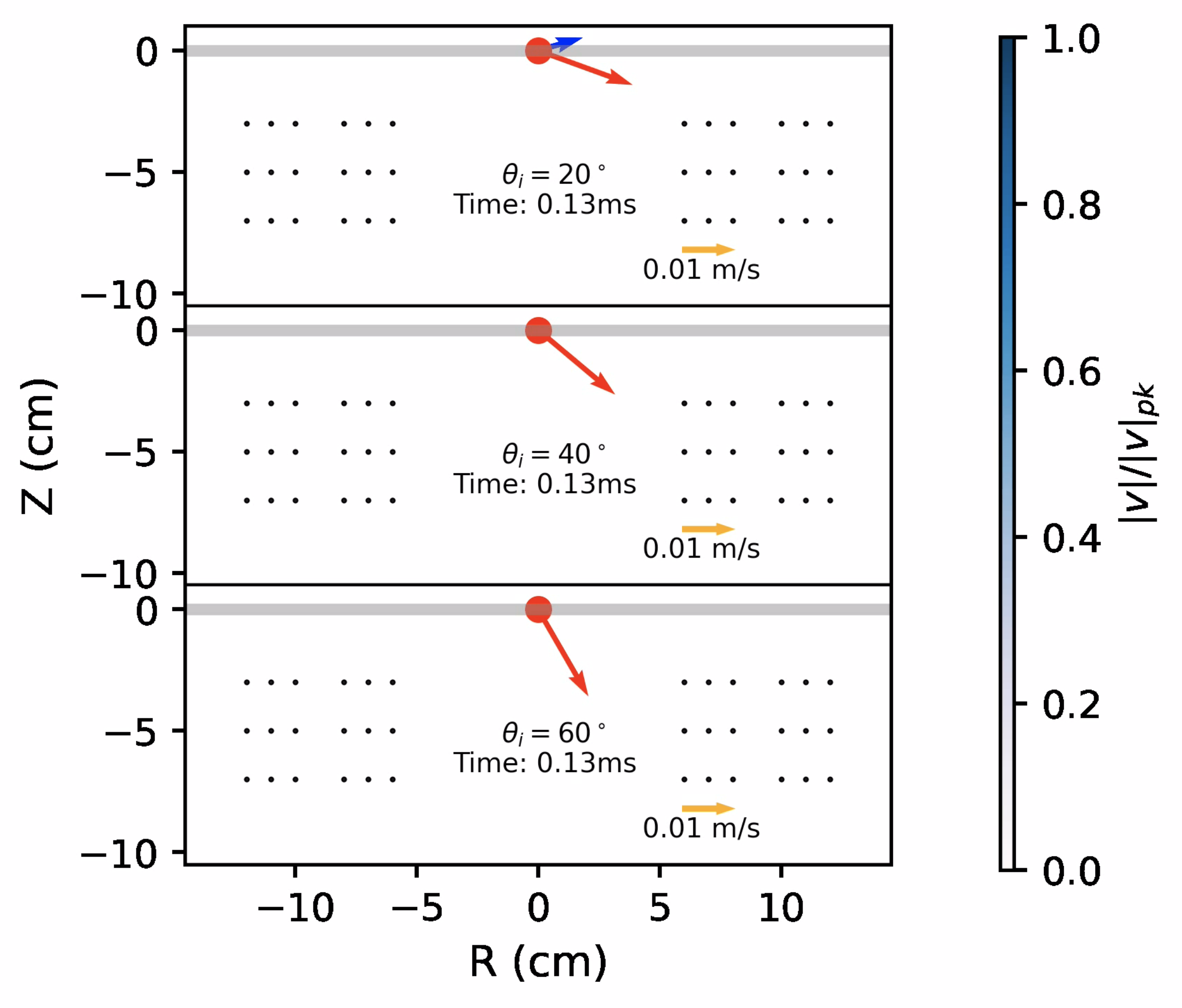}
\caption{
Supplemental video \texttt{ray\_angles.mp4}.
This video shows the velocity direction of subsurface pulses with a ray angle and as a function of time during impacts at impact angles of 20, 40 and $60^\circ$. This video animates the impact angles shown at three different times in Figure \ref{fig:ob_ray}. 
Get this movie here: \url{https://drive.google.com/file/d/1N-7OCYzAclfmmM2Efku5BsJmx7ptreOw/view?usp=sharing}.
}
\end{figure}

\bibliographystyle{elsarticle-harv}
\bibliography{refs.bib}

\begin{thebibliography}{66}
\expandafter\ifx\csname natexlab\endcsname\relax\def\natexlab#1{#1}\fi
\providecommand{\url}[1]{\texttt{#1}}
\providecommand{\href}[2]{#2}
\providecommand{\path}[1]{#1}
\providecommand{\DOIprefix}{doi:}
\providecommand{\ArXivprefix}{arXiv:}
\providecommand{\URLprefix}{URL: }
\providecommand{\Pubmedprefix}{pmid:}
\providecommand{\doi}[1]{\href{http://dx.doi.org/#1}{\path{#1}}}
\providecommand{\Pubmed}[1]{\href{pmid:#1}{\path{#1}}}
\providecommand{\bibinfo}[2]{#2}
\ifx\xfnm\relax \def\xfnm[#1]{\unskip,\space#1}\fi
\bibitem[{Agarwal et~al.(2021)Agarwal, Karsai, Goldman and
  Kamrin}]{Agarwal_2021}
\bibinfo{author}{Agarwal, S.}, \bibinfo{author}{Karsai, A.},
  \bibinfo{author}{Goldman, D.I.}, \bibinfo{author}{Kamrin, K.},
  \bibinfo{year}{2021}.
\newblock \bibinfo{title}{Efficacy of simple continuum models for diverse
  granular intrusions}.
\newblock \bibinfo{journal}{Soft Matter} \bibinfo{volume}{17},
  \bibinfo{pages}{7196--7209}.
\newblock \URLprefix \url{http://dx.doi.org/10.1039/D1SM00130B},
  \DOIprefix\doi{10.1039/D1SM00130B}.
\bibitem[{Anderson and Schultz(2006)}]{Anderson_2006}
\bibinfo{author}{Anderson, J.}, \bibinfo{author}{Schultz, P.H.},
  \bibinfo{year}{2006}.
\newblock \bibinfo{title}{Flow-field center migration during vertical and
  oblique impacts}.
\newblock \bibinfo{journal}{International Journal of Impact Engineering}
  \bibinfo{volume}{33}, \bibinfo{pages}{35--44}.
\bibitem[{{Anderson} et~al.(2003){Anderson}, {Schultz} and
  {Heineck}}]{Anderson_2003}
\bibinfo{author}{{Anderson}, J.L.B.}, \bibinfo{author}{{Schultz}, P.H.},
  \bibinfo{author}{{Heineck}, J.T.}, \bibinfo{year}{2003}.
\newblock \bibinfo{title}{{Asymmetry of ejecta flow during oblique impacts
  using three-dimensional particle image velocimetry}}.
\newblock \bibinfo{journal}{Journal of Geophysical Research (Planets)}
  \bibinfo{volume}{108}, \bibinfo{pages}{5094}.
\newblock \DOIprefix\doi{10.1029/2003JE002075}.
\bibitem[{{Anderson} et~al.(2004){Anderson}, {Schultz} and
  {Heineck}}]{Anderson_2004}
\bibinfo{author}{{Anderson}, J.L.B.}, \bibinfo{author}{{Schultz}, P.H.},
  \bibinfo{author}{{Heineck}, J.T.}, \bibinfo{year}{2004}.
\newblock \bibinfo{title}{{Experimental ejection angles for oblique impacts:
  Implications for the subsurface flow-field}}.
\newblock \bibinfo{journal}{Meteoritics \& Planetary Science}
  \bibinfo{volume}{39}, \bibinfo{pages}{303--320}.
\newblock \DOIprefix\doi{10.1111/j.1945-5100.2004.tb00342.x}.
\bibitem[{{Ballouz} et~al.(2021){Ballouz}, {Walsh}, {S{\'a}nchez}, {Holsapple},
  {Michel}, {Scheeres}, {Zhang}, {Richardson}, {Barnouin}, {Nolan}, {Bierhaus},
  {Connolly}, {Schwartz}, {{\c{C}}elik}, {Baba} and {Lauretta}}]{Ballouz_2021}
\bibinfo{author}{{Ballouz}, R.L.}, \bibinfo{author}{{Walsh}, K.J.},
  \bibinfo{author}{{S{\'a}nchez}, P.}, \bibinfo{author}{{Holsapple}, K.A.},
  \bibinfo{author}{{Michel}, P.}, \bibinfo{author}{{Scheeres}, D.J.},
  \bibinfo{author}{{Zhang}, Y.}, \bibinfo{author}{{Richardson}, D.C.},
  \bibinfo{author}{{Barnouin}, O.S.}, \bibinfo{author}{{Nolan}, M.C.},
  \bibinfo{author}{{Bierhaus}, E.B.}, \bibinfo{author}{{Connolly}, H.C.},
  \bibinfo{author}{{Schwartz}, S.R.}, \bibinfo{author}{{{\c{C}}elik}, O.},
  \bibinfo{author}{{Baba}, M.}, \bibinfo{author}{{Lauretta}, D.S.},
  \bibinfo{year}{2021}.
\newblock \bibinfo{title}{{Modified granular impact force laws for the
  OSIRIS-REx touchdown on the surface of asteroid (101955) Bennu}}.
\newblock \bibinfo{journal}{Monthly Notices of the Royal Astronomical Society}
  \bibinfo{volume}{507}, \bibinfo{pages}{5087--5105}.
\newblock \DOIprefix\doi{10.1093/mnras/stab2365}.
\bibitem[{Bester et~al.(2019)Bester, Cox, Zheng and Behringer}]{Bester_2019}
\bibinfo{author}{Bester, C.S.}, \bibinfo{author}{Cox, N.},
  \bibinfo{author}{Zheng, H.}, \bibinfo{author}{Behringer, R.P.},
  \bibinfo{year}{2019}.
\newblock \bibinfo{title}{Dynamics of oblique impact in a photoelastic granular
  medium}.
\newblock \bibinfo{journal}{arXiv preprint arXiv:1904.11077} .
\bibitem[{{Bottke} et~al.(2000){Bottke}, {Love}, {Tytell} and
  {Glotch}}]{Bottke_2000}
\bibinfo{author}{{Bottke}, W.F.}, \bibinfo{author}{{Love}, S.G.},
  \bibinfo{author}{{Tytell}, D.}, \bibinfo{author}{{Glotch}, T.},
  \bibinfo{year}{2000}.
\newblock \bibinfo{title}{{Interpreting the Elliptical Crater Populations on
  Mars, Venus, and the Moon}}.
\newblock \bibinfo{journal}{Icarus} \bibinfo{volume}{145},
  \bibinfo{pages}{108--121}.
\newblock \DOIprefix\doi{10.1006/icar.1999.6323}.
\bibitem[{{Brisset} et~al.(2022){Brisset}, {S{\'a}nchez}, {Cox}, {Corraliza},
  {Hatchitt}, {Madison} and {Miletich}}]{Brisset_2022}
\bibinfo{author}{{Brisset}, J.}, \bibinfo{author}{{S{\'a}nchez}, P.},
  \bibinfo{author}{{Cox}, C.}, \bibinfo{author}{{Corraliza}, D.},
  \bibinfo{author}{{Hatchitt}, J.}, \bibinfo{author}{{Madison}, A.},
  \bibinfo{author}{{Miletich}, T.}, \bibinfo{year}{2022}.
\newblock \bibinfo{title}{{Asteroid regolith strength: Role of grain size and
  surface properties}}.
\newblock \bibinfo{journal}{Planetary and Space Science} \bibinfo{volume}{220},
  \bibinfo{pages}{105533}.
\newblock \DOIprefix\doi{10.1016/j.pss.2022.105533}.
\bibitem[{Carrigy(1970)}]{Carrigy_1970}
\bibinfo{author}{Carrigy, M.A.}, \bibinfo{year}{1970}.
\newblock \bibinfo{title}{Experiments on the angles of repose of granular
  materials 1}.
\newblock \bibinfo{journal}{Sedimentology} \bibinfo{volume}{14},
  \bibinfo{pages}{147--158}.
\bibitem[{\c{C}elic et~al.(2022)\c{C}elic, Ballouz, Scheeres and
  Kawakatsu}]{Celik_2022}
\bibinfo{author}{\c{C}elic, O.}, \bibinfo{author}{Ballouz, R.L.},
  \bibinfo{author}{Scheeres, D.L.}, \bibinfo{author}{Kawakatsu, Y.},
  \bibinfo{year}{2022}.
\newblock \bibinfo{title}{A numerical simulation approach to the crater-scaling
  relationships in low-speed impacts under microgravity}.
\newblock \bibinfo{journal}{Icarus} \bibinfo{volume}{377},
  \bibinfo{pages}{114882}.
\bibitem[{{{\c{C}}elik} et~al.(2019){{\c{C}}elik}, {Baresi}, {Ballouz},
  {Ogawa}, {Wada} and {Kawakatsu}}]{Celik_2019}
\bibinfo{author}{{{\c{C}}elik}, O.}, \bibinfo{author}{{Baresi}, N.},
  \bibinfo{author}{{Ballouz}, R.L.}, \bibinfo{author}{{Ogawa}, K.},
  \bibinfo{author}{{Wada}, K.}, \bibinfo{author}{{Kawakatsu}, Y.},
  \bibinfo{year}{2019}.
\newblock \bibinfo{title}{{Ballistic deployment from quasi-satellite orbits
  around Phobos under realistic dynamical and surface environment
  constraints}}.
\newblock \bibinfo{journal}{Planetary and Space Science} \bibinfo{volume}{178},
  \bibinfo{pages}{104693}.
\newblock \DOIprefix\doi{10.1016/j.pss.2019.06.010}.
\bibitem[{{Chapman} and {McKinnon}(1986)}]{Chapman_1986}
\bibinfo{author}{{Chapman}, C.R.}, \bibinfo{author}{{McKinnon}, W.B.},
  \bibinfo{year}{1986}.
\newblock \bibinfo{title}{{Cratering of planetary satellites.}}, in:
  \bibinfo{editor}{{Burns}, J.A.}, \bibinfo{editor}{{Matthews}, M.S.} (Eds.),
  \bibinfo{booktitle}{IAU Colloq. 77: Some Background about Satellites}, pp.
  \bibinfo{pages}{492--580}.
\bibitem[{{Cline} and {Cintala}(2022)}]{Cline_2022}
\bibinfo{author}{{Cline}, C.J.}, \bibinfo{author}{{Cintala}, M.J.},
  \bibinfo{year}{2022}.
\newblock \bibinfo{title}{{The effects of target density, porosity, and
  friction on impact crater morphometry: Exploratory experimentation using
  various granular materials}}.
\newblock \bibinfo{journal}{Meteoritics \& Planetary Science}
  \bibinfo{volume}{57}, \bibinfo{pages}{1441--1459}.
\newblock \DOIprefix\doi{10.1111/maps.13886}.
\bibitem[{{Collins} et~al.(2011){Collins}, {Elbeshausen}, {Davison}, {Robbins}
  and {Hynek}}]{Collins_2011}
\bibinfo{author}{{Collins}, G.S.}, \bibinfo{author}{{Elbeshausen}, D.},
  \bibinfo{author}{{Davison}, T.M.}, \bibinfo{author}{{Robbins}, S.J.},
  \bibinfo{author}{{Hynek}, B.M.}, \bibinfo{year}{2011}.
\newblock \bibinfo{title}{{The size-frequency distribution of elliptical impact
  craters}}.
\newblock \bibinfo{journal}{Earth and Planetary Science Letters}
  \bibinfo{volume}{310}, \bibinfo{pages}{1--8}.
\newblock \DOIprefix\doi{10.1016/j.epsl.2011.07.023}.
\bibitem[{Cooper et~al.(1974)Cooper, Kovach and Watkins}]{cooper74}
\bibinfo{author}{Cooper, M.R.}, \bibinfo{author}{Kovach, R.L.},
  \bibinfo{author}{Watkins, J.S.}, \bibinfo{year}{1974}.
\newblock \bibinfo{title}{Lunar near-surface structure}.
\newblock \bibinfo{journal}{Reviews of Geophysics and Space Sciences}
  \bibinfo{volume}{12}, \bibinfo{pages}{291--308}.
\bibitem[{Croft(1981)}]{Croft_1981}
\bibinfo{author}{Croft, S.K.}, \bibinfo{year}{1981}.
\newblock \bibinfo{title}{The excavation stage of basin formation: A
  qualitative model in multi-ring basins}, in: \bibinfo{editor}{Schultz, P.H.},
  \bibinfo{editor}{Merrill, R.B.} (Eds.), \bibinfo{booktitle}{Multi-ring
  basins: Formation and evolution: Proceedings of the Lunar and Planetary
  Science Conference, Houston, TX, November 10-12, 1980},
  \bibinfo{publisher}{Pergamon Press, New York.}. pp.
  \bibinfo{pages}{207--225}.
\bibitem[{Dahl and Schultz(2001)}]{Dahl_2001}
\bibinfo{author}{Dahl, J.M.}, \bibinfo{author}{Schultz, P.H.},
  \bibinfo{year}{2001}.
\newblock \bibinfo{title}{Measurement of stress wave asymmetries in
  hypervelocity projectile impact experiments}.
\newblock \bibinfo{journal}{International Journal of Impact Engineering}
  \bibinfo{volume}{26}, \bibinfo{pages}{145--155}.
\newblock \URLprefix
  \url{https://www.sciencedirect.com/science/article/pii/S0734743X0100077X},
  \DOIprefix\doi{https://doi.org/10.1016/S0734-743X(01)00077-X}.
\bibitem[{Dalal and Triggs(2005)}]{Dalal_2005}
\bibinfo{author}{Dalal, N.}, \bibinfo{author}{Triggs, B.},
  \bibinfo{year}{2005}.
\newblock \bibinfo{title}{Histograms of oriented gradients for human
  detection}, in: \bibinfo{booktitle}{2005 {IEEE} Computer Society Conference
  on Computer Vision and Pattern Recognition ({CVPR} 05)},
  \bibinfo{publisher}{{IEEE}}. pp. \bibinfo{pages}{886--893}.
\newblock \URLprefix \url{https://doi.org/10.1109%2Fcvpr.2005.177},
  \DOIprefix\doi{10.1109/cvpr.2005.177}.
\bibitem[{{Elbeshausen} et~al.(2009){Elbeshausen}, {W{\"u}nnemann} and
  {Collins}}]{Elbeshausen_2009}
\bibinfo{author}{{Elbeshausen}, D.}, \bibinfo{author}{{W{\"u}nnemann}, K.},
  \bibinfo{author}{{Collins}, G.S.}, \bibinfo{year}{2009}.
\newblock \bibinfo{title}{{Scaling of oblique impacts in frictional targets:
  Implications for crater size and formation mechanisms}}.
\newblock \bibinfo{journal}{Icarus} \bibinfo{volume}{204},
  \bibinfo{pages}{716--731}.
\newblock \DOIprefix\doi{10.1016/j.icarus.2009.07.018}.
\bibitem[{Elbeshausen et~al.(2013)Elbeshausen, W{\"u}nnemann and
  Collins}]{Elbeshausen_2013}
\bibinfo{author}{Elbeshausen, D.}, \bibinfo{author}{W{\"u}nnemann, K.},
  \bibinfo{author}{Collins, G.S.}, \bibinfo{year}{2013}.
\newblock \bibinfo{title}{The transition from circular to elliptical impact
  craters}.
\newblock \bibinfo{journal}{Journal of Geophysical Research: Planets}
  \bibinfo{volume}{118}, \bibinfo{pages}{2295--2309}.
\bibitem[{Farah et~al.(2016)Farah, Anderson and Langer}]{Farah_2016}
\bibinfo{author}{Farah, S.}, \bibinfo{author}{Anderson, D.G.},
  \bibinfo{author}{Langer, R.}, \bibinfo{year}{2016}.
\newblock \bibinfo{title}{Physical and mechanical properties of {PLA}, and
  their functions in widespread applications — a comprehensive review}.
\newblock \bibinfo{journal}{Advanced Drug Delivery Reviews}
  \bibinfo{volume}{107}, \bibinfo{pages}{367--392}.
\newblock \URLprefix
  \url{https://www.sciencedirect.com/science/article/pii/S0169409X16302058},
  \DOIprefix\doi{https://doi.org/10.1016/j.addr.2016.06.012}.
\bibitem[{{Gault} and {Wedekind}(1978)}]{Gault_1978}
\bibinfo{author}{{Gault}, D.E.}, \bibinfo{author}{{Wedekind}, J.A.},
  \bibinfo{year}{1978}.
\newblock \bibinfo{title}{{Experimental Studies of Oblique Impact}}, in:
  \bibinfo{booktitle}{Lunar and Planetary Science Conference}, pp.
  \bibinfo{pages}{374--376}.
\bibitem[{Gilbert(1893)}]{Gilbert_1893}
\bibinfo{author}{Gilbert, G.}, \bibinfo{year}{1893}.
\newblock \bibinfo{title}{The {Moon’s} face. a study of the origin of its
  features}.
\newblock \bibinfo{journal}{Bulletin XII (Philosophical Society of Washington)}
  , \bibinfo{pages}{241--292}.
\bibitem[{Goldman and Umbanhowar(2008)}]{goldman08}
\bibinfo{author}{Goldman, D.I.}, \bibinfo{author}{Umbanhowar, P.},
  \bibinfo{year}{2008}.
\newblock \bibinfo{title}{Scaling and dynamics of sphere and disk impact into
  granular media}.
\newblock \bibinfo{journal}{Physics Review E} \bibinfo{volume}{77},
  \bibinfo{pages}{021308}.
\bibitem[{{Gudkova} et~al.(2011){Gudkova}, {Lognonn{\'e}} and
  {Gagnepain-Beyneix}}]{Gudkova_2011}
\bibinfo{author}{{Gudkova}, T.V.}, \bibinfo{author}{{Lognonn{\'e}}, P.},
  \bibinfo{author}{{Gagnepain-Beyneix}, J.}, \bibinfo{year}{2011}.
\newblock \bibinfo{title}{{Large impacts detected by the Apollo seismometers:
  Impactor mass and source cutoff frequency estimations}}.
\newblock \bibinfo{journal}{Icarus} \bibinfo{volume}{211},
  \bibinfo{pages}{1049--1065}.
\newblock \DOIprefix\doi{10.1016/j.icarus.2010.10.028}.
\bibitem[{{G\"uldermeister} and {W\"unnemann}(2017)}]{Guldermeister_2017}
\bibinfo{author}{{G\"uldermeister}, N.}, \bibinfo{author}{{W\"unnemann}, K.},
  \bibinfo{year}{2017}.
\newblock \bibinfo{title}{Quantitative analysis of impact-induced seismic
  signals by numerical modeling}.
\newblock \bibinfo{journal}{Icarus} \bibinfo{volume}{296},
  \bibinfo{pages}{15--27}.
\bibitem[{Hirabayashi et~al.(2020)Hirabayashi, Trowbridge and
  Bodewits}]{Hirabayashi_2020}
\bibinfo{author}{Hirabayashi, M.}, \bibinfo{author}{Trowbridge, A.J.},
  \bibinfo{author}{Bodewits, D.}, \bibinfo{year}{2020}.
\newblock \bibinfo{title}{The mysterious location of maryland on 2014 mu69 and
  the reconfiguration of its bilobate shape}.
\newblock \bibinfo{journal}{The Astrophysical Journal Letters}
  \bibinfo{volume}{891}, \bibinfo{pages}{L12}.
\newblock \URLprefix \url{https://dx.doi.org/10.3847/2041-8213/ab3e74},
  \DOIprefix\doi{10.3847/2041-8213/ab3e74}.
\bibitem[{Holsapple(1993)}]{holsapple93}
\bibinfo{author}{Holsapple, K.A.}, \bibinfo{year}{1993}.
\newblock \bibinfo{title}{The scaling of impact processes in planetary
  sciences}.
\newblock \bibinfo{journal}{Annual Review of Earth and Planetary Sciences}
  \bibinfo{volume}{21}, \bibinfo{pages}{333--373}.
\bibitem[{Housen and Holsapple(2011)}]{housen11}
\bibinfo{author}{Housen, K.R.}, \bibinfo{author}{Holsapple, K.A.},
  \bibinfo{year}{2011}.
\newblock \bibinfo{title}{Ejecta from impact craters}.
\newblock \bibinfo{journal}{Icarus} \bibinfo{volume}{211},
  \bibinfo{pages}{856--875}.
\bibitem[{Housen et~al.(1983)Housen, Schmidt and Holsapple}]{Housen_1983}
\bibinfo{author}{Housen, K.R.}, \bibinfo{author}{Schmidt, R.M.},
  \bibinfo{author}{Holsapple, K.A.}, \bibinfo{year}{1983}.
\newblock \bibinfo{title}{Crater ejecta scaling laws: Fundamental forms based
  on dimensional analysis}.
\newblock \bibinfo{journal}{J. Geophys. Res.} \bibinfo{volume}{88},
  \bibinfo{pages}{2485--2499}.
\bibitem[{{Keane} et~al.(2022){Keane}, {Porter}, {Beyer}, {Umurhan},
  {McKinnon}, {Moore}, {Spencer}, {Stern}, {Bierson}, {Binzel}, {Hamilton},
  {Lisse}, {Mao}, {Protopapa}, {Schenk}, {Showalter}, {Stansberry}, {White},
  {Verbiscer}, {Parker}, {Olkin}, {Weaver} and {Singer}}]{Keane_2022}
\bibinfo{author}{{Keane}, J.T.}, \bibinfo{author}{{Porter}, S.B.},
  \bibinfo{author}{{Beyer}, R.A.}, \bibinfo{author}{{Umurhan}, O.M.},
  \bibinfo{author}{{McKinnon}, W.B.}, \bibinfo{author}{{Moore}, J.M.},
  \bibinfo{author}{{Spencer}, J.R.}, \bibinfo{author}{{Stern}, S.A.},
  \bibinfo{author}{{Bierson}, C.J.}, \bibinfo{author}{{Binzel}, R.P.},
  \bibinfo{author}{{Hamilton}, D.P.}, \bibinfo{author}{{Lisse}, C.M.},
  \bibinfo{author}{{Mao}, X.}, \bibinfo{author}{{Protopapa}, S.},
  \bibinfo{author}{{Schenk}, P.M.}, \bibinfo{author}{{Showalter}, M.R.},
  \bibinfo{author}{{Stansberry}, J.A.}, \bibinfo{author}{{White}, O.L.},
  \bibinfo{author}{{Verbiscer}, A.J.}, \bibinfo{author}{{Parker}, J.W.},
  \bibinfo{author}{{Olkin}, C.B.}, \bibinfo{author}{{Weaver}, H.A.},
  \bibinfo{author}{{Singer}, K.N.}, \bibinfo{year}{2022}.
\newblock \bibinfo{title}{{The Geophysical Environment of (486958)
  Arrokoth{\textemdash}A Small Kuiper Belt Object Explored by New Horizons}}.
\newblock \bibinfo{journal}{Journal of Geophysical Research (Planets)}
  \bibinfo{volume}{127}, \bibinfo{pages}{e07068}.
\newblock \DOIprefix\doi{10.1029/2021JE007068}.
\bibitem[{{Kleinhans} et~al.(2011){Kleinhans}, {Markies}, {de Vet}, {in't Veld}
  and {Postema}}]{Kleinhans_2011}
\bibinfo{author}{{Kleinhans}, M.G.}, \bibinfo{author}{{Markies}, H.},
  \bibinfo{author}{{de Vet}, S.J.}, \bibinfo{author}{{in't Veld}, A.C.},
  \bibinfo{author}{{Postema}, F.N.}, \bibinfo{year}{2011}.
\newblock \bibinfo{title}{{Static and dynamic angles of repose in loose
  granular materials under reduced gravity}}.
\newblock \bibinfo{journal}{Journal of Geophysical Research (Planets)}
  \bibinfo{volume}{116}, \bibinfo{pages}{E11004}.
\newblock \DOIprefix\doi{10.1029/2011JE003865}.
\bibitem[{Kurosawa(2019)}]{Kurosawa_2019}
\bibinfo{author}{Kurosawa, K.}, \bibinfo{year}{2019}.
\newblock \bibinfo{title}{Impact cratering mechanics: A forward approach to
  predicting ejecta velocity distribution and transient crater radii}.
\newblock \bibinfo{journal}{Icarus} \bibinfo{volume}{317},
  \bibinfo{pages}{135--147}.
\bibitem[{{Lherm} and {Deguen}(2023)}]{Lherm_2023}
\bibinfo{author}{{Lherm}, V.}, \bibinfo{author}{{Deguen}, R.},
  \bibinfo{year}{2023}.
\newblock \bibinfo{title}{{Velocity field and cavity dynamics in drop impact
  experiments}}.
\newblock \bibinfo{journal}{Journal of Fluid Mechanics} \bibinfo{volume}{962},
  \bibinfo{pages}{A21}.
\newblock \DOIprefix\doi{10.1017/jfm.2023.297},
  \href{http://arxiv.org/abs/2305.03709}{{\tt arXiv:2305.03709}}.
\bibitem[{{Luo} et~al.(2022){Luo}, {Zhu} and {Ding}}]{Luo_2022}
\bibinfo{author}{{Luo}, X.Z.}, \bibinfo{author}{{Zhu}, M.H.},
  \bibinfo{author}{{Ding}, M.}, \bibinfo{year}{2022}.
\newblock \bibinfo{title}{{Ejecta Pattern of Oblique Impacts on the Moon From
  Numerical Simulations}}.
\newblock \bibinfo{journal}{Journal of Geophysical Research (Planets)}
  \bibinfo{volume}{127}, \bibinfo{pages}{e2022JE007333}.
\newblock \DOIprefix\doi{10.1029/2022JE007333}.
\bibitem[{{Mao} et~al.(2021){Mao}, {McKinnon}, {Singer}, {Keane}, {Beyer},
  {Greenstreet}, {Robbins}, {Schenk}, {Moore}, {Stern}, {Weaver}, {Spencer} and
  {Olkin}}]{Mao_2021}
\bibinfo{author}{{Mao}, X.}, \bibinfo{author}{{McKinnon}, W.B.},
  \bibinfo{author}{{Singer}, K.N.}, \bibinfo{author}{{Keane}, J.T.},
  \bibinfo{author}{{Beyer}, R.A.}, \bibinfo{author}{{Greenstreet}, S.},
  \bibinfo{author}{{Robbins}, S.J.}, \bibinfo{author}{{Schenk}, P.M.},
  \bibinfo{author}{{Moore}, J.M.}, \bibinfo{author}{{Stern}, S.A.},
  \bibinfo{author}{{Weaver}, H.A.}, \bibinfo{author}{{Spencer}, J.R.},
  \bibinfo{author}{{Olkin}, C.B.}, \bibinfo{year}{2021}.
\newblock \bibinfo{title}{{Collisions of Small Kuiper Belt Objects With
  (486958) Arrokoth: Implications for Its Spin Evolution and Bulk Density}}.
\newblock \bibinfo{journal}{Journal of Geophysical Research (Planets)}
  \bibinfo{volume}{126}, \bibinfo{pages}{e06961}.
\newblock \DOIprefix\doi{10.1029/2021JE006961}.
\bibitem[{Matsue et~al.(2020)Matsue, Yasui, Arakawa and Hasegawa}]{Matsue_2020}
\bibinfo{author}{Matsue, K.}, \bibinfo{author}{Yasui, M.},
  \bibinfo{author}{Arakawa, M.}, \bibinfo{author}{Hasegawa, S.},
  \bibinfo{year}{2020}.
\newblock \bibinfo{title}{Measurements of seismic waves induced by
  high-velocity impacts: Implications for seismic shaking surrounding impact
  craters on asteroids}.
\newblock \bibinfo{journal}{Icarus} \bibinfo{volume}{338},
  \bibinfo{pages}{113520}.
\bibitem[{{Maurel} et~al.(2018){Maurel}, {Michel}, {Biele}, {Ballouz} and
  {Thuillet}}]{Maurel_2018}
\bibinfo{author}{{Maurel}, C.}, \bibinfo{author}{{Michel}, P.},
  \bibinfo{author}{{Biele}, J.}, \bibinfo{author}{{Ballouz}, R.L.},
  \bibinfo{author}{{Thuillet}, F.}, \bibinfo{year}{2018}.
\newblock \bibinfo{title}{{Numerical simulations of the contact between the
  lander MASCOT and a regolith-covered surface}}.
\newblock \bibinfo{journal}{Advances in Space Research} \bibinfo{volume}{62},
  \bibinfo{pages}{2099--2124}.
\newblock \DOIprefix\doi{10.1016/j.asr.2017.05.029}.
\bibitem[{Maxwell(1977)}]{Maxwell_1977}
\bibinfo{author}{Maxwell, D.E.}, \bibinfo{year}{1977}.
\newblock \bibinfo{title}{Simple {Z} model of cratering, ejection, and the
  overturned flap}, in: \bibinfo{editor}{Roddy, D.J.}, \bibinfo{editor}{Pepin,
  R.O.}, \bibinfo{editor}{Merrill, R.B.} (Eds.), \bibinfo{booktitle}{Planetary
  and terrestrial implications: proceedings of the Symposium on Planetary
  Cratering Mechanics, Flagstaff, Arizona, September 13-17, 1976},
  \bibinfo{publisher}{New York: Pergamon Press}. pp.
  \bibinfo{pages}{1003--1008}.
\bibitem[{McGarr et~al.(1969)McGarr, Latham and Gault}]{mcgarr69}
\bibinfo{author}{McGarr, A.}, \bibinfo{author}{Latham, G.},
  \bibinfo{author}{Gault, D.}, \bibinfo{year}{1969}.
\newblock \bibinfo{title}{Meteoroid impacts as sources of seismicity on the
  {Moon}}.
\newblock \bibinfo{journal}{Journal of Geophysical Research}
  \bibinfo{volume}{74}, \bibinfo{pages}{5981--5994}.
\bibitem[{{McKinnon} et~al.(2022){McKinnon}, {Mao}, {Schenk}, {Singer},
  {Robbins}, {White}, {Beyer}, {Porter}, {Keane}, {Britt}, {Spencer}, {Grundy},
  {Moore}, {Stern}, {Weaver} and {Olkin}}]{McKinnon_2022}
\bibinfo{author}{{McKinnon}, W.B.}, \bibinfo{author}{{Mao}, X.},
  \bibinfo{author}{{Schenk}, P.M.}, \bibinfo{author}{{Singer}, K.N.},
  \bibinfo{author}{{Robbins}, S.J.}, \bibinfo{author}{{White}, O.L.},
  \bibinfo{author}{{Beyer}, R.A.}, \bibinfo{author}{{Porter}, S.B.},
  \bibinfo{author}{{Keane}, J.T.}, \bibinfo{author}{{Britt}, D.T.},
  \bibinfo{author}{{Spencer}, J.R.}, \bibinfo{author}{{Grundy}, W.M.},
  \bibinfo{author}{{Moore}, J.M.}, \bibinfo{author}{{Stern}, S.A.},
  \bibinfo{author}{{Weaver}, H.A.}, \bibinfo{author}{{Olkin}, C.B.},
  \bibinfo{year}{2022}.
\newblock \bibinfo{title}{{Snow Crash: Compaction Craters on (486958) Arrokoth
  and Other Small KBOs, With Implications}}.
\newblock \bibinfo{journal}{Geophysical Research Letters} \bibinfo{volume}{49},
  \bibinfo{pages}{e98406}.
\newblock \DOIprefix\doi{10.1029/2022GL098406}.
\bibitem[{Melosh(1985)}]{Melosh_1985}
\bibinfo{author}{Melosh, H.J.}, \bibinfo{year}{1985}.
\newblock \bibinfo{title}{Impact cratering mechanics: Relationship between the
  shock wave and excavation flow}.
\newblock \bibinfo{journal}{Icarus} \bibinfo{volume}{62},
  \bibinfo{pages}{339--343}.
\bibitem[{{Melosh}(1989)}]{Melosh_1989}
\bibinfo{author}{{Melosh}, H.J.}, \bibinfo{year}{1989}.
\newblock \bibinfo{title}{{Impact cratering : a geologic process}}.
\newblock \bibinfo{publisher}{New York: Oxford University Press}.
\bibitem[{Melosh and Ivanov(1999)}]{Melosh_1999}
\bibinfo{author}{Melosh, J.}, \bibinfo{author}{Ivanov, B.},
  \bibinfo{year}{1999}.
\newblock \bibinfo{title}{Impact crater collapse}.
\newblock \bibinfo{journal}{Annual Review of Earth and Planetary Sciences}
  \bibinfo{volume}{27}, \bibinfo{pages}{385--415}.
\bibitem[{Melosh and Pierazzo(2000)}]{Melosh_2000}
\bibinfo{author}{Melosh, J.}, \bibinfo{author}{Pierazzo, E.},
  \bibinfo{year}{2000}.
\newblock \bibinfo{title}{Understanding oblique impacts from experiments,
  observations, and modeling}.
\newblock \bibinfo{journal}{Annu. Rev. Earth Planet. Sci.} ,
  \bibinfo{pages}{141--167}.
\bibitem[{{Michikami} et~al.(2017){Michikami}, {Hagermann}, {Morota},
  {Haruyama} and {Hasegawa}}]{Michikami_2017}
\bibinfo{author}{{Michikami}, T.}, \bibinfo{author}{{Hagermann}, A.},
  \bibinfo{author}{{Morota}, T.}, \bibinfo{author}{{Haruyama}, J.},
  \bibinfo{author}{{Hasegawa}, S.}, \bibinfo{year}{2017}.
\newblock \bibinfo{title}{{Oblique impact cratering experiments in brittle
  targets: Implications for elliptical craters on the Moon}}.
\newblock \bibinfo{journal}{Planetary and Space Science} \bibinfo{volume}{135},
  \bibinfo{pages}{27--36}.
\newblock \DOIprefix\doi{10.1016/j.pss.2016.11.004}.
\bibitem[{Miklavcic et~al.(2022)Miklavcic, Askari, Sanchez, Quillen and
  Wright}]{Miklavcic_2022}
\bibinfo{author}{Miklavcic, P.M.}, \bibinfo{author}{Askari, H.},
  \bibinfo{author}{Sanchez, P.}, \bibinfo{author}{Quillen, A.C.},
  \bibinfo{author}{Wright, E.}, \bibinfo{year}{2022}.
\newblock \bibinfo{title}{Subsurface dynamics in oblique granular impacts}.
\newblock \bibinfo{journal}{Icarus} \bibinfo{volume}{385},
  \bibinfo{pages}{115089}.
\bibitem[{Miklavcic et~al.(2023)Miklavcic, Tokar, Wright, Sanchez, Glade,
  Quillen and Askari}]{Miklavcic_2023}
\bibinfo{author}{Miklavcic, P.M.}, \bibinfo{author}{Tokar, E.},
  \bibinfo{author}{Wright, E.}, \bibinfo{author}{Sanchez, P.},
  \bibinfo{author}{Glade, R.}, \bibinfo{author}{Quillen, A.},
  \bibinfo{author}{Askari, H.}, \bibinfo{year}{2023}.
\newblock \bibinfo{title}{Froude number scaling unifies impact trajectories
  into granular media across gravitational conditions}.
\newblock \bibinfo{journal}{\url{https://arxiv.org/pdf/2307.10998}} .
\bibitem[{Neiderbach et~al.(2023)Neiderbach, Suo, Wright, Quillen, Lee,
  Miklavcic, Askari and S\'anchez}]{Neiderbach_2023}
\bibinfo{author}{Neiderbach, M.}, \bibinfo{author}{Suo, B.},
  \bibinfo{author}{Wright, E.}, \bibinfo{author}{Quillen, A.},
  \bibinfo{author}{Lee, M.}, \bibinfo{author}{Miklavcic, P.},
  \bibinfo{author}{Askari, H.}, \bibinfo{author}{S\'anchez, P.},
  \bibinfo{year}{2023}.
\newblock \bibinfo{title}{Surface particle motions excited by a low velocity
  normal impact into a granular medium}.
\newblock \bibinfo{journal}{Icarus} \bibinfo{volume}{390},
  \bibinfo{pages}{115301}.
\bibitem[{Quillen et~al.(2022)Quillen, Neiderbach, Suo, South, Wright,
  Skerrett, S\'anchez, {C\'u\~nez}, Miklavcic and Askari}]{Quillen_2022}
\bibinfo{author}{Quillen, A.C.}, \bibinfo{author}{Neiderbach, M.},
  \bibinfo{author}{Suo, B.}, \bibinfo{author}{South, J.},
  \bibinfo{author}{Wright, E.}, \bibinfo{author}{Skerrett, N.},
  \bibinfo{author}{S\'anchez, P.}, \bibinfo{author}{{C\'u\~nez}, F.D.},
  \bibinfo{author}{Miklavcic, P.}, \bibinfo{author}{Askari, H.},
  \bibinfo{year}{2022}.
\newblock \bibinfo{title}{Propagation and attenuation of pulses driven by low
  velocity normal impacts in granular media}.
\newblock \bibinfo{journal}{Icarus} \bibinfo{volume}{386},
  \bibinfo{pages}{115139}.
\bibitem[{{Raducan} et~al.(2022){Raducan}, {Davison} and
  {Collins}}]{Raducan_2022}
\bibinfo{author}{{Raducan}, S.D.}, \bibinfo{author}{{Davison}, T.M.},
  \bibinfo{author}{{Collins}, G.S.}, \bibinfo{year}{2022}.
\newblock \bibinfo{title}{{Ejecta distribution and momentum transfer from
  oblique impacts on asteroid surfaces}}.
\newblock \bibinfo{journal}{Icarus} \bibinfo{volume}{374},
  \bibinfo{pages}{114793}.
\newblock \DOIprefix\doi{10.1016/j.icarus.2021.114793},
  \href{http://arxiv.org/abs/2105.01474}{{\tt arXiv:2105.01474}}.
\bibitem[{Rivkin et~al.(2021)Rivkin, Chabot, Stickle, Thomas, Richardson,
  Barnouin, Fahnestock, Ernst, Cheng, Chesley, Naidu, Statler, Barbee, Agrusa,
  Moskovitz, Daly, Pravec, Scheirich, Dotto, Corte, Michel, K{\"u}ppers,
  Atchison and Hirabayashi}]{Rivkin_2021}
\bibinfo{author}{Rivkin, A.S.}, \bibinfo{author}{Chabot, N.L.},
  \bibinfo{author}{Stickle, A.M.}, \bibinfo{author}{Thomas, C.A.},
  \bibinfo{author}{Richardson, D.C.}, \bibinfo{author}{Barnouin, O.},
  \bibinfo{author}{Fahnestock, E.G.}, \bibinfo{author}{Ernst, C.M.},
  \bibinfo{author}{Cheng, A.F.}, \bibinfo{author}{Chesley, S.},
  \bibinfo{author}{Naidu, S.}, \bibinfo{author}{Statler, T.S.},
  \bibinfo{author}{Barbee, B.}, \bibinfo{author}{Agrusa, H.},
  \bibinfo{author}{Moskovitz, N.}, \bibinfo{author}{Daly, R.T.},
  \bibinfo{author}{Pravec, P.}, \bibinfo{author}{Scheirich, P.},
  \bibinfo{author}{Dotto, E.}, \bibinfo{author}{Corte, V.D.},
  \bibinfo{author}{Michel, P.}, \bibinfo{author}{K{\"u}ppers, M.},
  \bibinfo{author}{Atchison, J.}, \bibinfo{author}{Hirabayashi, M.},
  \bibinfo{year}{2021}.
\newblock \bibinfo{title}{The double asteroid redirection test ({DART}):
  Planetary defense investigations and requirements}.
\newblock \bibinfo{journal}{The Planetary Science Journal} \bibinfo{volume}{2},
  \bibinfo{pages}{173}.
\newblock \DOIprefix\doi{10.3847/psj/ac063e}.
\bibitem[{{S{\'a}nchez} et~al.(2022){S{\'a}nchez}, {Scheeres} and
  {Quillen}}]{Sanchez_2022}
\bibinfo{author}{{S{\'a}nchez}, P.}, \bibinfo{author}{{Scheeres}, D.J.},
  \bibinfo{author}{{Quillen}, A.C.}, \bibinfo{year}{2022}.
\newblock \bibinfo{title}{{Transmission of a Seismic Wave Generated by Impacts
  on Granular Asteroids}}.
\newblock \bibinfo{journal}{Planetary Science Journal} \bibinfo{volume}{3},
  \bibinfo{pages}{245}.
\newblock \DOIprefix\doi{10.3847/PSJ/ac960c},
  \href{http://arxiv.org/abs/2209.11353}{{\tt arXiv:2209.11353}}.
\bibitem[{{Scheeres} et~al.(2010){Scheeres}, {Hartzell}, {S{\'a}nchez} and
  {Swift}}]{Scheeres_2010}
\bibinfo{author}{{Scheeres}, D.J.}, \bibinfo{author}{{Hartzell}, C.M.},
  \bibinfo{author}{{S{\'a}nchez}, P.}, \bibinfo{author}{{Swift}, M.},
  \bibinfo{year}{2010}.
\newblock \bibinfo{title}{{Scaling forces to asteroid surfaces: The role of
  cohesion}}.
\newblock \bibinfo{journal}{Icarus} \bibinfo{volume}{210},
  \bibinfo{pages}{968--984}.
\newblock \DOIprefix\doi{10.1016/j.icarus.2010.07.009},
  \href{http://arxiv.org/abs/1002.2478}{{\tt arXiv:1002.2478}}.
\bibitem[{Shoemaker(1961)}]{shoemaker_1961}
\bibinfo{author}{Shoemaker, E.M.}, \bibinfo{year}{1961}.
\newblock \bibinfo{title}{Interpretation of {Lunar} craters}.
  \bibinfo{publisher}{Academic Press, Elsevier Inc.}.
  chapter~\bibinfo{chapter}{8}.
\newblock pp. \bibinfo{pages}{283--359}.
\bibitem[{{Spencer} et~al.(2020){Spencer}, {Stern}, {Moore}, {Weaver},
  {Singer}, {Olkin}, {Verbiscer}, {McKinnon}, {Parker}, {Beyer}, {Keane},
  {Lauer}, {Porter}, {White}, {Buratti}, {El-Maarry}, {Lisse}, {Parker},
  {Throop}, {Robbins}, {Umurhan}, {Binzel}, {Britt}, {Buie}, {Cheng},
  {Cruikshank}, {Elliott}, {Gladstone}, {Grundy}, {Hill}, {Horanyi},
  {Jennings}, {Kavelaars}, {Linscott}, {McComas}, {McNutt}, {Protopapa},
  {Reuter}, {Schenk}, {Showalter}, {Young}, {Zangari}, {Abedin},
  {Beddingfield}, {Benecchi}, {Bernardoni}, {Bierson}, {Borncamp}, {Bray},
  {Chaikin}, {Dhingra}, {Fuentes}, {Fuse}, {Gay}, {Gwyn}, {Hamilton},
  {Hofgartner}, {Holman}, {Howard}, {Howett}, {Karoji}, {Kaufmann}, {Kinczyk},
  {May}, {Mountain}, {P{\"a}tzold}, {Petit}, {Piquette}, {Reid}, {Reitsema},
  {Runyon}, {Sheppard}, {Stansberry}, {Stryk}, {Tanga}, {Tholen}, {Trilling}
  and {Wasserman}}]{Spencer_2020}
\bibinfo{author}{{Spencer}, J.R.}, \bibinfo{author}{{Stern}, S.A.},
  \bibinfo{author}{{Moore}, J.M.}, \bibinfo{author}{{Weaver}, H.A.},
  \bibinfo{author}{{Singer}, K.N.}, \bibinfo{author}{{Olkin}, C.B.},
  \bibinfo{author}{{Verbiscer}, A.J.}, \bibinfo{author}{{McKinnon}, W.B.},
  \bibinfo{author}{{Parker}, J.W.}, \bibinfo{author}{{Beyer}, R.A.},
  \bibinfo{author}{{Keane}, J.T.}, \bibinfo{author}{{Lauer}, T.R.},
  \bibinfo{author}{{Porter}, S.B.}, \bibinfo{author}{{White}, O.L.},
  \bibinfo{author}{{Buratti}, B.J.}, \bibinfo{author}{{El-Maarry}, M.R.},
  \bibinfo{author}{{Lisse}, C.M.}, \bibinfo{author}{{Parker}, A.H.},
  \bibinfo{author}{{Throop}, H.B.}, \bibinfo{author}{{Robbins}, S.J.},
  \bibinfo{author}{{Umurhan}, O.M.}, \bibinfo{author}{{Binzel}, R.P.},
  \bibinfo{author}{{Britt}, D.T.}, \bibinfo{author}{{Buie}, M.W.},
  \bibinfo{author}{{Cheng}, A.F.}, \bibinfo{author}{{Cruikshank}, D.P.},
  \bibinfo{author}{{Elliott}, H.A.}, \bibinfo{author}{{Gladstone}, G.R.},
  \bibinfo{author}{{Grundy}, W.M.}, \bibinfo{author}{{Hill}, M.E.},
  \bibinfo{author}{{Horanyi}, M.}, \bibinfo{author}{{Jennings}, D.E.},
  \bibinfo{author}{{Kavelaars}, J.J.}, \bibinfo{author}{{Linscott}, I.R.},
  \bibinfo{author}{{McComas}, D.J.}, \bibinfo{author}{{McNutt}, R.L.},
  \bibinfo{author}{{Protopapa}, S.}, \bibinfo{author}{{Reuter}, D.C.},
  \bibinfo{author}{{Schenk}, P.M.}, \bibinfo{author}{{Showalter}, M.R.},
  \bibinfo{author}{{Young}, L.A.}, \bibinfo{author}{{Zangari}, A.M.},
  \bibinfo{author}{{Abedin}, A.Y.}, \bibinfo{author}{{Beddingfield}, C.B.},
  \bibinfo{author}{{Benecchi}, S.D.}, \bibinfo{author}{{Bernardoni}, E.},
  \bibinfo{author}{{Bierson}, C.J.}, \bibinfo{author}{{Borncamp}, D.},
  \bibinfo{author}{{Bray}, V.J.}, \bibinfo{author}{{Chaikin}, A.L.},
  \bibinfo{author}{{Dhingra}, R.D.}, \bibinfo{author}{{Fuentes}, C.},
  \bibinfo{author}{{Fuse}, T.}, \bibinfo{author}{{Gay}, P.L.},
  \bibinfo{author}{{Gwyn}, S.D.J.}, \bibinfo{author}{{Hamilton}, D.P.},
  \bibinfo{author}{{Hofgartner}, J.D.}, \bibinfo{author}{{Holman}, M.J.},
  \bibinfo{author}{{Howard}, A.D.}, \bibinfo{author}{{Howett}, C.J.A.},
  \bibinfo{author}{{Karoji}, H.}, \bibinfo{author}{{Kaufmann}, D.E.},
  \bibinfo{author}{{Kinczyk}, M.}, \bibinfo{author}{{May}, B.H.},
  \bibinfo{author}{{Mountain}, M.}, \bibinfo{author}{{P{\"a}tzold}, M.},
  \bibinfo{author}{{Petit}, J.M.}, \bibinfo{author}{{Piquette}, M.R.},
  \bibinfo{author}{{Reid}, I.N.}, \bibinfo{author}{{Reitsema}, H.J.},
  \bibinfo{author}{{Runyon}, K.D.}, \bibinfo{author}{{Sheppard}, S.S.},
  \bibinfo{author}{{Stansberry}, J.A.}, \bibinfo{author}{{Stryk}, T.},
  \bibinfo{author}{{Tanga}, P.}, \bibinfo{author}{{Tholen}, D.J.},
  \bibinfo{author}{{Trilling}, D.E.}, \bibinfo{author}{{Wasserman}, L.H.},
  \bibinfo{year}{2020}.
\newblock \bibinfo{title}{{The geology and geophysics of Kuiper Belt object
  (486958) Arrokoth}}.
\newblock \bibinfo{journal}{Science} \bibinfo{volume}{367},
  \bibinfo{pages}{aay3999}.
\newblock \DOIprefix\doi{10.1126/science.aay3999},
  \href{http://arxiv.org/abs/2004.00727}{{\tt arXiv:2004.00727}}.
\bibitem[{{Takizawa} and {Katsuragi}(2020)}]{Takizawa_2020}
\bibinfo{author}{{Takizawa}, S.}, \bibinfo{author}{{Katsuragi}, H.},
  \bibinfo{year}{2020}.
\newblock \bibinfo{title}{{Scaling laws for the oblique impact cratering on an
  inclined granular surface}}.
\newblock \bibinfo{journal}{Icarus} \bibinfo{volume}{335},
  \bibinfo{pages}{113409}.
\newblock \DOIprefix\doi{10.1016/j.icarus.2019.113409},
  \href{http://arxiv.org/abs/1904.11636}{{\tt arXiv:1904.11636}}.
\bibitem[{{Thuillet} et~al.(2021){Thuillet}, {Zhang}, {Michel}, {Biele},
  {Kameda}, {Sugita}, {Tatsumi}, {Schwartz} and {Ballouz}}]{Thuillet_2021}
\bibinfo{author}{{Thuillet}, F.}, \bibinfo{author}{{Zhang}, Y.},
  \bibinfo{author}{{Michel}, P.}, \bibinfo{author}{{Biele}, J.},
  \bibinfo{author}{{Kameda}, S.}, \bibinfo{author}{{Sugita}, S.},
  \bibinfo{author}{{Tatsumi}, E.}, \bibinfo{author}{{Schwartz}, S.R.},
  \bibinfo{author}{{Ballouz}, R.L.}, \bibinfo{year}{2021}.
\newblock \bibinfo{title}{{Numerical modeling of lander interaction with a
  low-gravity asteroid regolith surface. II. Interpreting the successful
  landing of Hayabusa2 MASCOT}}.
\newblock \bibinfo{journal}{Astronomy \& Astrophysics,} \bibinfo{volume}{648},
  \bibinfo{pages}{A56}.
\newblock \DOIprefix\doi{10.1051/0004-6361/201936128}.
\bibitem[{Tsimring and Volfson(2005)}]{tsimring05}
\bibinfo{author}{Tsimring, L.}, \bibinfo{author}{Volfson, D.},
  \bibinfo{year}{2005}.
\newblock \bibinfo{title}{Modelling of impact cratering in granular media}.
\newblock \bibinfo{journal}{Powder Technology} \bibinfo{volume}{2},
  \bibinfo{pages}{1215--1223}.
\bibitem[{Turtle et~al.(2005)Turtle, Pierazzo, Collins, Osinski, Melosh, Morgan
  and Reimold}]{Turtle_2005}
\bibinfo{author}{Turtle, E.}, \bibinfo{author}{Pierazzo, E.},
  \bibinfo{author}{Collins, G.}, \bibinfo{author}{Osinski, G.},
  \bibinfo{author}{Melosh, H.}, \bibinfo{author}{Morgan, J.},
  \bibinfo{author}{Reimold, W.}, \bibinfo{year}{2005}.
\newblock \bibinfo{title}{{Impact structures: What does crater diameter
  mean?}}, in: \bibinfo{booktitle}{{Large Meteorite Impacts III}}.
  \bibinfo{publisher}{Geological Society of America}. volume
  \bibinfo{volume}{Special Paper 384}, pp. \bibinfo{pages}{1--24}.
\newblock \URLprefix \url{https://doi.org/10.1130/0-8137-2384-1.1},
  \DOIprefix\doi{10.1130/0-8137-2384-1.1}.
\bibitem[{{Wallis} et~al.(2005){Wallis}, {Burchell}, {Cook}, {Solomon} and
  {McBride}}]{Wallis_2005}
\bibinfo{author}{{Wallis}, D.}, \bibinfo{author}{{Burchell}, M.J.},
  \bibinfo{author}{{Cook}, A.C.}, \bibinfo{author}{{Solomon}, C.J.},
  \bibinfo{author}{{McBride}, N.}, \bibinfo{year}{2005}.
\newblock \bibinfo{title}{{Azimuthal impact directions from oblique impact
  crater morphology}}.
\newblock \bibinfo{journal}{Monthly Notices of the Royal Astronomical Society}
  \bibinfo{volume}{359}, \bibinfo{pages}{1137--1149}.
\newblock \DOIprefix\doi{10.1111/j.1365-2966.2005.08978.x}.
\bibitem[{Wang et~al.(2012)Wang, Ye and Zheng}]{Wang_2012}
\bibinfo{author}{Wang, D.}, \bibinfo{author}{Ye, X.}, \bibinfo{author}{Zheng,
  X.}, \bibinfo{year}{2012}.
\newblock \bibinfo{title}{The scaling and dynamics of a projectile obliquely
  impacting a granular medium}.
\newblock \bibinfo{journal}{The European Physical Journal E}
  \bibinfo{volume}{35}, \bibinfo{pages}{7}.
\bibitem[{Wright et~al.(2022)Wright, Quillen, Sanchez, Schwartz, Nakajima,
  Askari and Miklavcic}]{Wright_2022}
\bibinfo{author}{Wright, E.}, \bibinfo{author}{Quillen, A.C.},
  \bibinfo{author}{Sanchez, P.}, \bibinfo{author}{Schwartz, S.R.},
  \bibinfo{author}{Nakajima, M.}, \bibinfo{author}{Askari, H.},
  \bibinfo{author}{Miklavcic, P.}, \bibinfo{year}{2022}.
\newblock \bibinfo{title}{Ricochets on asteroids {II}: Sensitivity of
  laboratory experiments of low velocity grazing impacts on substrate grain
  size}.
\newblock \bibinfo{journal}{Icarus} \bibinfo{volume}{376},
  \bibinfo{pages}{114868}.
\bibitem[{Wright et~al.(2020)Wright, Quillen, South, Nelson, S{\'{a}}nchez,
  Siu, Askari, Nakajima and Schwartz}]{Wright_2020b}
\bibinfo{author}{Wright, E.}, \bibinfo{author}{Quillen, A.C.},
  \bibinfo{author}{South, J.}, \bibinfo{author}{Nelson, R.C.},
  \bibinfo{author}{S{\'{a}}nchez, P.}, \bibinfo{author}{Siu, J.},
  \bibinfo{author}{Askari, H.}, \bibinfo{author}{Nakajima, M.},
  \bibinfo{author}{Schwartz, S.R.}, \bibinfo{year}{2020}.
\newblock \bibinfo{title}{Ricochets on asteroids: Experimental study of low
  velocity grazing impacts into granular media}.
\newblock \bibinfo{journal}{Icarus} \bibinfo{volume}{351},
  \bibinfo{pages}{113963}.
\newblock \DOIprefix\doi{10.1016/j.icarus.2020.113963}.
\bibitem[{{Yamamoto} et~al.(2006){Yamamoto}, {Wada}, {Okabe} and
  {Matsui}}]{Yamamoto_2006}
\bibinfo{author}{{Yamamoto}, S.}, \bibinfo{author}{{Wada}, K.},
  \bibinfo{author}{{Okabe}, N.}, \bibinfo{author}{{Matsui}, T.},
  \bibinfo{year}{2006}.
\newblock \bibinfo{title}{{Transient crater growth in granular targets: An
  experimental study of low velocity impacts into glass sphere targets}}.
\newblock \bibinfo{journal}{Icarus} \bibinfo{volume}{183},
  \bibinfo{pages}{215--224}.
\newblock \DOIprefix\doi{10.1016/j.icarus.2006.02.002}.
\bibitem[{Yasui et~al.(2015)Yasui, Matsumoto and Arakawa}]{yasui15}
\bibinfo{author}{Yasui, M.}, \bibinfo{author}{Matsumoto, E.},
  \bibinfo{author}{Arakawa, M.}, \bibinfo{year}{2015}.
\newblock \bibinfo{title}{Experimental study on impact-induced seismic wave
  propagation through granular materials}.
\newblock \bibinfo{journal}{Icarus} \bibinfo{volume}{260},
  \bibinfo{pages}{320--331}.

\end{thebibliography}

\if \ispreprint 1
\appendix
\section{Nomenclature}

\begin{table*}[ht]
    \centering
        \caption{Nomenclature}
    \begin{tabular}{lll}
    \hline
      Impact velocity & $v_{imp}$ \\
      Impact angle & $\theta_I$ \\
      Projectile mass & $M_p$ \\
      Projectile radius & $R_p$  \\
      Substrate bulk density & $\rho_s$\\
      Projectile density & $\rho_p$ \\
      Distance from site of impact & $r$ \\
      Crater volume & $V_{cr}$ \\
    Cartesian coordinates & $(x,y,z)$ \\
      Cylindrical coordinates & $(R,\phi,z)$ \\
      Spherical coordinates & $(r,\vartheta,\phi)$ \\
      Dimensionless scaling parameters & $\pi_2$, $\pi_3$, $\pi_4$, $\pi_V$ \\
      Froude number & $Fr$ \\
      Gravitational acceleration (Earth) & $g$ \\
      Crater formation time & $\tau_{ex} $ \\ 
    Crater semi-major, semi-minor axes & $a_{cr}, b_{cr}$ \\
       Ricochet angle & $\theta_{Ric}$ \\
       Ricochet velocity & $v_{Ric}$ \\
       Surface slope & $s$ \\
       Distance between impact site and crater center & $d_{ai}$ \\
       Maxwell model exponent & $Z$ \\
       Ejecta angle & $\vartheta_{ej}$\\
       Subsurface velocity vector & ${\bf u}$ \\ 
       Subsurface acceleration vector & ${\bf a}$ \\
        Peak velocity magnitude & $|v|_{pk}$ \\
       Peak acceleration magnitude  & $|a|_{pk}$ \\
       The horizontal component of momentum lost by projectile & $\Delta p_x$ \\
       Change in the vertical component of momentum of projectile & $
       \Delta p_z$  \\
       Kinetic energy lost by projectile & $\Delta E$\\
         \hline
    \end{tabular}
    \label{tab:nomen}
\end{table*}
\fi

\end{document}